\documentclass[12pt]{article}

\usepackage{natbib}
\usepackage[english]{babel}
\usepackage[utf8]{inputenc}
\usepackage[colorinlistoftodos]{todonotes}
\usepackage{amsthm}

\usepackage[top=1.3in, bottom=1.3in, left=1.1in, right=1.1in]{geometry}
\usepackage{amsmath}
\usepackage{mathtools}
\usepackage{booktabs}
\usepackage{amssymb}
\usepackage{array}
\usepackage{graphicx}
\usepackage{listings}
\usepackage{tikz}
\usepackage{enumerate}
\usepackage{enumitem}
\usepackage{bbm}
\usepackage{multirow}
\usepackage[bottom]{footmisc}
\usepackage{geometry}

\usepackage[T1]{fontenc}
\usepackage{lipsum}
\usepackage{multicol}
\usepackage{changepage}

\usepackage[hidelinks, hypertexnames=false]{hyperref} 

\usepackage[12pt]{moresize}
\usepackage{caption}
\usepackage{subfig}

\usepackage[]{titlesec}

\newtheorem{theorem}{Theorem}[section]
\newtheorem{lemma}[theorem]{Lemma}
\newtheorem{proposition}[theorem]{Proposition}
\newtheorem{corollary}[theorem]{Corollary}
\newtheorem{assumption}{Assumption}

\newtheorem{definition}{Definition}

\newcommand{\indep}{\perp \!\!\! \perp}

\usepackage{setspace}

\author{Th\'{e}o Durandard}

\title{\centering {\scshape Dynamic delegation in promotion contests}\thanks{Email: theo.durandard@kellogg.northwestern.edu. I am deeply grateful to Luis Rayo and Bruno Strulovici for constant guidance and support. I additionally thank Daniel Barron, Isaias Chaves, Alessandro Pavan, and Eran Shmaya for many detailed comments and suggestions. This project has also benefited from conversations with Nemanja Antic, Matteo Camboni, Henrique Castro-Pires, Piotr Dworczak, Andres Espitia, Benjamin Friedrich, Yingni Guo, Gaston Illanes, Harry Pei, Mike Powell, Federico Puglisi, James Schummer, Ludvig Sinander, Caroline Thomas, Udayan Vaidya, Boli Xu, and the comments of seminar participants at Northwestern. All mistakes are my own.}\\
	\vspace{15pt}\href{https://www.dropbox.com/s/hv2ousb8szt2gd3/JMP_Theo_Durandard.pdf?dl=0}{\textcolor{blue}{Click here for the latest version.}}
}
\date{\today}

\begin{document}
	
	\maketitle
	
	\begin{abstract}
		I study how organizations assign tasks to identify the best candidate to promote among a pool of workers. Task allocation and workers' motivation interact through the organization's promotion decisions. The organization designs the workers' careers to both screen and develop talent. When only non-routine tasks are informative about a worker's type and non-routine tasks are scarce, the organization's preferred promotion system is an \emph{index contest}. Each worker is assigned a number that depends only on his \emph{own type}. The principal delegates the non-routine task to the worker whose current index is the highest and promotes the first worker whose type exceeds a threshold. Each worker's threshold is independent of the other workers' types. Competition is mediated by the allocation of tasks: who gets the opportunity to prove themselves is a determinant factor in promotions. Finally, features of the optimal promotion contest rationalize the prevalence of fast-track promotion, the role of seniority, or when a group of workers is systemically advantaged.
	\end{abstract}

	\section*{Introduction}\label{sec:Introduction}
	
	\paragraph{} Matching tasks to the right workers is crucial to an organization's success. First, productive efficiency requires that more talented workers perform more complex, non-routine tasks. Second, workers' success in their current tasks is informative: Organizations also allocate tasks to learn about the workers and improve future matches. Third, a worker's assignment determines what he learns on the job. Assigning the right worker to the right task is then especially important when the organization seeks to identify and develop talented workers.\footnote{Former Xerox CEO Anne Mulcahy insists in \cite{mulcahy2010how} that it is crucial to identify candidates for promotion early and give them ``developmental responsibilities'' to develop strong workers and test their abilities.} Non-routine tasks are opportunities for workers to prove themselves. Workers understand that their career trajectories in the organization depend on the opportunities to showcase their talents. Task allocation and workers' motivation then interact through the organization's promotion decisions. Designing a good promotion system is thus both challenging and essential for the organization's success.\footnote{For example, \cite{rosen1982authority} insists on the importance of selecting the right person to lead an organization as they set its course, and their decisions are ``magnifies'' many times. \cite{rohmanHowyoupromote2018} note that when employees believe that promotion decisions are efficient, they are more likely to exert effort and follow the organization's leaders' directions and recommendations. The same authors also point out that ``stock returns are nearly three times the market average, voluntary turnover is half that of industry peers, and metrics for innovation, productivity, and growth consistently outperform competitors'' at companies that manage promotion effectively.} 
	
	A sound promotion system must include the task allocation process and the promotion rule. It must also balance exploitation (delegating non-routine tasks to a worker known to be good and eventually promoting him) and exploration (giving responsibilities to new workers). Ignoring exploration, one misses an essential part of the story: who gets the opportunity to prove themselves is a determining factor in promotions. In this paper, I ask how the organization optimally designs the promotion system to motivate workers and tackle the exploration-exploitation trade-off. I also address the following questions: How does incentive provision affect the allocation of tasks and the promotion decision? Can the allocation of opportunities exacerbate initial differences to induce significant disparities over time? What characteristics of a worker increase his chance of being promoted? 
	
	I explore these questions in a \emph{centralized dynamic contest model}. A principal (she) has one prize to award, the promotion. She decides how to allocate a non-routine task sequentially to $N$ workers (he/they) and when to give the prize. Each worker has a type represented by a stochastic process, and the processes are independent. When the principal allocates the non-routine task to a worker and the worker exerts effort, the principal gets a reward that depends on the worker's type. The worker's type also evolves, and the evolution of types could reflect the principal learning about the worker's or the worker acquiring new skills on the job. The other workers are assigned uninformative routine tasks, and their types remain frozen. Finally, the principal can only use the promise of future promotion to motivate the workers. In particular, in the baseline model, I rule out transfers to focus on the interaction between the two classical and conflicting\footnote{As famously illustrated by the Peter's principle described in \cite{peter1969peter}.} purposes of promotions: to ``assign people to the roles where they can contribute the most to the organization's performance'' and to provide ``incentives and rewards'' (\cite{roberts1992economics}).
	
	My model builds on the canonical multi-armed bandit model but departs in one critical aspect: the arms are workers who exert effort. Hence the arms are strategic. In the classic bandit model, when the decision maker pulls an arm, she gets a reward drawn stochastically from some fixed distribution. In some contexts, this assumption fits the behavior of the problem inputs: In clinical trials, it is natural to think that the patients will comply. However, in my problem, each arm corresponds to a worker whose incentives differ from the principal's. So, the arms are \emph{strategic}. When the principal allocates the non-routine task, her reward and the flow of information are controlled by the workers' choices of effort, and the workers exert effort only when compensated for it by the promise of future promotion. To incentivize effort, the principal must eventually promote a worker, stopping exploration.
	
	\paragraph{} In this setting, I characterize the principal-optimal promotion contest. Solving the principal's problem is challenging for three reasons. As in bandit models, the first one is the problem's dimensionality. The set of feasible promotion contests is large. Second, the principal's promotion decision can depend on all workers' types and their effort histories. So, it introduces a degree of dependence among the workers that complicates the problem.\footnote{The stoppable bandit models studied in the literature were so under the (exogenous) restriction that the decision to freeze an arm can only depend on the state of this arm. See, for example, \cite{glazebrook1979stoppable} or \cite{fershtman2022searching}.}\label{footnote:indexabilitystoppablebanditsmodel} Finally, the workers are strategic. Not all delegation and promotion rules incentivize effort. The set of ``implementable'' promotion contests is complex.
	
	Nevertheless, the optimal contest is simple. I prove that, as in the canonical model, \emph{indexability} holds: The optimal promotion contest takes the form of an \emph{index contest}. Each worker is assigned a number (his \emph{index}) that depends only on his type and the cost of incentive provision. The principal sequentially delegates the non-routine task to the worker whose \emph{index} is the highest. Eventually, she promotes the \emph{first} worker whose type exceeds a threshold. Both the worker's index and promotion threshold are \emph{independent} of the successes and failures of the other workers. Finally, I also show that it is optimal to promote one worker only when the principal can design the prize-sharing rule, i.e., decide to promote multiple workers. The optimal contest is a winner-take-all contest.
	
	\paragraph{} In the index contest, the delegation rule mediates the competition for promotion between the workers. To understand the determinants of promotions, it is crucial to consider the factors that affect the allocation and timing of opportunities. This has two significant consequences: (i) for the contractual arrangements between the principal and the workers and (ii) for the effect of initial differences on promotion decisions (especially when thinking about discrimination in promotion practices). 
	
	First, no mention of competition needs to appear in the contractual arrangement between each worker and the principal. One interpretation of the index contest is the following. The principal successively offers short-term individual trial contracts to one worker at a time. Each trial contract specifies a target and a (potentially stochastic) deadline. The worker gets the promotion if he achieves the target before the deadline. Otherwise, the manager offers a new trial contract to one of the other workers until a worker eventually succeeds. The trial contracts do not rely on relative performance measures: they are independent of what the other workers do \footnote{This appears consistent with some evidence that contracts and promotion guidelines rarely mention relative performance explicitly. For example, both \cite{bretz1989comparing} and \cite{bretz1992current} find that less than a third of organization uses rankings explicitly. Even among organizations that use ranking measures, they generally "supplement other performance appraisal methods" such as the management by objectives approach (that relies on absolute performance, see \cite{drucker1954f}). More recently, \cite{campbell2008nonfinancial} provides evidence that promotion decisions are made on an absolute measure of performance in the fast food industry. Finally, anecdotal evidence suggests that organizations that used explicit ranking appear to be abandoning it, see \cite{oconnor_2021_why}. Unfortunately, more evidence and data on contracts and promotion guidelines are seldom available.} The principal uses contracts that incentivize workers separately: the promotion thresholds do not condition on indicators of relative performance (on the other workers' types). This may be surprising: why would not the principal use relative performance to compare the candidates and select the best of them? However, it should not be. Who gets the opportunity already summarizes the relevant ranking information. In the \emph{index contest}, other workers' efforts and successes only affect the likelihood that the worker will be delegated the non-routine task and, hence, get the chance to prove themselves. It is irrelevant to the promotion decision once the worker gets the opportunity.
	
	Second, the principal delegates the non-routine task sequentially to the workers in the index contest. This generates significant path dependence in promotion decisions: a worker who is not given a chance initially may never get the opportunity to prove himself, hence, will not be promoted. In particular, early successes have an outsized impact on the probability of promotion. They increase both the likelihood of being promoted before any of the other workers gets the opportunity to showcase their talent and the likelihood that the worker is allocated the non-routine task again later. One should therefore be careful where to look to identify discrimination in promotion practices. In particular, a firm may always promote the most qualified candidate and yet discriminates. That is because the principal may also discriminate through the allocation of opportunities. In the context of my model, the index delegation rule may treat different groups very differently. For example, minor differences among workers in learning speed or the cost of effort may lead the principal to delegate to one first. In \emph{reinforcing environments},\footnote{This includes bad news Poisson learning or realistic representations of on-the-job learning. See Definition \ref{def:reinforcingenvironments} in Section \ref{sec:Applications}.} this dramatically affects their promotion chance and expected time to promotion. There, the early assignment of the non-routine task largely determines the promotion decision. If the principal delegation rule is biased toward one group, workers from the other group will never get an opportunity to be promoted. Moreover, at the time of promotion, they will also appear less qualified than the promoted worker. Their type will be lower than that of the promoted worker, and they will not have worked on non-routine tasks as much. This is an instance of what has been described outside of economics as systemic discrimination: discrimination based on systemic group differences in observable characteristics or treatment (see \cite{bohren2022systemic} for a treatment of systemic discrimination within economics).
	
	I also obtain further predictions for organizational design from my characterization. A notable feature of the index contest is that the promotion thresholds decrease over time. So, a worker's type, when promoted, decreases with time. A first consequence is that fast tracks (i.e., that a quickly promoted worker often gets another promotion soon after, see \cite{baker1994internal} and \cite{ariga1999fast}) should then not be surprising. When a worker is promoted quickly, his type upon promotion is higher. Hence, he starts from a better place when entering a potential new promotion contest at the next level. So, his expected time to promotion decreases: the worker is expected to be promoted again soon. Second, the type of a worker at the time of promotion decreases the longer he stays in his current position. So, faster-promoted workers should perform better upon promotion than more slowly-promoted workers. Third, in the \emph{index contest}, seniority is not explicit but still confers an advantage. It becomes easier to be promoted for a worker as time passes. The \emph{index contest} backloads incentives, implicitly putting weight on seniority.\footnote{Seniority has been widely used as an explicit promotion criterion (especially in public administrations), and, although it has fallen from favor since the 1980s, it is still seen as an important determinant of promotions, see \cite{dobson1988seniority} or \cite{phelan2001promotion}.}
	
	\paragraph{} Finally, I study multiple extensions: I relax some of the assumptions made in the model. I show that the winner-take-all index context is optimal when the principal can design the prize. I consider the possibility of transfers, and I study different information structures. In my setting, if transfers are unrestricted, the manager can incentivize effort at no cost, and the first best is achieved. However, if wages only depend on the workers' current types and the workers are protected by limited liability, the principal cannot freely punish a worker who decides to shirk. So, intertemporal distortions like the ones absent transfers are reintroduced, and the \emph{index contest} (with adapted indices and promotion thresholds) is optimal. Secondly, in the baseline model, information is symmetric. All the players observe the types of all workers. If only the workers observe their types and can credibly communicate them to the principal, the \emph{index contest} is still implementable (and optimal). In particular, workers do not need to observe who has been in charge of the project and how successful other workers were.
	
	\paragraph{} Besides promotion decisions, my results apply to various problems in which a principal owns an asset and wants to allocate it to the best agent among a pool of candidates of unknown ability. This includes outsourcing and procurement decisions by firms, a venture capitalist's investment decision between multiple start-ups, or the CERN research board deciding which team of researchers can use the colliders and when an experiment should be abandoned, for example. When the principal earns rewards and learns by delegating the asset, the optimal mechanism is an \emph{index contest}. 
	
	\paragraph{} The rest of the article is organized as follows. The related literature is discussed in the remainder of the introduction. In Section \ref{sec:Model}, I formally describe the environment. In Section \ref{sec:MainResults}, I introduce the \emph{index contest} and presents its properties. In Section \ref{sec:Applications}, I study the implications of my findings on discrimination. In Section \ref{sec:Proofofmaintheorem}, I present an outline of the proof of my main result: the optimality of the \emph{index contest}. In Section \ref{sec:extensions}, I consider extensions. I conclude in Section \ref{sec:conclusion} with a brief discussion of the results and lines of future research. All proofs not in the main text are in the Appendices \ref{app:appendixA}, \ref{app:appendixB}, and Online Appendix \ref{app:onlineapp}.
	
	\paragraph{Literature:} This paper studies a dynamic contest for experimentation and characterizes the optimal promotion contest when the principal can control the learning process. I then use this model to study a big under-studied topic in personnel economics: how the allocation of tasks affects promotions and worker careers. So, my paper builds on several streams of literature. 
	
	First, as mentioned above, I build on the canonical bandit problem solved in \cite{gittins1974dynamic}. \cite{gittins2011multi} offer a textbook treatment. \cite{bergemann2006bandit} is a good survey on bandit problems (in economics). Here, the authors solve for the optimal delegation rule when the arms are passive. Other papers in this literature consider learning about multiple alternatives before making an (irreversible) decision. Examples includes \cite{austen2018optimal}, \cite{fudenberg2018speed}, \cite{ke2019optimal}, \cite{ke2016search}, and \cite{che2019optimal}. More closely related is the study of stoppable bandit models in \cite{glazebrook1979stoppable}. Glazebrook considers a multi-armed bandit model in which the decision maker decides which arm to pull every period but can also choose to freeze an arm and play it forever. He gives conditions under which indexability is preserved. Again, all these papers are concerned with a decision problem: the arms or alternatives are passive, and they study the optimal way to allocate attention before making a decision absent incentive consideration. This is fundamentally different from my model. I am interested in how organizations allocate tasks when the arms are \emph{strategic}. To solve this problem, I identify a new condition under which indexability holds in bandit superprocess problems. I then use this condition to show that the problem's separability is preserved. Despite the strategic nature of the problem, the optimal delegation rule is an index rule.
	
	A few other papers also look at bandit problems with strategic arms. The key distinction between these papers and mine is that, in my paper, the principal is constrained in her ability to provide incentives. In particular, there are no transfers (or only limited transfers), and promotions are scarce. So the agents compete for the prize, which creates a strategic dependence between arms. This strategic dependence is largely absent in the other papers in this literature. In \cite{bergemann1996learning} and \cite{felli1996learning,felli2018firm}, a principal allocates an asset between two strategic agents over time. The values from allocating to each agent for the principal are initially unknown but can be learned over time. Their models, like mine, can be understood as bandit models with strategic arms. However, in \cite{bergemann1996learning,felli1996learning,felli2018firm}, transfers are unrestricted and only affect the ``cost of utilization'' of each arm, not the information the players get. This implies that all (Markov Perfect) equilibria are efficient (or pairwise efficient in \cite{felli2018firm}). So, the allocation policy is undistorted in any (Markov Perfect) equilibrium.\footnote{Although, in \cite{felli2018firm}, the players' investment decisions may be distorted.} Since the classic Gittins index policy maximizes total surplus, the principal always chooses the agent whose associated Gittins index is the highest. The question is then how surplus is allocated between players. On the other hand, in my framework, the conflict of interest between the principal and the workers prevents allocative efficiency. This is also the case in \cite{kakade2013optimal}, \cite{pavan2014dynamic}, \cite{bardhi2020early}, or \cite{fershtman2022matching}, who study strategic bandit models in which the experimentation outcomes are privately observed; or in \cite{guo2016dynamic} and \cite{mcclellan2017experimentation}, who study versions of a $1^{\frac{1}{2}}$-arm strategic bandit model when the principal has limited instruments. For example, in \cite{kakade2013optimal}, \cite{pavan2014dynamic}, \cite{bardhi2020early}, or \cite{fershtman2022matching}, to incentivize disclosure, the principal needs to pay rents to the agents. The latter creates dissonance between the principal's and the agents' value for experimentation and, hence, changes the relative value of pulling one arm rather than another. In these papers, the indices are, therefore, also distorted. Contrary to my paper, however, there is no strategic dependency between the arms in the above papers. The presence of transfers allows them to abstract from any linkage of incentive problems across employees. The allocation maximizes the total virtual value and therefore follows from the standard Gittins characterization applied to the ``virtual value processes''. In my setting, this linkage of incentives is central as promotions, hence incentives, are scarce. The principal has to promise an eventual promotion for which the workers compete. Promise-keeping then distorts the future delegation process. In particular, the set of implementable delegation rules in the continuation game depends on the history. The classical approach to indexability therefore fails. Nevertheless, I show that indexability still holds. The indices reflect the strategic nature of the problem and the constraints it places on both learning and exploitation. I then focus on how incentives provision distorts the delegation and promotion rules.
	
	My paper is related to a last stream of works on multi-agent experimentation. See, for instance, \cite{bolton1999strategic}, \cite{keller2005strategic}, \cite{bonatti2011collaborating}, and \cite{halac2017contests}. However, the fundamental trade-offs are different. In these papers, the agents experiment on a common bandit machine and therefore have incentives to free-ride on each others' costly experimentation. Free-riding is absent from my model, as each arm is a separate agent and the agent's types are independent. So, there is no positive externality across workers. The central trade-off in my paper is between retaining the option value of experimentation and motivating workers. Two other papers on multi-agent experimentation are related. \cite{de2021selecting} and \cite{deb2020fostering} also study how to select the best agent to execute a task when the agents only care about being selected. They focus on different trade-offs than I do. \cite{deb2020fostering} look at the trade-off between retaining option value via competition and harnessing gains from collaboration, while \cite{de2021selecting} are interested in mechanisms guaranteeing that the agents willingly display their private information, ensuring efficiency. My paper is complementary to theirs. It illustrates how the principal-optimal allocation rule responds to a different environment and trade-off.
	
	In particular, I show that the optimal allocation rule is an \emph{index contest}. So, my paper also contributes to the growing literature on the design of dynamic contests pioneered by \cite{taylor1995digging} and extended by \cite{halac2017contests}, \cite{benkert2020designing}, or \cite{ely2021optimal}. The critical difference between my paper and the rest of the dynamic contest literature is that, in my model, the contest is centralized; i.e., the principal controls the assignment of tasks. Therefore, the set of participants at every point is endogenous and chosen by the principal. This is crucial for my application. Organizations control the allocation of tasks. Therefore, the results I derive are qualitatively different from those in the rest of the dynamic contest literature, where the set of participants is exogeneous. Moreover, the optimal contest in my model is a winner-take-all contest and not a prize-sharing contest as in \cite{halac2017contests} or \cite{ely2021optimal}, for example. Comparing my results to these papers can also help us understand when a more meritocratic (winner-take-all) system or a more equal (prize-sharing) system helps the principal.
	
	Finally, my paper contributes to the extensive literature on personnel economics that studies careers in organizations (see \cite{prendergast1999provision} for a survey). I consider an environment where learning about workers shapes their career trajectories and hence generates career-concern incentives (\cite{harris1982wagedynamics,holmstrom1999managerial}). \cite{macdonald1982information,macdonald1982market}, or \cite{gibbons1999theory} also emphasize the importance of learning and task assignments in shaping career dynamics, which \cite{pastorino2019careers} empirically documents. In these papers, tasks are equally informative. So the players' choices do not affect learning. Instead, I focus on a setting where tasks vary in the information they generate, as in \cite{antonovics2012experimentation}, \cite{canidio2019task}, or \cite{madsen2022incentive}. These papers, however, focus on the distortionary effects of promotions and career concerns on risk-taking when the workers control their occupational choices. In contrast, in my paper, the principal controls the allocation of tasks. So my paper complements their findings. In particular, I study a trade-off that arises in task allocation problems when the principal is primarily concerned with alleviating the time inconsistency problem of promotions, as in \cite{waldman2003ex}, which is absent in their papers. Since promotions reward past effort and sort workers, a sound promotion system should do both. Moreover, the optimal way to incentivize effort may be suboptimal for selection. Indeed, as mentioned above, the optimal index contest vastly differs from the optimal dynamic contest to incentivize effort in \cite{ely2021optimal}. I show how the principal-optimal task allocation balances incentives provision and selection. This trade-off is also absent from other papers that look at how firms assign tasks and learn, such as \cite{pastorino2004optimal} or \cite{bar2022motivating}, in which the principal can incentivize each worker's effort separately. Finally, in all the previous papers that study learning through task allocation, the employer faces no constraints on learning. She can assign all workers to non-routine tasks simultaneously. On the contrary, I assume that non-routine tasks are scarce, reflecting that not all workers can simultaneously lead a team, for example. So, my paper complements their works by studying how firms design careers to screen and develop workers when learning opportunities are limited. In particular, I show that the delegation process is sequential, meritocratic, and creates a significant path dependence in promotion decisions: The principal first delegates opportunities to the best workers as measured by their \emph{index} and immediately promotes them in case of success. So, my paper also relates to the analysis of turnover in a leadership position. This question has been studied by, among others, \cite{mortensen1994job}, \cite{atkeson2005dynamic}, and \cite{garrett2012managerial}. As in these papers, seniority matters for promotion decisions, and I extend such finding to a multi-agent setting. Here the main difference is my focus on the dynamic process of experimentation that leads to the promotion decision.
	
	\section{Model}\label{sec:Model}
	
	\paragraph{} Let $(\Omega, \bar{\mathcal{F}}, \mathbb{P})$ be a probability space rich enough to accommodate all the objects defined below. A principal (she) and $N$ workers (he/they) interact in an infinite-horizon continuous-time stochastic game. All players discount the future at a common discount rate $r>0$.\footnote{This assumption can be relaxed: the analysis can also accommodate for random discount factors for example.} The principal has to decide how to delegate one non-routine task and many routine tasks among the workers to maximize her continuation value. When the non-routine task is delegated to one of the workers, the principal gets a flow rewards that depends on the current type of the worker and whether the worker exerts effort. If he does, his type also evolves (stochastically). To motivate the workers to exert effort when delegated the non-routine task, the principal can allocate an indivisible prize that the wokers value; i.e., she can decide to promote one of them. 
	
	\subsection{Actions}\label{subsec:players}
	
\begin{figure}
		\begin{tikzpicture}[xscale=2, yscale =2]
		\draw [thick] (-0.3,0) -- (4.7, 0);
		\draw [thick, ->] (0.1,-0.8) -- (0.1,-0.02);
		\draw [thick, ->] (1.1,0.8) -- (1.1,0.02);
		\draw [thick, ->] (2.1,-0.8) -- (2.1,-0.02);
		\draw [thick, ->] (3.1,0.8) -- (3.1,0.02);
		\draw [thick, ->] (4.1,-0.8) -- (4.1,-0.02);
		\draw [fill] (0.1,0) circle [radius=0.035];
		\draw [fill] (1.1,0) circle [radius=0.035];
		\draw [fill] (2.1,0) circle [radius=0.035];
		\draw [fill] (3.1,0) circle [radius=0.035];
		\draw [fill] (4.1,0) circle [radius=0.035];
		\node[align=center, below] at (0, -0.8)%
		{$P$ delegates to \\ $i \in \left\{ 1, 2\right\}$};
		\node[align=center, above] at (1, 0.8)%
		{$i$ decides whether to \\ exert effort $e^i \in \{0,1\}$};
		\node[align=center, below] at (1.95, -0.8)%
		{$P$ gets flow payoff \\ $e^i \pi^i(X^i_t)$};
		\node[align=center, above] at (3.2, 0.8)%
		{$i$'s type evolves: \\ $X^i_{t+dt} = X^i_t + e^i dX^i_t$};
		\node[align=left, below] at (4.7, -0.8)%
		{$P$ decides whether to: \\ (i) continue the game, or \\ (ii) promote $j \in \{1, ,\dots, N\}$, or \\ (iii) take outside option $W$};
		\node[right] at (5, 0)%
		{Period $t+dt$};
		\node[left] at (-0.4, 0)%
		{Period $t$};
		\end{tikzpicture}
		\caption{Heuristic ``stage game''}
		\label{fig:stagegame}
	\end{figure}
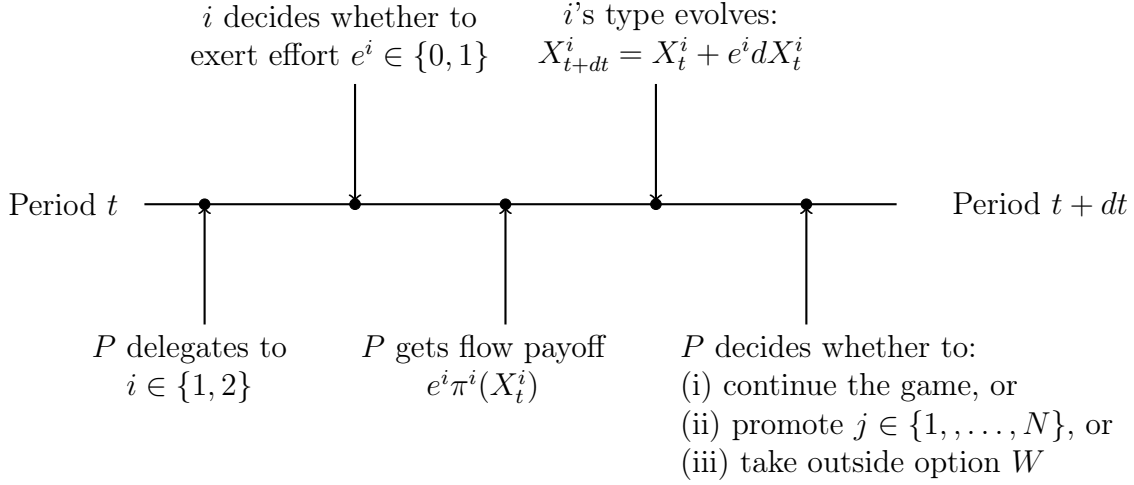

	\paragraph{} Heuristically, at each time $t$, the principal and the workers play in the ``stage game'' depicted in figure \ref{fig:stagegame}. Within each period $[t, t+dt)$, the principal chooses who to delegate the non-routine task to. The other workers are allocated routine tasks. When she delegates the non-routine task to worker $i \in \left\{1,\dots, N\right\}$, worker $i$ then decides whether to exert effort to complete the task. If worker $i$ exerts effort, the principal learns about worker $i$, gets a reward $\pi^i\left( x^i \right)$ that depends on worker $i$'s current type $x^i$, and worker $i$'s type evolves (stochastically). If he does not, the principal gets no reward and worker $i$'s type stay the same. The principal then decides whether to (i) continue to experiment before allocating the prize, (ii) promote one of the workers, or (iii) allocate the prize to an external worker (i.e. take her outside option $W$). If she chooses to continue to experiment, the next ``period'', $[t+dt, t+2dt)$ the players play the same ``stage game''. If she chose to allocate the prize, her only decision in the continuation game is who to delegate the non-routine task to. The workers then decides whether to exert effort to complete the task. I assume that the principal can commit at time zero to an history contingent sequence of plays, while the workers cannot.
	
	\paragraph{} Formally, at time $t=0$, the principal commits to a history-dependent promotion contest comprising of (i) a promotion time $\tau$ specifying when the promotion is allocated; (ii) a promotion decision $d$ specifying which of the workers is promoted; and (iii) a delegation rule $\alpha$ that assigns at every instant the non-routine task to some worker.
	
	The promotion time, $\tau$, is a $\bar{\mathcal{F}}$-measurable mapping from $\Omega$ to $\mathbb{R}$. The promotion decision is a (stochastic) process $d = \left( d^0 = \left\{ d^0_t \right\}_{t\geq 0}, \dots, d^N = \left\{ d^N_t \right\}_{t\geq 0} \right)$, which takes value in $\{0,1\}^{N+1} \cap \Delta^{N+1}$, where $\Delta^{N+1}$ is the $N+1$-dimensional simplex. If $d^i_\tau =1$, worker $i$ is promoted at time $\tau$. $d^0_{\tau} =1$ stands for the principal's decision to take her outside option. Finally, the delegation rule $\alpha = \left( \alpha^1 = \left\{ \alpha^1_t \right\}_{t\geq 0}, \dots, \alpha^N = \left\{ \alpha^N_t \right\}_{t\geq 0} \right)$ is a (stochastic) process which takes value in the $N$-dimensional simplex, $\Delta^N$. $\alpha^i_t$ is the share of the non-routine task worker $i$ is responsible for at each instant $t \geq 0$. The process $t \to \alpha_t$ is (at least) Borel measurable $\mathbb{P}$-a.s..\footnote{The filtration to which $\alpha$, $\tau$, and $d$ are adapted to is not defined yet. This is deferred to Section \ref{subsec:information}, after the dynamics of the workers' types and the information structure are introduced.}
	
	\paragraph{} Workers cannot commit. Each instant, they decide whether to exert effort when delegated (a positive share of) the non-routine task. $a^i_t \in \{0,1\}$ denotes the effort decision of worker $i$ at time $t\geq0$. The effort process generated by the decisions of worker $i$ is $a^i = \left\{ a^i_t\right\}_{t\geq 0}$. $t \to a^i_t$ is required to be Borel measurable $\mathbb{P}$-a.s..\footnote{As above, a more precise measurability requirement is postponed until Section \ref{subsec:information}.}
	
	\subsection{Workers' types}\label{subsec:dynamics}
	
	\paragraph{} Together, the choices of effort and the delegation rule determine the evolution of the workers' types. To describe the state dynamics, I follow the multi-parameter approach pioneered by Mandelbaum in discrete time in \cite{mandelbaum1986discrete} and in continuous time in \cite{mandelbaum1987continuous}.
	
	For all $i \in \{1, \dots, N\}$, let $\mathcal{F}^i \coloneqq \left\{ \mathcal{F}^i_t \right\}_{t\geq 0}$ be a filtration in $\bar{\mathcal{F}}$ and $X^i = \left\{ X^i_t \right\}_{t\geq 0}$ be a $\mathcal{F}^i$-adapted process with values in the interval $\mathcal{X}^i \subseteq \mathbb{R}$. For simplicity, assume that either $X^i$ does not reach the boundary of the set $\mathcal{X}^i$ or the boundary is absorbing. Define
	\begin{align}\label{eq:strategictimechange}
	T^i(t) \coloneqq \int_{0}^t a^i_s \alpha^i_s ds, \, \forall \, 0\leq t < \infty.
	\end{align}
	$T^i(t)$ is the amount of time worker $i$ worked on the non-routine task. At time $t$, the type of worker $i$ is $X^i_{T^i(t)}$. So worker $i$'s type evolves (stochastically) only when he exerts effort. When he does not, his type is frozen. Intuitively, one can think of the evolution of the type as follows: Nature first draw a path $X^i = \left\{ X^i_s \right\}_{s\geq 0}$ for worker $i$. The delegation rule and the worker's choices of effort then jointly control ``the passage of time'', $T^i(t)$, i.e., the speed at which the worker's type moves along the path $X^i$.
	
	Define the delegation process $T = \left( T^1 = \left\{ T^1(t) \right\}_{t\geq 0}, \dots, T^N = \left\{ T^N(t) \right\}_{t\geq 0} \right)$. The state of the game at time $t$ is
	\begin{align*}
	X_{T(t)} = \left( X^1_{T^1(t)}, \dots, X^N_{T^N(t)} \right).
	\end{align*}
	$\left\{ X_{T(t)} \right\}_{t\geq 0}$ is a multi-parameter process adapted to the multi-parameter filtration
	\begin{align*}
	\mathcal{F} \coloneqq \left\{ \mathcal{F}_{\bar{t}} \coloneqq \bigvee_{i=1}^N \mathcal{F}^i_{t^i}, \quad \bar{t} = (t^1, \dots, t^N) \in [0,\infty)^N \right\}
	\end{align*}
	defined on the orthant $[0,\infty)^N$. 
	
	\paragraph{} I make the following assumptions on the types' processes.
	\begin{assumption}\label{assumption:filtrationsindependenceandregularity}
		The filtrations $\mathcal{F}^i$, $i \in \{1,\dots N\}$, are mutually independent and they satisfy the usual hypothesis of right-continuity and completeness.\footnote{See, e.g., \cite{protter2005stochastic}.}
	\end{assumption}
	Assumption \ref{assumption:filtrationsindependenceandregularity} implies that the manager does not learn anything about the type of one worker by observing the type of another.
	\begin{assumption}\label{assumption:Feller}
		The processes $\left( X^i, \mathcal{F}^i \right)$, $i =1,\dots, N$, are Feller.\footnote{Recall that any Feller process admits a c\`{a}dl\`{a}g modification. So I always assume that $X^i$ is c\`{a}dl\`{a}g.}
	\end{assumption}
	Assumption \ref{assumption:Feller} is made to guarantee that the type process has the strong Markovian property: the distribution of future realizations only depends on the current value of the process. The Feller property is however stronger: it also guarantees that the expectation operator conditional on the value of the type process is continuous. This second property is not needed, but simplifies the analysis.
	\begin{assumption}\label{assumption:comparisontheorem}
		For all $i \in \left\{ 1,\dots, N\right\}$, if $X^i_0 = x^i \geq \tilde{X}^i_0 = \tilde{x}^i$, then, for all $s \geq 0$, $X^i_s \geq \tilde{X}^i_s$ $\mathbb{P}$-a.s..
	\end{assumption}
	Assumption \ref{assumption:comparisontheorem} states that if a worker's initial type increases, so does his type at any instant $t \geq 0$. Because Feller processes are time-homogeneous, it also implies that, if a worker's type is higher at time $t$ along one path than along another, so is his type at any instant $s\geq t$.
	\begin{assumption}\label{assumption:onesidedjumps}
		For each $i \in \left\{ 1,\dots, N\right\}$, if $t \to X^i_{t}$ is not continuous, then either (i) $X^i_{t^-} - X^i_t >0$ at all discontinuity points $t \in \mathbb{R}$; or (ii) $X^i_{t^-} - X^i_t <0$ at all discontinuity points $t \in \mathbb{R}$.
	\end{assumption}
	Assumption \ref{assumption:onesidedjumps} is a restriction on the jump of the processes $X^i$.The jumps must be``one-sided''; i.e., if the process $X^i$ jumps up, it cannot jump down, and conversely. In particular, if $X^i$ is a continuous process, Assumption \ref{assumption:onesidedjumps} holds trivially.
	\begin{assumption}\label{assumption:potentialforimprovment}
		For all $i \in\left\{ 1,\dots, N \right\}$ and for all $x \in \mathcal{X}^i$, $\mathbb{P}_x\left( \{\tau^i_{(x, \infty)} =0\}\right) = 1$, where $\tau^i_{(x, \infty)} \coloneqq \inf\left\{ t \geq 0 \, : \, X^i_t \in (x, \infty) \right\}$. Moreover, if $X^i$ jumps down, for all $\kappa, \epsilon>0$, there exists $\delta>0$ such that, for all $x \in \mathcal{X}^i$, $\mathbb{E}_x\left[\tau_{(x-{\delta}, x+\delta)} \right]<\epsilon$.
	\end{assumption}
	Assumption \ref{assumption:potentialforimprovment} states that any worker can always become more productive. It simplifies the arguments and guarantees the existence of a solution to the principal's problem. The second part of Assumption \ref{assumption:potentialforimprovment} strengthens the first part for the case in which $X^i$ jumps down. In particular, it guarantees that the expected time $X^i$ stays in any small interval is small. I relax this assumption in Section \ref{subsec:relaxingassumptionstypeprocess}.
	
	In Appendix \ref{app:jumpdiffusion}, I show that my framework accommodates all jump-diffusion processes that satisfy mild regularity and monotonicity conditions. In particular, it includes the commonly studied cases in which workers can be either good or bad and the principal learns whether the worker is good or bad according to a Brownian signal or a bad news Poisson signal. In these examples, worker $i$'s type at time $t$, $X^i_{T^i(t)}$, is the belief that worker $i$ is good after he has worked for $T^i(t)$ unit of time on the project.
	
	\subsection{Information and Strategies}\label{subsec:information}
	
	\paragraph{Information:} The principal and the workers perfectly observe the delegation rule chosen by the principal, and the effort decisions and types of all the workers. Information is symmetric, but incomplete.\footnote{Alternative information structures are discussed in Section \ref{sec:extensions}.}
	
	\paragraph{Workers' strategies:} It is well-known that perfect monitoring in continuous-time games can come with complications.\footnote{Continuous time is not well-ordered, and, therefore, seemingly well-defined promotion contests and effort strategies can fail to uniquely determine the outcome of the game. For a more detailed discussion, see \cite{simon1989extensive}, \cite{bergin1993continuous}, or \cite{park2020extensive} for deterministic games, and \cite{durandard2022strategies} for stochastic games.} To avoid the issue, I take a reduced form approach. 
	\begin{definition}\label{def:delegationprocess}
		A \textbf{dynamic delegation process} is a process
		\begin{align*}
			T = \left\{ T(t) = \left( T^1(t), \dots, T^N(t) \right), \, t\geq 0 \right\}
		\end{align*}
		taking values in $[0,\infty)^N$ such that, for all $i \in \{1,\dots, N\}$,
		\begin{enumerate}
			\item $\left\{T(t) \leq \bar{t} \right\} = \bigcap_{i=1}^N \left\{ T^i(t) \leq t^i \right\} \in \mathcal{F}_{\bar{t}}$ for all $\bar{t} = (t^1,\dots, t^N) \in [0,\infty)^N, \, t\geq 0$,
			\item $T^i(\cdot)$ is nondecreasing, right-continuous, with $T^i(0)  = 0$,
			\item $\sum_{i=1}^{N} \left(T^i(t) - T^i(u)\right) \leq  t-u, \quad \forall t \geq u\geq 0$. 
		\end{enumerate}
		Denote by $\mathcal{D}$ the set of all dynamic delegation processes.\footnote{In the theory of multi-parameter processes, $T(t)$ is a stopping point in $[0,\infty)^N$ and a delegation rule $T$ is called an optional increasing path (Walsh 1981, \cite{walsh1981optional}). It can be thought of as a multi-parameter time change.}
	\end{definition}
	Condition 1. in Definition \ref{def:delegationprocess} ensures that delegation processes are adapted to the multi-parameter filtration $\mathcal{F}$. So they are non-anticipative: they do not depend on future events.
	
	Given a dynamic delegation process $T \in \mathcal{D}$, define the one parameter filtration $\mathcal{G}^T = \left\{ \mathcal{G}^T_t \right\}_{t\geq 0}$ as follows. Let $\nu: \Omega \to [0,\infty)^N$. $\nu$ is a stopping point of $\mathcal{F}$ if $\left\{ \nu \leq \bar{t} \right\} \in \mathcal{F}_{\bar{t}}$ for all $\bar{t} \in [0,\infty)^N$. For any stopping point $\nu$, define the sigma-field
	\begin{align*}
	\mathcal{F}(\nu) \coloneqq \left\{ A \in \bar{\mathcal{F}} \, :\, A \cap \left\{ \nu \leq \bar{t} \right\} \in \mathcal{F}_{\bar{t}}, \, \forall \bar{t} \in  [0,\infty)^N \right\}.
	\end{align*}
	Then, for all $0\leq t< \infty$, let $\mathcal{G}^{T}_t \coloneqq \mathcal{F}(T(t))$.
	
	In the remaining of the paper, with a small abuse of notation, I will redefine promotion contests as:
	\begin{definition}\label{def:promotioncontest}
		A promotion contest $\left( T, \tau, d\right)$ consists of a dynamic delegation process $T$, a $\mathcal{G}^T$-stopping time $\tau$, and a $\mathcal{G}^T$-optional promotion decision rule $d$, such that $\mathbb{P}$-a.s.
		\begin{align*}
		T^i(t) \coloneqq \int_{0}^t a^i_s \alpha^i_s ds,
		\end{align*}
		for all $0 \leq t \leq \tau$ and all $i = 1,\dots, N$.
		
		Denote by $\mathcal{P}$ the set of all promotion contests.
	\end{definition}
	Finally, a strategy profile is \textbf{admissible} if it uniquely defines a promotion contest after all histories $\mathbb{P}$-a.s.. I will require that the space of strategies is such that (i) any strategy profile in which all workers change their effort decision at most once and the principal can adjust the contest upon observing such changes is included, and (ii) if a strategy profile belongs to the strategy space, then any $h_t$-``truncated'' strategy profile does too. The $h_t$-``truncated'' strategy profile is the strategy profile that coincides with the original profile for any history that does not contain $h_t$ and such that all players play a Markov continuation strategy after history $h_t$. Both conditions (i) and (ii) are richness conditions on the strategy space. They are satisfied for example by the space of semi-Markov strategies or the strategy space defined in \cite{durandard2022strategies}.
	
	In particular, condition (i) guarantees that any promotion contest can be obtained as the outcome of an \textbf{admissible} strategy profile. By definition of admissibility, the set of continuation values at any instant $t\geq 0$ that can be generated in the game coincides with the set of values generated by the set of promotion contests.
	
	\subsection{Payoffs and objective}\label{subsec:payoffs}
	
	\paragraph{} At time $t\geq 0$, when worker $i$ is delegated a share $\alpha^i_t$ of the project and exerts effort $a^i_t$, the principal gets a flow reward $\alpha^i_t a^i_t \pi^i\left(X^i_{T^i(t)}\right)$. Worker $i$ incurs a flow cost $\alpha^i_t a^i_t c^i\left(X^i_{T^i(t)}\right)$, proportional to the fraction of the task he is responsible for. Upon promotion (at time $\tau$), worker $i$ gets a payoff, $g^i>0$, and is now compensated for working on the non-routine task: he gets a flow payoff $\alpha^i_t a^i_t c^i\left(X^i_{T^i(t)}\right)$. When the principal takes the outside option, i.e., allocate the promotion to an external worker, she gets $W>0$.
	
	I make the following assumption on the principal's flow rewards and the workers' flow costs.
	\begin{assumption}\label{assumption:candpi}
		(i) $\pi^i : \mathcal{X}^i \to \mathbb{R}$ is upper semicontinuous, nondecreasing, nonnegative, and such that
		\begin{align*}
		\mathbb{E}\left[ \int_{0}^{\infty} e^{-rt} \pi^i(X^i_t) dt \mid X^i_0 = x \right] < \infty
		\end{align*} 
		for all $x \in \mathcal{X}$.
		(ii) $c^i: \mathcal{X}^i \to \mathbb{R}$ is lower semicontinuous, nonincreasing, and nonnegative. 
	\end{assumption}
	
	\paragraph{} So, given a promotion contest $\left( T, \tau, d \right)$, the principal's expected payoff is
	\begin{align*}
	\Pi^M\left(T, \tau, d; W \right) \coloneqq \mathbb{E} \left[ \sum_{i=1}^N \int_{0}^\tau e^{-rt} \pi^i(X^i_{T^i(t)})dT^i(t) + e^{-r\tau} \bar{\pi}\left( X_{T(\tau)}, d_{\tau} \right) \right],
	\end{align*}
	The workers' expected payoffs are
	\begin{align*}
	U^i\left( T,\tau, d \right)\coloneqq \mathbb{E} \left[ e^{-r\tau} g d^i_{\tau} - \int_{0}^{\infty} e^{-rt} (1-d^i_{\tau} \mathbbm{1}_{\{t\geq \tau\}}) c^i\left(X^i_{T^i(t)}\right) dT^i(t) \right].
	\end{align*}
	Define also the workers' continuations payoff at time $t \geq 0$ as
	\begin{align*}
	U^i_t\left( T,\tau, d \right) \coloneqq \mathbb{E} \left[ e^{-r(\tau-t)} g^i d^i_{\tau} \mathbbm{1}_{t\leq \tau} - \int_{t}^{\infty} e^{-r(s-t)} (1-d^i_{\tau} \mathbbm{1}_{\{s\geq \tau\}})c^i\left(X^i_{T^i(s)}\right) dT^i(s) \mid \mathcal{G}^T_t \right].
	\end{align*}
	
	\begin{definition}\label{def:implementablepromotioncontests}
		A promotion contest $(T,\tau, d)$ is \textbf{implementable} if there exists a promotion contest $\left(\alpha, \tau, d\right)$ such that (i) there exists a (weak) Perfect Bayesian Nash equilibrium with effort processes $a$ in the game defined by $\left( \alpha, \tau, d \right)$ played by the workers, and (ii) such that, for all $i \in \left\{ 1,\dots, N\right\}$,
		\begin{align*}
		T^i(t) = \int_{0}^{t}\alpha^i_s a^i_s ds, \quad 0 \leq t \leq \tau, \, \mathbb{P}\text{-a.s..}
		\end{align*}
		
		Denote by $\mathcal{P}^I$ be the set of all implementable promotion contests.
	\end{definition}
	
	\paragraph{} The principal designs the promotion contest to maximize her total expected payoff among all implementable promotion contests:
	\begin{align}\label{eq:maximizationprogram}
	\Pi^M \coloneqq \underset{(T,\tau,d) \in \mathcal{P}^I}{\sup } \mathbb{E}\left[ \sum_{i=1}^N \int_{0}^\tau e^{-rt} \pi^i(X^i_{T^i(t)})dT^i(t) + e^{-r\tau} \bar{\pi}\left( X_{T(\tau)}, d_{\tau} \right) \right]. \tag{Obj}
	\end{align}
	
	Finally, I make the following assumption.
	\begin{assumption}\label{assumption:nontrivial}
		For all $i \in \left\{ 1,\dots, N\right\}$, there exists $\left(T, \tau, d \right) \in \mathcal{P}^I$, with $T^i(t) =t$ for all $t \geq 0$, such that
		\begin{align*}
		\mathbb{E}& \left[ \int_{0}^{\tau} e^{-rt} \pi^i\left( X^i_t \right) dt + e^{-r\tau} \left( (1-d^0_{\tau}) \int_{\tau}^{\infty} e^{-r(t-\tau)}\pi^i\left( X^i_t\right)dt +d^0_{\tau} W \right) \right]\\
		& \qquad > \mathbb{E}\left[ \int_{0}^{\infty} e^{-rt} \pi^i\left( X^i_t \right) dt \right].
		\end{align*}
	\end{assumption}
	Assumption \ref{assumption:nontrivial} guarantees that the principal's problem when worker $i$ is the only candidate is not trivial, i.e., she can do better than promote worker $i$ immediately. It is not needed, but it simplifies some of the arguments by restricting the number of cases to consider.
	
	\subsection{Discussion of the model}\label{sec:discussion}
	
	\paragraph{} Before moving to the analysis, I comment on several features of the model.
	
	\paragraph{A constrained multi-armed bandit model:} As mentioned in the introduction, the model is a bandit problem with strategic arms. At each instant, the principal chooses which arm to pull (which worker to delegate) or takes her outside option. As in bandit models, the workers' types only evolve when they work on the project. For example, the principal learns about a worker's fixed but unknown potential. Implicit here is that the principal allocates her attention only to the worker delegated the non-routine task or that performing other jobs is not informative for the promotion. So learning is conditional on delegation. Another example corresponds to the acquisition of new skills and on-the-job learning: the workers' skills improve when responsible for the non-routine task. 
	
	Moreover, I assume that the workers' types are independent. The performance of one of the workers when delegated the task is uninformative about the potential of another worker. In particular, the workers do not learn from one another. I also focus on environments in which the workers do not need to cooperate: in my model, there is no payoff externality. The workers' efforts are substitutes, and the reward the principal obtains only depends on who is in charge of the non-routine task (and not on the types of other workers).
	
	These assumptions are crucial. As in classic bandit models, very little can be said when the workers' types are correlated or when a worker's type evolves when the principal does not delegate the project to him.\footnote{One could relax the last assumption (the absence of payoff externalities), following the analysis in \cite{nash1980generalized} or \cite{eliaz2021optimal}. They prove that indexability (with Nash indices) holds in multi-armed bandit problems in which the reward from pulling an arm also depends on the states of the other arms. \cite{nash1980generalized} consider the case when arms are complements, while \cite{eliaz2021optimal} consider both the cases when arms are substitutes and complements.}

	\paragraph{Multi-parameter formulation:} To describe the types' dynamics, I adopt the multi-parameter approach pioneered by Mandelbaum in \cite{mandelbaum1986discrete} for the multi-armed bandit model. This is critical to guarantee that the types' processes can be defined on a fixed (exogenous) probability space. It also simplifies the analysis. It also allows capturing many dynamic contracting environments with one unified approach. The alternative method would be to assume that the type of each worker is defined as the solution of a stochastic differential equation with drift, diffusion, and jump coefficients equal to zero when the workers do not exert effort. However, such stochastic differential equations would be unlikely to admit strong solutions.\footnote{See \cite{karatzas1984gittins}, and the discussion in \cite{mandelbaum1987continuous}.} By taking the multi-parameter approach, I do not need to work with multiple (endogenous) probability measures.
	
	\paragraph{Only value of promotions is strategic:} Another assumption of the model is the absence of direct value in promoting someone for the principal. The flow payoff from the non-routine task is the same whether the worker completing it has been promoted. One can think that a given non-routine task is associated with an opening position in the organization, for example, bringing a new product to the market. The principal allocates this same task whether or not she has already promoted a worker. So, the promotion has \emph{no direct payoff effect}. It has, however, a strategic role. Workers value promotion. Hence, the principal uses the promises of a future promotion to motivate the workers. In particular, upon promotion, the workers get an \emph{exogenous} prize and are compensated for their effort when working on the non-routine task.\footnote{In Section \ref{subsec:prizedesign}, I relax this assumption and allow the principal to design the prize.} This is for simplicity. It reflects the idea that the organization designs the position so that the promoted worker willingly exerts effort and obtains a strictly positive rent. It guarantees that the model remains tractable and allows me to focus on the interaction between the allocation of tasks and the promotion decision. 
	
	\section{Main Result}\label{sec:MainResults}
	
	\paragraph{} Lemma \ref{lemma:nonnegativecontinuations} below characterizes the set of implementable promotion contests $\mathcal{P}^I$. In particular, it shows that it is nonempty and, hence, that the value of the principal is finite (by Assumption \ref{assumption:candpi} (i)).
	\begin{lemma}\label{lemma:nonnegativecontinuations}
		A promotion contest is implementable if and only if the continuation value of each worker is nonnegative after any possible history: For all $i \in \left\{ 1,\dots, N\right\}$ and all $t \geq 0$, $U^i_t \geq 0$.
	\end{lemma}
	Its proof is in Appendix \ref{app:nonnegativecontinuations}. It follows from Lemma \ref{lemma:nonnegativecontinuations} that, in any implementable promotion contest, the non-routine task is allocated to the promoted worker forever once the promotion decision is made. The principal already spent her incentive capital. The best thing she can do is then to delegate the task to the promoted worker. So, with a small abuse of notation, redefine the continuation value the principal obtains upon promotion as
	\begin{align*}
	\bar{\pi}\left( X_{T(\tau)},d \right) \coloneqq d^0_{\tau} W + \sum_{i=1}^N d^i_{\tau} \int_{\tau}^{\infty} e^{-rt} \pi^i\left(X^i_{T^i(\tau) +t}\right) dt.
	\end{align*}	
	The principal's problem \eqref{eq:maximizationprogram} is then equivalent to:
	\begin{align*}
	\Pi^M \coloneqq \underset{(T,\tau,d) \in \mathcal{P}}{\sup } \mathbb{E}\left[ \sum_{i=1}^N \int_{0}^\tau e^{-rt} \pi^i(X^i_{T^i(t)})dT^i(t) + e^{-r\tau} \bar{\pi}\left( X_{T(\tau)}, d_{\tau} \right) \right],
	\end{align*}
	subject to the dynamic participation constraints: for all $i$ and all possible histories $h_t$ with $t\leq \tau$,
	\begin{align*}
	\mathbb{E}\left[ e^{-r(\tau-t)} g^i d^i_{\tau} - \int_{0}^{\tau} e^{-r(s-t)} c^i\left( X^i_{T^i(s)} \right) dT^i(s) \mid h_t \right] \geq 0.
	\end{align*}
	
	\subsection{Benchmark}
	
	\paragraph{}A natural benchmark is when the principal does not need to incentivize the workers to exert effort (which corresponds to $c^i(\cdot) =0$ for all $i$). The problem then reduces to the standard multi-armed bandit problem (with passive arms):
	\begin{align}
	\underset{(T, \tau)\in \mathcal{D}\times \mathcal{T}}{\sup } \mathbb{E}\left[ \sum_{i=1}^N \int_{0}^\tau e^{-rt} \pi^i\left(X^i_{T^i(t)}\right)dT^i(t) + e^{-r\tau} W\right] \tag{Bm}.
	\end{align}
	Hence, when $c^i(\cdot) =0$, all promotion contests give a nonnegative continuation payoff to the worker $i$ after any possible history. Since the flow rewards the principal obtains when worker $i$ performs the non-routine task are the same before and after promotion, promoting worker $i$ has no direct value. It also has no strategic value when $c^i(\cdot) =0$. However, it has a cost: it restricts the principal's future options. So the principal always wants to postpone the promotion. When the workers do not need to be incentivized, any rationale for promotion disappears, and it is never optimal to promote a worker.
	
	The solution of this problem is well-known. It is the index rule associated with the the (classic) Gittins' index. Both index rules and the Gittins' indices are defined now.
	\begin{definition}\label{def:index}
		A delegation process $T$ is called an \textbf{index rule} if, for all $i \in \left\{ 1,\dots, N\right\}$, there exists a $\mathcal{F}^i$-adapted processes $\Gamma^i \coloneqq \left\{ \Gamma^i_t \right\}_{t\geq 0}$ such that $T^i(t)$ is flat off the set 
		\begin{align*}
		\left\{ t \geq 0 \, : \, \underline{\Gamma}^i_{T^i(t)} = \bigvee_{j=1}^N \underline{\Gamma}^j_{T^j(t)} \right\} \, \mathbb{P}\text{-a.s.,}
		\end{align*}
		where $\underline{\Gamma}^i_t = \underset{0\leq s\leq t}{\inf } \, \Gamma^i_s$. 
		
		The process $\Gamma^i$ is \textbf{worker $i$'s index}.
	\end{definition}
	In continuous time, the existence of index rules is not obvious. It is proved (by construction) in \cite{mandelbaum1987continuous}, \cite{el1994dynamic}, or \cite{el1997synchronization}. For completeness, in Appendix \ref{app:strategicindex}, I reproduce the construction in \cite{el1997synchronization} to obtain an index delegation rule associated with (arbitrary) indices $\left( \Gamma^1, \dots, \Gamma^N \right)$, as I will need properties specific to this construction.
	
	\begin{definition}
		The (classic) Gittins' index $\Gamma^{g,i} \coloneqq \left\{\Gamma^{g,i}_t \right\}_{t\geq 0}$ associated with worker (arm) $i$ is defined by, for all $t\geq 0$,
		\begin{align}\label{eq:Gittinsindex}
		r \Gamma^{g,i}_t \coloneqq \underset{\tau > t }{\sup } \, \frac{\mathbb{E} \left[ \int_{t}^{\tau} e^{-rs} \pi^i(X^i_s) ds \mid \mathcal{F}^i_t \right] }{\mathbb{E} \left[ \int_{t}^{\tau} e^{-rs}ds \mid \mathcal{F}^i_t \right]}, \tag{GI}
		\end{align}
		with the convention that $\frac{0}{0}=-\infty$
	\end{definition}
	$\Gamma^{g,i}_t$ is the maximal constant price the principal is willing to pay to include worker $i$ in the pool of candidates up to time $t + \tau^*$; where $\tau^*$ is the optimal stopping time in \eqref{eq:Gittinsindex}. $\Gamma^i_t$ captures both the payoff from exploiting arm $i$ up to time $t+\tau^*$ and the value of the information the principal obtains.
	
	\begin{proposition}\label{prop:Gittinsindexpassivearms}
		The index rule associated with the Gittin's indices is optimal in the multi-armed bandit problem (with passive arms).
	\end{proposition}
	Proposition \ref{prop:Gittinsindexpassivearms} restates the well-known optimality of the Gittins' index rule for the multi-armed bandit problem. It is obtained as a special case of the main Theorem \ref{theorem:indexability} below. Its proof is in Appendix \ref{app:Gittinsindexpassivearms}. 
	
	\paragraph{} Proposition \ref{prop:Gittinsindexpassivearms} formally establishes that giving the prize to any of the candidates is never optimal when they do not need to be incentivized. Hence, it confirms that the value of the promotion is purely strategic in my model. When the workers do not need to be incentivized, the principal never promotes them. However, she still takes her outside option (i.e., hire externally) when she becomes too pessimistic any of the workers is good. 
	
	As a result, the principal only has to balance exploration and exploitation: delegating to a new worker to learn about him or to a worker known to be good to enjoy the higher reward obtained from the non-routine task. The Gittins' index rule addresses this trade-off. To see this, suppose that every time the principal delegates to worker $i$, she has to pay $\underline{\Gamma}^i_{T^i(t)} \coloneqq \underset{0\leq s\leq t}{\inf} \Gamma^i_{T^i(s)}$. By definition, it is the maximal flow price the principal would pay to delegate to worker $i$ from time $t$ to $t+\tau^*$. So the principal is indifferent between allocating the task to worker $i$ and stopping the game: her value from delegating to worker $i$ is zero. Following the Gittins index rule guarantees that her continuation value at all times is zero. If she, however, were to choose a different strategy, her value would be negative. So, given such prices, the index rule is optimal: it maximizes the profit collected by the bandit machine as it moves up the use of the more costly arms and postpones the use of the less costly ones. Since the prices are set to be the greatest possible to ensure the principal participation, they maximize the profit of the bandit machine among all possible prices. The index rule, therefore, maximizes total surplus and hence is optimal. This intuition was developed by Weber in his proof of indexability in \cite{weber1992gittins}.
	
	However, because the Gittins' rule never promotes any of the workers, it is not implementable: the continuation value of each worker when delegated the non-routine task is strictly negative. Hence the need to incentivize the workers to exert effort constrains the principal ability to learn. So the Gittins' ``prices'' are too high in the index contest: the principal would not delegate to the workers at these prices. In the next Section, I solve the multi-armed bandit problem with strategic arms.
	
	\subsection{The index contest}\label{subsec:banditcontest}
	
	\paragraph{} The strategic index rule will play a crucial role. To define it formally, I need to introduce the promotion thresholds, $P^i(\cdot)$'s, and promotion times, $\tau^{s,i}$'s, first. Define $\tau^i_{(\underline{x}, \bar{x})} \coloneqq \inf \left\{ t\geq 0 \, :\ , X^i_t \not \in (\underline{x}, \bar{x})  \right\}$. For all $i \in \left\{ 1,\dots, N\right\}$, for all $\underline{x} \leq x \leq \bar{x} \in \mathcal{X}^i$, let 
	\begin{align*}
	U^i(x, \underline{x}, \bar{x}) \coloneqq \mathbb{E}\left[  e^{-r\tau^i_{(\underline{x}, \bar{x})}} g^i \mathbbm{1}_{\{ X^i_{\tau} \geq \bar{x}  \}} - \int_{0}^{\tau^i_{(\underline{x}, \bar{x})}} e^{-rt} c^i(X^i_t) dt \mid X^i_0 = x \right].
	\end{align*}
	$ U^i(x, \underline{x}, \bar{x})$ is $i$'s continuation value when his current type is $x$ and he exerts effort until his type exceeds $\bar{x}$ and he is promoted or his type falls below $\underline{x}$ and he ``quits'' and gets payoff $0$. Define then $i$'s promotion threshold as:
	\begin{align*}
	\bar{P}^i(\underline{x}) = \sup\Big\{ \bar{x} \geq \underline{x} \, :\, \underset{x \to \underline{x}}{\lim }\, U^i\left( {x}, \underline{x}, \bar{x} \right) \geq 0 \Big\}.
	\end{align*}
	$\bar{P}^i(\underline{x})$ is the largest promotion threshold for which worker $i$ is willing to stay in the game as long as his type does not fall below $\underline{x}$. Moreover $\bar{P}^i(\cdot)$ is \emph{increasing}.
	
	Define also the stopping time $\tau^{s,i} \coloneqq \inf \left\{ t \geq 0 \, :\, X^i_t \geq \bar{P}^i\left( \underline{X}^i_t \right) \right\}$, where $\underline{X}^i_t \coloneqq \underset{0\leq s\leq t}{\inf} X^i_s$ is the running minimum of $X^i$. Theorem \ref{theorem:singleagentoptimal} in Section \ref{subsec:singlearmproblem} shows that $\tau^{s,i}$ is the optimal promotion time when worker $i$ is the only worker. Next define the $\mathcal{F}^i$-adapted process $h^i$ as
	\begin{align*}
	h^{s,i}_t \coloneqq \pi^i\left(X^i_t\right) \mathbbm{1}_{\{ t < \tau^{s,i} \}} + \bar{\pi}^i\left( X^i_{\tau^{s,i}} \right) \mathbbm{1}_{\{ t \geq \tau^{s,i} \}}, \quad t\geq 0,
	\end{align*}
	where
	\begin{align*}
	\bar{\pi}^i\left(x \right) \coloneqq r \mathbb{E}\left[ \int_{0}^{\infty} e^{-rt} \pi^i\left( X^i_t \right) dt \mid X^i_0 =x\right].
	\end{align*}
	
	The \textbf{strategic index} of worker $i$ is defined by
	\begin{align*}
	\Gamma^{s,i}_t \coloneqq \inf \left\{ W \geq 0 \, : \, V^i(t;W) \leq W \right\},
	\end{align*}
	where
	\begin{align*}
	V^i(t;W) \coloneqq \underset{\tau \geq t}{ \sup } \, \mathbb{E}\left[ \int_{0}^{\tau} e^{-r(s-t)} h^{s,i}_t dt + e^{-r(\tau - t)} W \mid \mathcal{F}^i_t\right].
	\end{align*}
	Worker $i$'s index is the ``equitable surrender value'', i.e., the smallest $W$ such that the principal prefers to take the outside option immediately rather than to delegate to worker $i$ for some time before making a decision (when worker $i$ is the only worker). Moreover, observe that, by assumption \ref{assumption:Feller}, the strategic index is a function of $X^i_t$ and $\underline{X}^i_t$ only: $\Gamma^{s,i}_t = \Gamma^{s,i} \left( X^i_t, \underline{X}^i_t \right)$.
	
	As in the classical bandit problem, it can be shown to be equal to
	\begin{align}\label{eq:alternativedefstrategicindex}
	r \Gamma^{s,i}_t = \underset{\tau > t }{\sup } \, \frac{\mathbb{E} \left[ \int_{t}^{\tau} e^{-rs} h^{s,i}_s ds \mid \mathcal{F}^i_t \right] }{\mathbb{E} \left[ \int_{t}^{\tau} e^{-rs}ds \mid \mathcal{F}^i_t \right]}, 
	\end{align}
	with the convention that $\frac{0}{0}=-\infty$. The strategic index coincides with the classical Gittins index for the modified payoff stream $\{h^i_s\}_{s\geq 0}$. In particular, $\Gamma^{s,i}_t$ is the maximum price the principal is willing to pay for the possibility of including the worker in the pool of candidates. Moreover, the second expression also makes it clear that the worker's \emph{strategic index} is similar to the classic Gittins' index \eqref{eq:Gittinsindex}. The difference resides in what information is optimally \emph{acquired}. 
	
	Here, the workers control both the rewards and the flow of information. Moreover, their incentives differ from the principal's. Worker $i$'s strategic index then takes into account the incentive provision problem. Since worker $i$ only exerts effort if it increases his chance of promotion, the principal has to motivate him to work by promising he will eventually get the prize. However, upon promotion, collecting information has no value for the principal, as she cannot incentivize other workers to work anymore. That's captured in the definition of the process $h^{s,i}$: after the promotion time, the flow reward is $\mathbb{E}\left[ \pi^i(X^i_t) \mid \mathcal{F}^i_{\tau} \right]$. It is as if no new information is obtained. Contrary to Gittins' index, the strategic index ignores the information generated after the promotion decision when assessing the value of delegating to a worker. Interestingly, when the cost of providing incentives goes to zero (when $c^i \to 0$), the strategic index process converges to the Gittins' index from below pointwise $\mathbb{P}$-a.s..
	
	The index delegation rule associated with the strategic index processes $\left( \Gamma^{s,1},\dots, \Gamma^{s,N} \right)$ is called the \textbf{strategic index rule}. 
	
	\begin{definition}\label{def:indexcontest}
		The \textbf{index contest} (i) follows the \textbf{strategic index rule}, (ii) promotes the \textbf{first} worker $i$ whose type reaches his \textbf{promotion threshold} $\bar{P}^i(\underline{X}^i_t)$, and (iii) takes the outside option at time $\tau^0 = \inf\left\{ t \geq 0 \, : \, \bigvee_{i=1}^N \underline{\Gamma}^i_{T^{s,i}(t)} \leq W \right\}$ if no worker was promoted before.
	\end{definition}
	Figure \ref{fig:maintheorem} illustrates the \textbf{index contest} with two workers. Each can be good or bad. Their types are the posterior beliefs that they are good, and the principal learns about them according to the Poisson arrival of bad news. Initially, worker $1$ is better, so the principal first delegates to worker $1$. However, too much bad news arrives. Therefore, she switches to worker $2$. Worker $2$ performs well and eventually gets the promotion.
	
	\begin{center}\label{fig:maintheorem}
		\includegraphics[width=100mm]{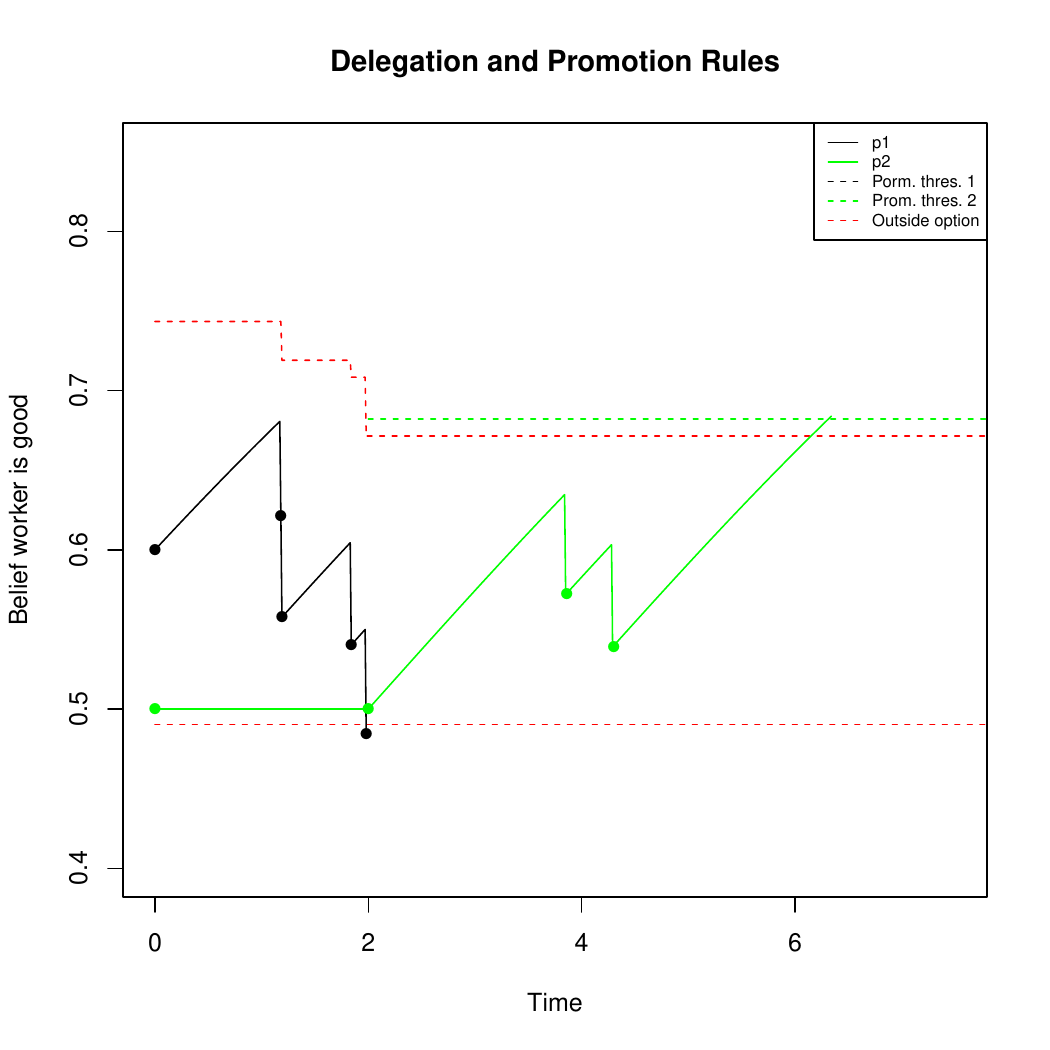}
	\end{center}
	
	\begin{proposition}\label{prop:strategicbanditcontest}
		The index contest can be \textbf{implemented} in a (weak) Perfect Bayesian equilibrium without commitment. The workers' strategies only depends on their own type $X^i_{T^i(t)}$, the current running minimum of their type $\underline{X}^i_{T^i(t)}$, and whether the principal promoted a worker. The principal's payoff is
		\begin{align}\label{eq:payoffequalitylowerenvelopeandprocesses}
		\mathbb{E} \left[ \int_{0}^{\infty} r e^{-rt} \bigvee_{i=1}^N\underline{\Gamma}^{s,i}_{T^{s,i}(t)} dt \right]. \tag{$\Pi^M$}
		\end{align}
	\end{proposition}
	The proof of Proposition \ref{prop:strategicbanditcontest} is in Appendix \ref{app:strategicbanditcontest}.
	
	\subsection{Optimality}\label{subsec:optimality}
	
	\paragraph{} The main result is the optimality of the \emph{index contest}: despite the agency frictions, indexability is preserved. When deciding who to delegate the non-routine task to and who to promote, the principal considers each worker separately. She delegates to and eventually promotes the best worker, as measured by the value of his associated strategic index.
	
	\begin{theorem}\label{theorem:indexability}
		The \textbf{index contest} maximizes the principal's payoff among all implementable promotion contests.
	\end{theorem}
	Theorem \ref{theorem:indexability} is proved in Section \ref{sec:Proofofmaintheorem}. Its proof requires to address two main challenges.
	
	First, implementable promotion contests have to balance the incentives of all workers. So, there is no reason a priori that it treats them separately. For example, if the principal promotes worker $1$ when his type exceeds that of worker $2$ by $\frac{1}{2}$, it creates a strategic dependence between the arms. The optimal delegation rule may not be an index rule, and the index of both workers $1$ and $2$ does not remain frozen when the other worker is in charge of the non-routine task.
	
	To overcome this problem, I show that the principal can treat the workers separately (when it comes to incentive provision), i.e., she chooses $N$ different promotion rules, each incentivizing one worker to exert effort. To do so, I study a relaxed problem in which participation constraints only hold in expectations (conditional on the workers' type). The relaxed problem coincides with the setting where each worker can only see the evolution of his own type. It pulls together many information sets. In that relaxed problem, workers have to be informed when promoted to maximize the length of the experimentation phase. Otherwise, their continuation payoff would be strictly less than the value they associate with the promotion. Therefore the principal could delegate the project a little longer before making her decision (which benefits the principal). So, the promotion time associated with each worker has to be measurable with respect to their own type. As a result, it is without loss of optimality for the principal to choose a delegation rule and $N$ individual promotion contracts (i.e., $N$ individual promotion time and promotion decision that depend only on the type of the worker).
	
	The solution, however, needs not be a solution to the original problem, as the delegation rule and individual promotion contracts may not be jointly implementable. Hence the principal may be unable to keep the independent promises she made to distinct workers. I will come back to this after describing the relaxed problem more carefully.
	
	Second, even if each worker's promotion time and decision depend only on his own type process, the problem is still not a standard bandit problem. To use the techniques developed in the bandit literature, I rewrite the flow payoff the principal gets upon promotion as the expected payoff from delegating to the worker, conditional on the information available at the time of promotion. Each arm is then a superprocess: each arm comprises multiple possible reward processes, one for each individual promotion contract. So, the principal chooses both which arm to pull and which contract to offer. In particular, when the principal pulls an arm, the flow payoff and the information depends on the ``promotion contract''.
	
	There is no guarantee that indexability holds for superprocesses. However, in the Markovian setting, there exists a condition for which it does, sometimes known as the Whittle condition or condition W (\cite{whittle1980multi}, \cite{glazebrook1982sufficient}). It says that the optimal action chosen in each state in the single-armed retirement problem is independent of the outside option W. If for some value of the outside option, it is optimal to choose an action rule, then it is also optimal to choose the same action rule for any other value of the outside option (as long as the arm is not retired). If the Whittle condition holds, the bandit problem with superprocesses is indexable. 
	
	In my setting, I show that a version of condition W for general (non-Markovian) superprocesses holds in the single-worker promotion problem. At time 0, the optimal promotion contract in the single-worker problem is independent of the outside option (before the principal takes her outside option). That is, provided that the principal has not taken her outside option yet, if the worker is promoted after some history, he is also promoted after this same history when the value of the outside option is smaller. I then show that this condition is sufficient for indexability to hold.
	
	The index contest is the solution to this problem. To build some intuition for this result, consider the case of two workers. Suppose that the strategic index of worker $1$ is initially higher than that of worker $2$. Suppose also that the value of the principal's outside option equals worker $2$'s index. If worker $2$ were the only worker, the principal would take the outside option immediately. The principal's problem then reduces to the problem in which she can only delegate to worker $1$, promote worker $1$, or take the outside option. The \emph{index contest} guarantees that the principal offers the optimal single agent promotion mechanism to worker $1$ (as Theorem \ref{theorem:singleagentoptimal} in Section \ref{subsec:singlearmproblem} below shows). Eventually, either worker $1$ is promoted, and the game ends, or his strategic index falls below the strategic index of worker $2$, hence, below the value of the outside option. In that case, the principal should take her outside option. Instead, imagine that, when it happens, the value of the outside option also falls to the level of worker $1$'s strategic index. In the new continuation problem, the principal never delegates to worker $1$. However, she is willing to delegate to worker $2$. In particular, she offers the single-player optimum promotion contract to worker $2$. The \emph{index contest} repeatedly plays the single-player optimal promotion mechanism for the best current worker (as measured by the strategic indices) until one worker is promoted or the principal takes her outside option. Promotion happens the first time a worker's type reaches $\bar{P}^i(\underline{X}^i_t)$. The principal takes the outside option when there is no benefit from experimentation anymore, i.e., when $\Gamma^{s,i} \leq W$ for all $i$. So, at any point in time, when one worker is delegated the project, his promotion threshold is equal to the optimal threshold in the single-agent problem.
	
	As mentioned above, the index contest needs not be implementable. Fortunately, it is, as Proposition \ref{prop:strategicbanditcontest} shows. Intuitively, when only one worker is allocated the task, the only promise-keeping constraint that matters is the one for the worker currently assigned the task. All other constraints are redundant and can be ignored.
	
	\paragraph{} The above intuition suggests the following interpretation of the \emph{index contest}. The principal approaches the workers successively. The indices' ranking determines the order in which the workers are approached. When the principal selects one worker, she offers him an \emph{individual trial contract}. It consists of a target: the promotion threshold, and a (potentially stochastic) deadline. If the worker achieves the target before the deadline, the principal promotes him. He is then in charge of the non-routine task forever. Otherwise, the principal approaches another worker until one succeeds, or the principal becomes too pessimistic and takes her outside option. Interestingly, if the principal could reoptimize at the end of each short-term contract (when she starts delegating to a new worker in the index contest), she would choose the same continuation promotion contest. Every time a new worker gets an opportunity to prove himself, the continuation mechanism is optimal for the principal.
	
	The above interpretation of the index contest is reminiscent of promotion practices described in the strategic management literature. For example, \cite{stumpf1981management} propose to evaluate the workers sequentially until the principal identifies a satisfactory one. More generally, the \emph{index contest} is also related to absolute merit-based promotion systems,\footnote{See \cite{phelan2001promotion}.} in which the first worker who meets a minimum performance target gets the promotion. My results suggest that one should expect organizations to use absolute merit-based promotion systems when it is important to fill the position with the right worker. On the other hand, when motivating effort is more important, other promotion systems, such as the classic winner-take-all promotion contest of \cite{lazear1981rank} or the cyclical egalitarian contest proposed by \cite{ely2021optimal} may be better, and, hence, more common. Intuitively, these promotion systems are very good at incentivizing effort but less so at ensuring that the promoted worker's potential is high. The \emph{index contest} guarantees that the non-routine task runs smoothly after the promotion decision is made. It balances incentives provision and selection.
	
	\subsection{Features of the index contest}\label{sec:featuresandcomparativestatics}
	
	\paragraph{No commitment:} Often, one may want to assume little commitment within an organization: most of the day-to-day activities are not governed by formal contracts, it is unlikely that the performance of a worker is verifiable\dots In my setting, the principal does not need any commitment power, as Proposition \ref{prop:strategicbanditcontest} shows. The \emph{index contest} is implementable even if the principal cannot commit to the delegation rule, delegation time, or promotion decision. Maybe even more interestingly, it does not require sophisticated coordinated punishments. It is implementable in a (weak) Perfect Bayesian equilibrium by ``grim trigger'' strategies. Moreover, each worker's strategy only depends on his current type, the running minimum of his type process, and whether the principal has promoted a worker yet.
	
	\paragraph{Fast track:} In the \textbf{index contest}, the promotion thresholds are decreasing over time (as increasing functions of the running minimums of the workers' types). So a worker's potential upon promotion decreases with time:
	\begin{proposition}[Speed and accuracy]\label{prop:fasttrack}
		If $\pi^i(\cdot) = \pi(\cdot)$ for all $i \in \left\{ 1,\dots, N \right\}$ and the processes $X^i$'s have the same law, then the promoted worker's type and the principal's continuation value upon promotion are nonincreasing over time $\mathbb{P}$-a.s..
	\end{proposition}
	Proposition \ref{prop:fasttrack} follows from the fact that the promotion threshold is $\mathbb{P}$-a.s. nonincreasing over time. The proof is omitted.
	
	Pushing the interpretation beyond the model, the above proposition suggests that fast tracks\footnote{I.e., that a quickly promoted worker often gets another promotion soon after. See \cite{baker1994internal} and \cite{ariga1999fast}.} should not be surprising. When a worker is promoted quickly, his type upon promotion is high. So, when entering a potential new contest for further promotion at the next level of the organization, he starts from a better position. In turn, it implies that his expected time to promotion is shorter and that the worker's chances to be promoted again soon are high.
	
	\paragraph{Seniority:} Finally, the decrease over time of the promotion thresholds also has an interesting implication for seniority. As time passes, it becomes easier for each worker to be promoted (conditional on his type). Proposition \ref{prop:seniority} formalizes this statements.
	\begin{proposition}\label{prop:seniority}
		In the index contest, worker $i$'s promotion probability, $\mathbb{E}\left[ d^i\mid \mathcal{G}^{T^s}_t \right]$ and continuation value, $U^i_t$, are non nondecreasing over time conditional on $X_{T^i(t)} = x$. His expected time to a promotion is nonincreasing in $t$, conditional on $X^i_{T^i(t)} =x$.
	\end{proposition}
	Proposition \ref{prop:seniority} also follows immediately from the promotion threshold being nonincreasing over time $\mathbb{P}$-a.s.. The proof is omitted.
	
	\paragraph{Convex compensation structure:} Learning is essential when the cost of promoting the wrong worker is high. Hence, the principal always benefits from a larger prize, as illustrated by the following proposition.
	\begin{proposition}[Value of the project]\label{prop:valueofproject}
		The principal's value increases with the value of the promotion $g = \left( g^1, \dots, g^N \right)$.
	\end{proposition}
	Proposition \ref{prop:strategicbanditcontest} is immediate: Let $ \bar{g} \geq \underline{g}$. Then any promotion contest feasible for the value vector $\bar{g}$ is also feasible for $\underline{g}$. 
	
	But, she should benefit from a larger prize especially when learning is paramount. That's because it makes incentivizing experimentation easier and helps the principal make a better decision. Proposition \ref{prop:convexcompensation} confirms this point and shows that the principal acquires more information about the promoted worker as $g^i$ increases. 
	\begin{proposition}\label{prop:convexcompensation}
		As $g^i$ increases, the principal learns more about worker $i$.
	\end{proposition}
	
	\begin{proof}[Proof of Proposition \ref{prop:convexcompensation}]
		Let $\bar{g}^i \geq \underline{g}^i$. Observe first that the index of worker $i$ is greater when the prize is $\bar{g}^i$; therefore, the principal acquires information about worker $i$ sooner. Moreover, in the index contest with reward $g^i\in \left\{ \bar{g}^i, \underline{g}^i\right\}$, worker $i$ is promoted after being responsible for the non-routine task for the time $\tau^i(g^i) = \inf\left\{ t \geq 0 \,: \, X^i_t \geq \bar{P}^i(\underline{X}^i_t; g^i) \right\}$. Note that $\bar{P}^i(\cdot, \bar{g}^i)\geq \bar{P}^i(\cdot, \underline{g}^i)$, and therefore $\tau^i\left( \bar{g}^i \right) \geq \tau^i\left( \underline{g}^i \right)$. Putting these two observations together concludes the proof.
	\end{proof}
	Intuitively, when workers value the promotion more (i.e., the prize is bigger), they are willing to exert effort for an extended time. So the principal can acquire more information and make a better promotion decision. 
	
	This can help understand why many organizations have a convex compensation structure (i.e., the bonuses paid upon promotion and the wage spread between positions increase when moving up in the hierarchy).\footnote{See \cite{devaro2006internal} for example.} At the top of the organization, the cost of promoting the wrong worker is potentially high. Extending the exploration phase is, therefore, valuable. A convex compensation structure achieves this. 
	
	However, how to measure the value of information here is not obvious. I propose to use the following definition:
	\begin{definition}\label{def:valueofinformation}
		The value of information in the promotion problem with $\tilde{\pi}^i\left( \cdot \right)$ is higher than the value of information in the promotion problem with ${\pi}^i\left( \cdot \right)$ if, for all $t\geq 0$,
		\begin{align*}
		\frac{\partial {\Gamma}^{s,i}_t(\tilde{\pi}^i)}{\partial \tau^{s,i}} \geq \frac{\partial {\Gamma}^{s,i}_t(\pi^i)}{\partial \tau^{s,i}},
		\end{align*}
		$\mathbb{P}-a.s.$, where ${\Gamma}^{s,i}_t({\pi})$ is the strategic index of worker $i$ when the flow payoff the principal gets when worker $i$ with type $x^i$ leads the project is $\pi\left( x^i \right)$.
	\end{definition}
	Intuitively, the above definition says that the benefit from waiting for one more instant before promoting worker $i$ is larger for $\tilde{\pi}^i$ than for $\pi^i$, i.e., there is more to gain from acquiring information as the cost of mistakes increases. It captures the extent to which marginal information is actionable: whether it helps the principal to make a better decision.
	\begin{proposition}\label{prop:convexcompensation2}
		Let $\bar{g} \geq \underline{g}$ and the value of information associated with $\tilde{\pi}^i$ be higher than the value of information associated with ${\pi}^i$, for all $i \in \left\{ 1,\dots, N\right\}$. Then 
		\begin{align}\label{eq:convexcompensationsupermodularity}
		\Pi^M(\bar{g},\tilde{\pi}) - \Pi^M(\bar{g},{\pi}) \geq \Pi^M(\underline{g},\tilde{\pi}) - \Pi^M(\underline{g},{\pi}).
		\end{align}
	\end{proposition}
	
	\begin{proof}[Proof of Proposition \ref{prop:convexcompensation2}]
		To prove \eqref{eq:convexcompensationsupermodularity}, it is enough to show that, for all $i \in \left\{ 1,\dots, N \right\}$,
		\begin{align*}
		\frac{\partial \Pi^M(g, \pi)}{\partial g^i} \text{ is increasing in } \pi \text{ for the order of Defintion \ref{def:valueofinformation}.}
		\end{align*}
		since $\Pi^M$ is nondecreasing and locally Lipschitz (by Assumption \ref{assumption:potentialforimprovment} and the definition of $\tau^{s,i}$ as ${\pi}^i(\cdot)$ is locally bounded) in $g^i$, hence, differentiable almost everywhere. 
		
		Consider $i \in \left\{1,\dots, N\right\}$. By the envelope theorem for dynamic optimization (e.g. Theorem 1 in \cite{lafrance1991envelope} and the discussion above),
		\begin{align*}
		\frac{\partial \Pi^M(g, c)}{\partial g^i} = \mathbb{E} \left[\int_{0}^{\infty} e^{-rt} \frac{\partial \underline{\Gamma}^{s,i}_{T^{s,i}(t)}}{\partial g^i} dT^{s,i}(t)\right].
		\end{align*}
		This follows immediately from my definition of an increase in the value of information, since
		\begin{align*}
		\frac{\partial \underline{\Gamma}^{s,i}_{T^{s,i}(t)}}{\partial g^i} = \frac{\partial \tau^{s,i}}{\partial g}  \frac{\partial \underline{\Gamma}^{s,i}_{T^{s,i}(t)}}{\partial \tau^{s,i}},
		\end{align*}
		$\tau^{s,i}$ is independent of $\pi^i$, and $\frac{\partial \tau^{s,i}}{\partial g} \geq 0$ by Proposition \ref{prop:convexcompensation}.
	\end{proof}
	Traditionally, contest theory has suggested that the convexity of the compensation structure in organizations results from the higher return of effort at higher positions in the hierarchy. My results offer a complimentary story: when the returns of selecting the right worker are high, larger bonuses let the principal experiment longer and promote a better worker.
	
	\section{Strategic amplification}\label{sec:Applications}
	
	\paragraph{} One of the initial questions I asked was whether the allocation of opportunities could exacerbate initial differences to produce significant disparities over time. Because in the \emph{index contest}, the principal delegates the project \emph{sequentially} and promotes the \emph{first} worker whose type reaches his promotion threshold, being delegated first is an advantage. This is especially true if, at every step of the index contest, i.e., during every trial contract, the probability that the worker leading the project reaches his target and hence gets the promotion is large. 
	
	In this section, I define a class of environments that I call \emph{reinforcing environments}, in which initial differences compound. In these environments, being delegated the project leads to a significant chance of promotion. This has two main implications: First, the timing of the first opportunity matters. A worker in charge of the non-routine task earlier is much more likely to be promoted. So, what determines the assignment of non-routine tasks early on is crucial to understanding who has a chance to be promoted. Secondly, initial differences lead to substantial differences during the exploration phase. To identify discrimination, conditioning on the potential of the workers upon promotion or their history of responsibilities in the organization may be a bad idea. Both depend on the endogenous delegation path. If discrimination occurs in the allocation of opportunities, it will remain undetected.
	
	\paragraph{} The following example illustrates the logic. Two symmetric workers compete for the promotion. Their types' processes $X^i$ keep track of their instantaneous (nonnegative) productivity. When they work on the project, their productivity drifts up at a constant speed $\mu$. This could reflect on-the-job learning. However, they can reach a dead end. Dead ends arrive according to a Poisson process with parameter $\lambda$. When a dead end comes, the worker needs to devise a new strategy and restart from scratch: his type jumps to zero. So, the type of each worker evolves according to the differential equation $dX^i_t = \mu dt$ if he does not reach a dead end and jumps down to zero if he does. The principal gets a flow payoff of $X^i_t$ when he delegates the project to worker $i \in \left\{1,2\right\}$. I assume that the workers' costs of effort are constant and equal to $c>0$ and that both associated value $g>0$ to the promotion. Finally, I assume the principal's outside option is small and, therefore, never taken.
	
	Let $\bar{t}$ be the unique solution of
	\begin{align*}
	\lambda c\int_{0}^{\bar{t}} e^{-(r+\lambda)t} dt = g. 
	\end{align*}
	The workers' promotion thresholds are given by $\bar{P}^i(\underline{X}^i_t) = \underline{X}^i_t + \mu \bar{t}$. The workers' indices can be taken to be the worker's types.\footnote{For $i \in \left\{ 1,2\right\}, \, \Gamma^{s,i}_t = \Gamma^s(X^i_t, \underline{X}^i_t)$ and the function $\Gamma^s(\cdot, \cdot)$ is increasing both variables. So the ranking of the indices at any instant is the same as the ranking of types when the principal plays the associated index delegation rule.}
	
	In the index contest, the first worker is promoted before the second worker even has a chance to lead the project with probability $(1- e^{-\lambda \bar{t}})$. Moreover, if the principal (lexicographically) prefers worker $1$, i.e., when indifferent, she delegates to worker $1$, then the probability that worker $2$ is promoted in this environment is $(1- e^{-\lambda \bar{t}}) e^{-\lambda \bar{t}}$. That is, worker $2$ is promoted if and only if worker $1$ does not succeed initially and worker $2$ does not hit a dead end the first (and only) time he works on the non-routine task. So, when $\bar{t}$ is either small or large, worker $2$'s promotion probability is close to zero. On the other hand, worker $1$'s promotion probability is close to one.
	
	\paragraph{} The above example is simple and clearly illustrates that the sequential nature of delegation exacerbates small differences in environments in which the workers' types (and, hence, their indices) tend to go up when they work. Under the condition below, the logic of the above example easily extends.
	\begin{definition}\label{def:reinforcingenvironments}
		An environment $\left( X^i, c^i(\cdot), g^i, \pi^i(\cdot) \right)_{i=1}^N$ is \textbf{reinforcing} if, there exists $\delta>0$ such that, for all $i \in \underset{j \in \left\{1,\dots, N\right\}}{\arg \max} \, \Gamma^{i,s}_0$,
		\begin{align}
		\mathbb{P}\left( \tau^i \leq \tau^i_{-}(X^i_0) \right) > \delta, \tag{RC}
		\end{align}
		where $\tau^i = \inf\left\{ t\geq 0 \, : \, X^i_t \geq \bar{P}^i(X^i_0) \right\}$ and $\tau^i_{-}(X^i_0) = \inf\left\{ t\geq 0 \, : \, X^i_t < \bar{P}^i(X^i_0) \right\}$.
	\end{definition}
	\begin{proposition}\label{prop:reinforcing}
		In a reinforcing environment, a worker $i \not \in \underset{j \in \left\{1,\dots, N\right\}}{\arg \max} \, \Gamma^{i,s}_0$'s probability to be promoted is bounded above by 
		\begin{align*}
		(1-\delta)^K.
		\end{align*}
		where $K$ is the cardinality of $\underset{j \in \left\{1,\dots, N\right\}}{\arg \max} \, \Gamma^{i,s}_0$.
	\end{proposition}
	
	\begin{proof}
		In the index contest, every worker $k \in \underset{j \in \left\{1,\dots, N\right\}}{\arg \max} \, \Gamma^{i,s}_0$ will be delegated the project before worker $i$. The probability that each of the worker $k \in \underset{j \in \left\{1,\dots, N\right\}}{\arg \max} \, \Gamma^{i,s}_0$ succeeds upon being delegated the project is greater than $\delta$. The result then follows.
	\end{proof}
	
	A direct consequence of Proposition \ref{prop:reinforcing} is that if $\delta$ is large, then the first worker gets the promotion with a considerable probability, and the other workers will not. Moreover, in large promotion contests with two different groups, each composed of initially identical workers, workers from the disadvantaged group face long odds when it comes to promotions. When the pool of candidates for promotion is large, any slight initial disadvantage is disqualifying. The logic here is reminiscent of \cite{cornell1996culture}.
	
	\paragraph{} My findings can help understand some of the mechanisms behind the ``promotion gaps'' documented in the literature (see, for example, \cite{bronson2019wage}, \cite{benson2021potential} and \cite{hospido2022gender}). This is especially important as wage growth is known to be closely related to job mobility, especially within firms (see \cite{baker1995internal}, \cite{lazear2007personnel}, or \cite{waldman2013handbook} and the references therein). The main point is that understanding and addressing the roots and causes of the different allocations of opportunities is crucial.
	
	\section{Proof of Theorem \ref{theorem:indexability}}\label{sec:Proofofmaintheorem}
	
	\paragraph{} The proof of Theorem \ref{theorem:indexability} is divided into the five following steps.
	\begin{itemize}
		\item In Section \ref{subsec:constrainedoptimization}, I relax the problem: in particular, each worker's dynamic participation constraint must only hold on expectation (conditional on the worker's own type), but not necessarily after all possible histories. 
		\item Section \ref{subsec:singlearmproblem} solves the problem with only one worker. Its solution is given in Theorem \ref{theorem:singleagentoptimal}. The argument adapts the logic of the proof of Theorem 1 in \cite{mcclellan2017experimentation} to our setting.\footnote{See also \cite{harris1982wagedynamics}, \cite{thomas1988self}, or \cite{grochulski2011optimal} for similar ideas.} 
		\item In Section \ref{subsec:measurablestopping}, I show that it is without loss of optimality to focus on promotion contests such that at most one worker is promoted and such that the promotion time of worker $i$ is a $\mathcal{F}^i$-stopping time. 
		\item Next, in Section \ref{subsec:upperboundonRP}, I derive an upper bound on the payoff the principal can get in any promotion contest that gives a nonnegative continuation value to all workers at all times, using the results from the three previous steps. Proposition \ref{prop:upperboundonRP} establishes that the principal's payoff in any implementable promotion contest is at most \eqref{eq:payoffequalitylowerenvelopeandprocesses}.
		\item Section \ref{subsec:Proofofmaintheorem} verifies that the \textbf{index contest} achieves the upper bound, hence proving Theorem \ref{theorem:indexability}. This follows from Proposition \ref{prop:strategicbanditcontest}.
	\end{itemize}
	
	\subsection{The Relaxed Program}\label{subsec:constrainedoptimization}
	
	\paragraph{} The principal solves the following optimization program:
	\begin{align*}
	\Pi^M \coloneqq \underset{(T,\tau,d) \in \mathcal{P}}{\sup } \mathbb{E}\left[ \sum_{i=1}^N \int_{0}^\tau e^{-rt} \pi^i(X^i_{T^i(t)})dT^i(t) + e^{-r\tau} \bar{\pi}\left( X_{T(\tau)}, d_{\tau} \right) \right], \tag{Obj}
	\end{align*}
	subject to the dynamic participation constraints: for all $i$ and all possible histories $h_t$ with $t\leq \tau$,
	\begin{align*}
	\mathbb{E}\left[ e^{-r(\tau-t)} g^i d^i_{\tau} - \int_{0}^{\tau} e^{-r(s-t)} c^i\left( X^i_{T^i(s)} \right) dT^i(s) \mid h_t \right] \geq 0.
	\end{align*}
	
	As a first step in the proof, I consider the relaxed problem in which the principal can randomize over possible stopping point. To introduce it formally, I need to define a number of new objects. For a filtration $\mathcal{H} = \left\{\mathcal{H}_t\right\}_{t\geq 0}$, define the set of $\mathcal{H}$-\textbf{randomized} stopping times as
	\begin{align*}
	\mathcal{S}\left(\mathcal{H}\right) \coloneqq \left\{ S \in \mathcal{N}_0^{\infty}(\mathcal{H}) \, : \, d S \in \mathcal{M}^{\infty}_{+}(\mathcal{H}), \, S_{0^-} =0, \, S_{\infty} \leq 1 \right\}.
	\end{align*}
	$\mathcal{N}_0^{\infty}\left( \mathcal{H} \right)$ is the set of $\mathcal{H}$-adapted process with values in $\left[0,\infty\right)$ such that $n \in \mathcal{N}_0^{\infty}\left( \mathcal{H} \right)$ if $n$ has nondecreasing paths $\mathbb{P}$-a.s.. $\mathcal{M}^{\infty}_{+}\left( \mathcal{H} \right)$ is the set of $\mathcal{H}$-optional random measure. Observe that any randomized stopping time is equivalent to a $\mathcal{F}_t \otimes \mathcal{B}([0,1])$-stopping time defined on the enlarged filtered probability space $(\Omega\times[0,1], \mathcal{H}\times \mathcal{B}([0,1]), \{\mathcal{H}_t \times \mathcal{B}([0,1]) \}_{t \geq 0}, \mathbb{P} \otimes \lambda)$, where $\lambda$ is the Lebesgue measure on $[0,1]$.\footnote{See, for example, \cite{camboni2022monotonestopping}.} 
	Finally, let $\mathcal{C}$ be the set of $\bar{\mathcal{F}}$-measurable promotion rule:
	\begin{align*}
	\mathcal{C} \coloneqq \left\{ d \, : \, \text{ for all } t\geq 0,\, d_t \text{ is } \bar{\mathcal{F}}\text{-measurable and } \sum_{i=0}^N d^i_t = 1 \, \,  \mathbb{P\text{-a.s.}} \right\},
	\end{align*} 
	and $\mathcal{C}^*$ be the set of nondecreasing promotion rule:
	\begin{align*}
	\mathcal{C}^* \coloneqq \left\{ d \in \mathcal{C} \, : \, d^i\text{'s paths are  c\`{a}dl\`{a}g and nondecreasing } \mathbb{P\text{-a.s. for }} i=1,\dots, N \right\}.
	\end{align*}
	The set of \textbf{randomized promotion contest} consists of all the promotion contests such that the promotion time $\tau$ is a randomized stopping time: $\tau \in \mathcal{S}(\mathcal{G}^T)$, and the decision rule $d$ belongs to $\mathcal{C}^*$. It is denoted by $\mathcal{P}^r$.
	
	Consider then the relaxed program:
	\begin{align}\label{eq:relaxedprogram}
	\Pi \coloneqq \underset{(T, \tau, d)\in \mathcal{P}^r}{\sup}\, & \mathbb{E}\Bigg[ \sum_{i=1}^N\int_{0}^\tau e^{-rt} \pi^i\left(X^i_{T^i(t)}\right)dT^i(t) + e^{-r\tau} \bar{\pi}\left( X_{T(\tau)}, d \right)\Bigg] \tag{RP}
	\end{align}
	subject to, for all $i \in \left\{ 1,\dots, N \right\}$, for all $t \geq 0$, $\mathbb{P}$-a.s.,
	\begin{align}\label{eq:DPC}
	\mathbb{E} \left[ e^{-r \left(\tau- \tau\wedge t\right)} g^i d^i_{\tau}  - \int_{\tau\wedge t}^\tau e^{-r(s-\tau\wedge t)} c^i\left(X^i_{T^i(s)}\right) d T^i(s)  \mid \mathcal{\mathcal{F}}^i_{T^i(t)} \right] \geq 0. \tag{DPC}
	\end{align}
	\begin{proposition}\label{prop:relaxation}
		The value of \eqref{eq:maximizationprogram} is weakly lower than the value of \eqref{eq:relaxedprogram}: $\Pi^M \leq \Pi$.
	\end{proposition}
	Proposition \ref{prop:relaxation} shows that the value of program \eqref{eq:relaxedprogram} is an upper bound on the principal's payoff for any implementable promotion contest. If an implementable promotion contest achieves this upper bound, this is the principal's preferred one. It relaxes \eqref{eq:maximizationprogram} in three ways. First, it replaces the feasibility set $\mathcal{P}^I$ by the set of all randomized promotion contests. This will allow to prove compactness. Secondly, it only requires that the workers have nonnegative continuation payoffs at all times $\mathbb{P}$-a.s. (and not necessarily after all possible histories). Thirdly, it pulls together all the $\mathcal{G}^T_t$ information sets that are not $\mathcal{F}^i_{T^i(t)}$ measurable, hence relaxing the constraints the principal faces. Its proof is in Appendix \ref{app:constrainedoptimization}.
	
	The remaining of Section \ref{sec:Proofofmaintheorem} is dedicated to the proof that the index contest achieves the optimum in \eqref{eq:relaxedprogram}.
	
	\subsection{The $1^{\frac{1}{2}}$-arm case}\label{subsec:singlearmproblem}
	
	\paragraph{} As in the classic bandit framework, the solution builds on the one arm problem. When there is only one worker, say worker $i$, the relaxed problem \eqref{eq:relaxedprogram} introduced above becomes
	\begin{align}\label{eq:relaxedproblemi}
	\Pi^i \coloneqq \underset{\left(\tau, d^i\right) \in \mathcal{P}^{r,i}}{\sup}\, & \mathbb{E}\Bigg[ \int_{0}^\tau e^{-rt} \pi^i\left(X^i_{t}\right)dt + e^{-r\tau} \left( d^i_{\tau} \int_{\tau}^{\infty} e^{-r(t -\tau)} \pi^i\left(X^i_t\right) dt + (1-d^i_{\tau}) W \right) \Bigg] \tag{RP$^i$}
	\end{align}
	subject to, for all $t \geq 0$, $\mathbb{P}$-a.s.,
	\begin{align}\label{eq:DPCi}
	\mathbb{E} \left[ e^{-r \left(\tau- \tau\wedge t\right)} g^i d^i_{\tau}  - \int_{\tau\wedge t}^\tau e^{-r(s-\tau\wedge t)} c^i\left(X^i_{s}\right) d s  \mid \mathcal{\mathcal{F}}^i_{t} \right] \geq 0. \tag{DPC$^i$}
	\end{align}
	$\mathcal{P}^{r,i}$ is the set of all pairs $(\tau, d^i)$ such that $\tau$ is a (randomized) $\mathcal{F}^i$-stopping time and $d^i$ is a $\mathcal{F}^i$-optional decision rule in $\mathcal{C}^*$. Define also $\mathcal{P}^{I,r,i}$: the set of all pairs $(\tau, d^i) \in \mathcal{P}^{r,i}$ that satisfy the constraints \eqref{eq:DPCi}.
	
	Recall that
	\begin{align*}
	U^i(x, \underline{x}, \bar{x}) \coloneqq \mathbb{E} \left[ e^{-r\tau} g^i d^i_{\tau} - \int_{0}^{\tau} e^{-rt} c^i\left(X^i_t\right)dt \mid x \right]
	\end{align*}
	is the continuation value of the worker with $X_0 =x$, $\tau = \inf\left\{ t\geq 0 \, : \, X^i_t \not \in (\underline{x},\bar{x})  \right\}$ and $d^i_{\tau} = \mathbbm{1}_{\{ X^i_{\tau} \geq \bar{x} \}}$. Define then 
	\begin{align*}
	{p}^i(P) \coloneqq \inf\left\{ x \in \mathcal{X}^i \, : \, \underset{p \in \mathcal{X}^i}{\sup} U^i(P,p,x) >0  \right\}.
	\end{align*}
	$p^i(P)$ is the smallest value of $x \in \mathcal{X}^i$ at which the worker is willing to keep working if he is promoted only when his type exceed $P$. Finally also 
	
	Recall also that worker $i$'s promotion threshold is given by the (nondecreasing) function $\bar{P}^i$ by
	\begin{align*}
	\bar{P}^i(\underline{x}) = \sup\Big\{ \bar{x} \geq \underline{x} \, :\, \underset{x \to \underline{x}}{\lim }\, U^i\left( {x}, \underline{x}, \bar{x} \right) \geq 0 \Big\}.
	\end{align*}
	Finally define $\underline{p}^i(W)$ as the unique solution of
	\begin{align*}
	\Gamma^{s,i}\left( \underline{p}^i, \underline{p}^i  \right) =W.
	\end{align*}
	
	\paragraph{} Theorem \ref{theorem:singleagentoptimal} characterizes the solution of the single worker promotion contest: \eqref{eq:relaxedproblemi}.
	\begin{theorem}\label{theorem:singleagentoptimal}
		The promotion contest
		\begin{align*}
		\tau \coloneqq \inf\left\{ t \geq 0 \, : \, X^i_t \not \in \left[ \underline{p}^i(W), \bar{P}^i\left( \underline{X}^i_t \right) \right) \right\}\text{ and } d^i_{\tau} \coloneqq \mathbbm{1}_{\{ X^i_{\tau} \geq \bar{P}^i\left( \underline{X}^i_{\tau} \right) \}}
		\end{align*}
		is optimal in the single worker problem \eqref{eq:relaxedproblemi}. 
	\end{theorem}
	
	Theorem \ref{theorem:singleagentoptimal} states that it is optimal to delegate to worker $1$ until his type either (i) reaches the promotion threshold $\bar{P}^i(\underline{X}^i_t)$, or (ii the principal becomes too pessimistic about him. To understand why that is, recall that the flow reward (conditional on worker $1$'s type) the principal obtains when worker $1$ operates the project is the same before and after promotion. So the principal always wants to postpone her decision, as she gets more information about the worker at no cost if she waits. Since the worker's type is strongly Markovian, a likely candidate for the promotion time is the first hitting time of a threshold as high as possible. In particular, if the cost of effort is zero, the principal promotes the worker when his type reaches the upper boundary of $\mathcal{X}^i$. However, when effort is costly, this threshold is too high. So, the principal chooses the highest threshold for which the worker is willing to exert effort instead. If the agent's type increases, the promotion threshold stays constant: the principal needs to keep her promises. On the other hand, when the worker's type decreases, the worker becomes more pessimistic about his promotion chances. The principal then has to lower the promotion threshold to motivate the worker. The logic is the same as in \cite{mcclellan2017experimentation}: the promotion threshold becomes laxer when the participation constraint binds.\footnote{See also \cite{harris1982wagedynamics}, \cite{thomas1988self}, or \cite{grochulski2011optimal}.} Because of the monotonicity of the problem, this constraint binds precisely when the worker's type decreases.
	
	Formally, the proof of Theorem \ref{theorem:singleagentoptimal} is based on the idea of the proof of Theorem 1 in \cite{mcclellan2017experimentation}. It follows from the five steps below:
	\begin{itemize}
		\item First consider a relaxation of problem \eqref{eq:relaxedproblemi} for which the constraint \eqref{eq:DPCi} only needs to hold for on a finite set of (stopping) times.
		\item Lemma \ref{lemma:lagrangian} derives the Lagrangian associated with the relaxed problem as an application of Theorem 1 in \cite{balzer2002duality}.
		\item In the third step, useful properties of the solution of the relaxed problem introduced in step 1 are established.
		\item The fourth step identifies a promotion contest that guarantees the principal a payoff of at least the value of the relaxed problem introduced in the first step. It is enough to focus on promotion contests that promote worker $i$ after good performances (as $X^i$ crosses an upper threshold from bellow) and take the outside option after bad outcomes (when $X^i$ crosses a lower threshold from above). 
		\item Putting everything together and letting the set of times at which \eqref{eq:DPCi} holds grow dense yields Theorem \ref{theorem:singleagentoptimal}.
	\end{itemize}
	Steps 1, 2, and 5 are essentially the same as in the proof of Theorem 1 in \cite{mcclellan2017experimentation}. Steps 3 and 4 are new and specific to our setting. The details are in Appendix \ref{app:singlearmproblemproof}. Supporting Lemmas are in Appendix \ref{app:singlearmproblemproofsupportinglemmas}.
	
	\begin{corollary}\label{corollary:domination}
		Let $\left( \tau, {d}^i \right)$ be feasible in the single worker problem \eqref{eq:relaxedproblemi} Then, for all $\bar{W}\geq W$,
		\begin{align*}
		\mathbb{E} &\left[ \int_{0}^{\tau} e^{-rt} {\pi}^i\left( X^i_t \right) d t + e^{-r \tau} \left( d^i_{\tau}\bar{\pi}^i\left(X^i_{\tau} \right) + (1- d^i_{\tau}) \bar{W}\right) \right] \\
		& \leq \mathbb{E} \left[ \int_{0}^{\tau^{s,i}\wedge \tau^i(\underline{p}^i(\bar{W}))} e^{-rt} \pi^i\left(X^i\right) dt + e^{-r\tau^{s,i}\wedge \tau^i(\underline{p}^i(\bar{W})}\left(\bar{\pi}^i\left(X^i_{\tau^{s,i}} \right) \mathbbm{1}_{\{\tau^{s,i} < \tau^i(\underline{p}^i(\bar{W}) \}} + \bar{W} \mathbbm{1}_{\{\tau^{s,i} \geq \tau^i(\underline{p}^i(\bar{W}) \}} \right)   \right].
		\end{align*}
	\end{corollary}
	
	\begin{proof}
		Observe that the set $\mathcal{P}^{I,r,i}$ is independent of $\bar{W}$ and that Assumption \ref{assumption:nontrivial} is satisfied for any $\bar{W}\geq W$. The result follows from Theorem \ref{theorem:singleagentoptimal}. 
	\end{proof}
	
	\subsection{Measurable stopping}\label{subsec:measurablestopping}
	
	\paragraph{} The main result of this section shows that it is enough to focus on a subset of the implementable promotion contests such that the decision to promote worker $i$ does not depend on the type of the other workers.
	
	\begin{proposition}\label{prop:measurablestopping}
		The supremum in \eqref{eq:relaxedprogram} is achieved by a (randomized) promotion contest $\left( T,\tau, d \right)$. Moreover, $\tau = \left( \bigwedge_{i=1}^N \tau^i \right) \wedge \tau^0$, where $\tau^i$ is a $\mathcal{F}^i$-stopping time, $\tau^0$ is a $\mathcal{G}^T$- randomized stopping time, and $d^i_{\tau} = 1$ only if $\tau^i \leq \tau = \left( \bigwedge_{i=1}^N \tau^i \right) \wedge \tau^0$. 
	\end{proposition}
	
	Proposition \ref{prop:measurablestopping} has two parts. The first part states that the supremum in \eqref{eq:relaxedprogram} is achieved by a a promotion contest. It follows from Theorem \ref{theorem:existencerelaxedprogram} in Appendix \ref{app:existence}.
	
	The second part characterizes the promotion time $\tau$. It is the minimum of $N$ $\mathcal{F}^i$-stopping times, $\tau^i$'s, and one $\mathcal{G}^T$-randomized stopping time $\tau^0$. It follows from Corollary \ref{corollary:stoppingismeasurable} in Appendix \ref{app:measurablestopping2}.
	
	\subsection{An upper bound on the value of \eqref{eq:relaxedprogram}}\label{subsec:upperboundonRP}
	
	\paragraph{} Proposition \ref{prop:upperboundonRP} derives an upper bound on the principal's payoff in any implementable promotion contest.
	\begin{proposition}\label{prop:upperboundonRP}
		The value of \eqref{eq:relaxedprogram} is bounded above by
		\begin{align*}
		\mathbb{E} \left[ \int_{0}^{\infty} r e^{-rt} \bigvee_{i=1}^N\underline{\Gamma}^{s,i}_{T^{s,i}(t)} dt \right] \tag{$\Pi^M$}.
		\end{align*}
	\end{proposition}
	To build some intuition, it is useful to go back to the proof of indexability for superprocesses.\footnote{See Chapter 4 in \cite{gittins2011multi} or \cite{durandard2022index}, for example.} Start with $N$ independent payoff processes $\tilde{\pi}^i_t$, one for each superprocess. To each of these payoff processes, associate the index process $\tilde{\Gamma}^i_t$ defined as the ``equitable surrender value'', i.e. the smallest $W$ such that 
	\begin{align*}
	W = \tilde{V}^i_t(W) \coloneqq \underset{\tau \geq t}{\sup } \, \mathbb{E}\left[ \int_{t}^{\tau}e^{-r(s-t)} \tilde{\pi}^i_s ds + e^{-r(\tau-t)} W \mid \mathcal{F}^i_t \right].
	\end{align*}
	This index process has the desirable property that, for all $\tilde{W}$,\footnote{See Proposition 3.2 in \cite{el1994dynamic}.}
	\begin{align*}
	\mathbb{E}\left[ \int_{0}^{\infty}e^{-rt} r\underline{\tilde{\Gamma}}^i_t\vee W dt  \right] = \underset{\tau}{\sup } \, \mathbb{E}\left[ \int_{0}^{\tau}e^{-rt} \tilde{\pi}^i_t dt + e^{-r\tau} W \right],
	\end{align*}
	where $\underline{\tilde{\Gamma}}^i_t$ is the lower envelope of $\tilde{\Gamma}^i_t$. Whittle's condition guarantees that one of the payoff processes within each superprocess is such that its associated index process $\tilde{\Gamma}^{i,*}$ dominates the associated index process of all other possible index processes associated with this superprocess,  i.e., for all $W$, 
	\begin{align*}
	\mathbb{E}\left[ \int_{0}^{\infty}e^{-rt} r\underline{\tilde{\Gamma}}^{i,*}_t\vee W dt  \right] \geq \mathbb{E}\left[ \int_{0}^{\infty}e^{-rt} r\underline{\tilde{\Gamma}}^i_t\vee W dt  \right].
	\end{align*}
	It is then possible to show that the optimal policy picks the dominating process for each superprocess and pulls the arm whose index is the highest at each instant $t$.
	
	Here, start with an implementable promotion contest $\left( T,\tau,d \right) \in \mathcal{P}^I$ and find $N$ single-arm implementable promotion contests $(\tau^i, d^i) \in \mathcal{P}^{I,i}$. This $N$ single-arm implementable promotion contests generates $N$ payoff processes for the principal:
	\begin{align*}
	h^i_t \coloneqq \pi^i(X^i_t) \mathbbm{1}_{\{ t < \tau^i \}} +  r \bar{\pi}^i\left( X^i_{\tau^i} \right) \mathbbm{1}_{\{ t \geq \tau^i \}}.
	\end{align*}
	By the results of Section \ref{subsec:measurablestopping}, each $h^i$ can be chosen to be $\mathcal{F}^i$-adapted. As in the proof of indexability for superprocesses, one would like to associate to each of these payoff processes an \emph{index process} $\Gamma^i_t$. However, the index process cannot be the ``equitable surrender value'' in the retirement problem:
	\begin{align*}
	\bar{V}^i_t(W) \coloneqq \underset{\tau \geq t}{\sup } \, \mathbb{E}\left[ \int_{t}^{\tau}e^{-r(s-t)} {h}^i_s ds + e^{-r(\tau-t)} W \mid \mathcal{F}^i_t \right].
	\end{align*}
	Intuitively, this would allow the principal to break her promises and take the outside option while the worker's promised continuation utility is \emph{strictly} positive. Hence, the retirement problem above does not take into account the worker's participation constraint. To overcome this issue, consider instead the optimal retirement problem in which the principal can take the outside option only on a set of decision times at which the continuation value of the worker is zero:
	\begin{align*}
	\tilde{V}^i\left(t, W;  \tau^i, d^i\right) \coloneqq \underset{\rho \in \mathcal{T}^s(t; {h}^i)}{ \sup} \, \mathbb{E} \left[ \int_{t}^{\rho} e^{-r(s-t)} {h}^i_{s} d s + e^{-r \rho} W \mid \mathcal{F}^i_{t} \right];
	\end{align*}
	where 
	\begin{align*}
	\mathcal{T}^{s}(t;  \tau^i, d^i) = \left\{ s \geq t \, : \, U^i_s(\tau^i, d^i) = 0 \right\},
	\end{align*}
	and $U^i_s(\tau^i, d^i)$ is worker $i$'s continuation value at time $s$ for the single-arm promotion contest $(\tau^i, d^i) \in \mathcal{P}^{I,i}$. The index process is then the ``equitable surrender value'' in this alternative retirement problem. One can therefore think of the problem as a multi-armed bandit problem in which the completion time of each task is the random duration between two times such that the worker's continuation is zero.
	
	Finally, Corollary \ref{corollary:domination} guarantees that each index process is dominated (in the sense of Whittle) by the strategic index process. The conclusion then follows from the same arguments as in the nonstrategic case. 
	
	The proof of Proposition \ref{prop:upperboundonRP} is in Appendix \ref{app:upperboundonRP}. The derivation of the index processes associated with our alternative retirement problem is in Appendix \ref{app:upperboundonRPsupportinglemmas}.
	
	\subsection{Proof of Theorem \ref{theorem:indexability}}\label{subsec:Proofofmaintheorem}
	
	\paragraph{} By Proposition \ref{prop:upperboundonRP}, any implementable promotion contest gives a payoff weakly smaller than 
	\begin{align*}
	\mathbb{E} \left[ \int_{0}^{\infty} r e^{-rt} \bigvee_{i=1}^N\underline{\Gamma}^{s,i}_{T^{s,i}(t)} dt \right].
	\end{align*}
	By Proposition \ref{prop:strategicbanditcontest}, the principal obtains an expected payoff of 
	\begin{align*}
	\mathbb{E} \left[ \int_{0}^{\infty} r e^{-rt} \bigvee_{i=1}^N\underline{\Gamma}^{s,i}_{T^{s,i}(t)} dt \right]
	\end{align*}
	in the index contest. Thus the index contest is optimal. \qed

	\section{Extensions}\label{sec:extensions}
	
	\paragraph{} In this section, I discuss multiple extensions.
	
	\subsection{Relaxing Assumptions \ref{assumption:potentialforimprovment} and \ref{assumption:nontrivial}}\label{subsec:relaxingassumptionstypeprocess}
	
	\paragraph{} Assumptions \ref{assumption:potentialforimprovment} and \ref{assumption:nontrivial} simplify the analysis but rule out potentially interesting settings. In particular, Assumption \ref{assumption:potentialforimprovment} excludes Poisson learning with good news, a case that has received a lot of attention in the economic literature, while Assumption \ref{assumption:nontrivial} excludes problems in which the principal has no outside option, i.e., in which the position has to fill internally.  
	
	However, both can be relaxed, as the Corollaries below establishes. Interestingly, both Corollaries rely on the continuity of the principal's value. Corollary \ref{corollary:nontrivial} uses that the value is continuous in the payoff from the outside option, $W,$ while Corollary \ref{corollary:potentialforimprovment} uses that the value is continuous in the process $X^i$ (in the appropriate topology). Their proofs are in Online Appendix \ref{app:relaxingassumptions}.
	
	\begin{corollary}\label{corollary:nontrivial}
		The index contest is still optimal when Assumption \ref{assumption:nontrivial} does not hold. The principal never takes the outside option.
	\end{corollary}
	
	Corollary replaces Assumption \ref{assumption:potentialforimprovment} with the following assumption:
	\begin{assumption}\label{assumption:nopotentialforimprovement}
		For all $i \in \left\{1,\dots, N\right\}$, there exists a sequence $\left( X^{i,n}\right) _{n\in \mathbb{N}}$ such that (i) $X^{i,n}$ satisfies Assumption \ref{assumption:potentialforimprovment}, (ii) $X^{i,n} -X^i$ is $\mathcal{F}^i$-adapted, and (iii) $X^i = \underset{n\to \infty}{\lim} X^{i,n}$ uniformly on compact sets $\mathbb{P}$-a.s..
	\end{assumption}
	Hence, any process satisfying Assumptions \ref{assumption:Feller}, \ref{assumption:comparisontheorem}, and \ref{assumption:onesidedjumps}, but not \ref{assumption:potentialforimprovment} can be approximated by a sequence of processes $X^{i,n}$ that satisfy \ref{assumption:potentialforimprovment}. Assumption \ref{assumption:nopotentialforimprovement} simply guarantees that this sequence is $\mathcal{F}^i$-adapted. In particular, if, for all $i$, the probability space $\left(\Omega, \bar{\mathcal{F}}, \mathbb{P} \right)$ contains a $\mathcal{F}^i$-Brownian motion, Assumption \ref{assumption:nopotentialforimprovement} is satisfied. Define
	\begin{align*}
	\tau^0 \coloneqq \inf\left\{t \geq 0 \, : \, \Gamma^{s,i}_{T^i(t)} \leq W \, \text{ for all } i \right\},
	\end{align*}
	and
	\begin{align*}
	\tau^i \coloneqq \inf\{t \geq 0 \, : \, T^i(t) > \bar{P}^i\left( \underline{X}^i_{T^i(t)}\right)  \} \wedge \tau^{p,i},
	\end{align*}
	where $\tau^{p,i}$ is the first tick of a Poisson clock that runs only on $\{X^i_t = \bar{P}^i(\underline{X}^i_t) \}$ which intensity is chosen to leave $i$ indifferent between exerting effort or not when promoted at time $\tau^{i}$.
	\begin{corollary}\label{corollary:potentialforimprovment}
		Suppose that Assumption \ref{assumption:potentialforimprovment} is replaced with Assumption \ref{assumption:nopotentialforimprovement} and that the $\pi^i$'s are continuous. Then the index contest associated with the strategic indices $\Gamma^{s,i}$ and the promotion time $\tau^* = \tau^0\wedge \bigwedge_{i=1}^N \tau^i$ is optimal.
	\end{corollary}
	Interestingly, Corollary \ref{corollary:potentialforimprovment} shows that when $\bar{P}^i(\underline{X}^i_t) = \underline{X}^i_t = X^i_t$ the strategic index associated to worker $i$ is equal to the expected value of promoting $i$ immediately: information has no value. This is the case, for example, if worker can be good or bad, the principal learns about worker $i$ through the Poisson arrival of good news, the probability that worker $i$ is good at time $t$ (hence his type $ X^i_{T^i(t)} \coloneqq \mathbb{P}\left( \{ i \text{ is good }\} \mid \mathcal{F}^i_{T^i(t)} \right)$) is too low.
	
	\subsection{Prize design}\label{subsec:prizedesign}
	
	\paragraph{} The main of this section establishes that when the principal can design the prize, the index contest is still optimal, i.e., the principal prefers to allocate the entire prize to one worker only. Moreover, there is no value in giving multiple ``smaller'' promotion to a worker.
	
	The model is identical to the one presented in Section \ref{sec:Model}, except for the following two differences: (i) the prize is divisible, and (ii) the principal chooses (potentially) multiple times at which to promote workers. Formally, at time $t=0$, the principal commits to a history-dependent promotion contest comprising of (i) a set of promotion time $\left\{ \tau_k \right\}_{k=1}^K$ (with $K \in \mathbb{N} \cup {\infty}$) specifying when a fraction of the prize is allocated; (ii) a promotion decision $d$ specifying which of the workers is promoted; and (iii) a delegation rule $\alpha$ that assigns at every instant the non-routine task to some worker. The promotion times, $\tau_k$'s, are ${\mathcal{G}^T}$-stopping time such that $\tau_0 = 0$ and $\tau_k < \tau_{k+1}$ $\mathbb{P}$-a.s.. The promotion decision is a $\mathcal{G}^T$-adapted (stochastic) process $d = \left( d^0 = \left\{ d^0_t \right\}_{t\geq 0}, \dots, d^N = \left\{ d^N_t \right\}_{t\geq 0} \right) \in \mathcal{C}^*$. Again $d^0$ stands for the principal's decision to take her outside option. Finally, the delegation rule $T = \left( T^1 = \left\{ T^1(t) \right\}_{t\geq 0}, \dots, T^N = \left\{ T^N(t) \right\}_{t\geq 0} \right) \in \mathcal{D}$ is a delegation process. The workers only decides to exert effort $a^i_t$ in $\{0,1\}$ when they are delegated the non-routine task.
	
	Finally, the following additional assumption is maintained in this section.
	\begin{assumption}\label{assumption:submartingalepayoffs}
		(i) For all $i \in \left\{ 1,\dots, N\right\}$, the process $\left\{ \pi^i\left( X^i_s \right) \right\}_{s\geq 0}$ is a submartingale.
		
		(ii) For all $i \in \left\{ 1,\dots, N\right\}$, the cost of effort is constant: $c^i\left( \cdot \right) \coloneqq c^i$.
	\end{assumption}
	Assumption \ref{assumption:submartingalepayoffs} (i) guarantees that upon promotion, the principal always wants there is no penalty from delegating the full project to the promoted worker. Assumption \ref{assumption:submartingalepayoffs} (ii) simplifies the argument. 
	
	So, given a promotion contest $\left( T, \left\{ \tau_k \right\}_{k=1}^K, d \right)$, the principal's expected payoff is
	\begin{align*}
	\Pi^M\left(T, \left\{ \tau_k \right\}_{k=1}^K, d; W \right) \coloneqq \mathbb{E} \left[ \sum_{k=1}^{K} \left( \sum_{i=1}^N \int_{\tau_{k-1}}^{\tau_{k}} e^{-rt} \pi^i(X^i_{T^i(t)})dT^i(t) + e^{-r\tau_k} \bar{\pi}\left( X_{T(\tau)}, d_{\tau_k} \right) \right) \right],
	\end{align*}
	where 
	\begin{align*}
	\bar{\pi}\left( x, d\right) \coloneqq d^0 W + \sum_{i=1}^N \mathbb{E}\left[ \int_{0}^{\infty} e^{-rt} \pi^i\left( X^i_{d^i t} \right)d \left( d^i t \right) \mid X^i_0 =x^i\right] .
	\end{align*}
	The workers' expected payoffs are
	\begin{align*}
	U^i\left( T,\left\{ \tau_k \right\}_{k=1}^K, d \right)\coloneqq \mathbb{E} \left[ \sum_{k=1}^{K} e^{-r\tau_k} g d^i_{\tau_k} - \int_{0}^{\infty} e^{-rt} (1- \sum_{k=1}^{K}d^i_{\tau_k} \mathbbm{1}_{\{t\geq \tau_k\}}) c^i dT^i(t) \right].
	\end{align*}
	The principal's objective is to design the promotion contest that maximizes her payoff among all implementable promotion contest. As above, this is equivalent to the maximization program:
	\begin{align}\label{eq:prizedesign}
	\Pi^M \coloneqq \underset{(T,\left\{ \tau_k \right\}_{k=1}^K,d) \in \mathcal{P}}{\sup } \mathbb{E}\left[  \sum_{k=1}^{K} \left( \sum_{i=1}^N \int_{\tau_{k-1}}^{\tau_{k}} e^{-rt} \pi^i(X^i_{T^i(t)})dT^i(t) + e^{-r\tau_k} \bar{\pi}\left( X_{T(\tau)}, d_{\tau_k} \right) \right) \right], \tag{Prize design}
	\end{align}
	subject to the dynamic participation constraints: for all $i$ and all possible histories $h_t$,
	\begin{align*}
	\mathbb{E}\left[ \sum_{k=1}^{K} e^{-r(\tau_k-t)} g d^i_{\tau_k} \mathbbm{1}_{\{ t\leq \tau_k \}} - \int_{t}^{\infty} e^{-rt} (1- \sum_{k=1}^{K}d^i_{\tau_k} \mathbbm{1}_{\{t\geq \tau_k\}}) c^i dT^i(t) \mid h_t \right] \geq 0.
	\end{align*}
	
	\begin{theorem}\label{theorem:winnertakeall}
		Suppose that Assumption \ref{assumption:submartingalepayoffs} holds. Then the \textbf{index contest} solves \eqref{eq:prizedesign}.
	\end{theorem}
	Theorem \ref{theorem:winnertakeall} shows that optimal promotion contest grants the \emph{entire prize to one worker at most}. The optimal contest is a winner-take-all. This is reminiscent of the classic result in \cite{moldovanu2001optimal} of the optimality of a single prize. In our dynamic setting, fully allocating the prize to only one worker is also optimal. The index contest is meritocratic: the worker who performs the best (upon getting the opportunity) is promoted. This contrasts from recent results in dynamic contest theory in which the optimal contest was shown to be more egalitarian (see \cite{halac2017contests} and \cite{ely2021optimal}, for example).
	
	In Online Appendix \ref{app:winnertakeall}, I indicate how to modify the proof of Theorem \ref{theorem:indexability} to obtain Theorem \ref{theorem:winnertakeall}. In particular, it follows the same steps. The only difference is in showing that one can focus on promotion contests in which the promotion times are measurable. Proposition \ref{prop:winnertakeall} in Online Appendix \ref{app:winnertakeall} replaces Proposition \ref{prop:measurablestopping}. The rest of the proof is identical.
	
	\subsection{Transfers}\label{subsec:transfers}
	
	\paragraph{} I ruled out transfers for three reasons in the main model. The first and most fundamental one was that I wanted to focus on the trade-off between the two classical promotion roles, i.e., incentives provision and sorting. The second reason for this restriction is empirical. In most organizations, compensation is promotion based.\footnote{\cite{baker1988compensation} find that ``[m]ost of the average increases in an employee's compensation can be traced to promotions and not to continued service in a particular position.''. See also \cite{gibbs1995incentive} and \cite{bernhardt1995strategic}. It is also consistent with the observed separation of roles: Compensation and benefits managers within the human resource department have authority over the compensation structure, while the assignment of responsibilities and tasks are made within each department by managers that can closely monitor and supervise their team.} This is the case in public administrations, where the salary grid is fixed, for example. Finally, the analysis developed in this article becomes intractable for general wages (although the main trade-off seems to be preserved when workers are protected by limited liability). So a complete analysis of transfers is well beyond the scope of my paper. Nevertheless, in this section, I point out how my model can accommodate restricted forms of transfers. 
	
	Suppose that the principal can only choose transfers that depend on the worker's current type and his effort decision (i.e., pay a flow wage $w^i_t = w^i(a^i_t, X^i_t)$ to worker $i$ at every instant $t\geq 0$) and that the workers are protected by limited liability (i.e. $w^i_t \geq 0$). Then the index context is still optimal, under Assumption \ref{assumption:onesidedjumps}.(ii) (when the workers' types can only jump down), as long as $\pi^i(\cdot) -w^i(1, \cdot)$ is nondecreasing. This can be seen from the proof of Theorem \ref{theorem:indexability} directly. 
	
	For example, if the wage paid to each of the workers is a constant fraction $\beta^i \in [0,1]$ of the flow payoff the principal obtains (i.e., $w^i_t = \beta^i \pi^i(X^i_{T^i(t)})dT^i(t)$), then the index contest is optimal. The strategic index are computed for the payoff process $(1-\beta^i)\pi^i(X^i(t))$, effort costs $c(X^i_t) - \beta^i\pi^i(X^i(t))$, and value of promotion $\tilde{g}^i\left( X^i_t \right) \coloneqq g^i + \beta^i \bar{\pi}^i(X^i_t)$ . One can then imagine that the principal engages in Nash bargaining with the workers (with threat points equal to their outside option) before the game starts to determine the $\beta^i$'s.
	
	\subsection{Different information structures}\label{subsec:differentinformation}
	
	\paragraph{} Finally, the workers' types are assumed to be observable by all the players: by both the other workers and the principal. Interestingly, the index promotion contest remains optimal if each worker only observes his type and the principal does not observe the evolution of the types, but the workers can reveal their current type to the principal credibly. Hence it is easily seen that it is weakly dominant for the workers to reveal their type to the principal when $X^i_t = \underline{X}^i_t$ or $X^i_t = \bar{P}^i\left( \underline{X}^i_{t} \right)$, which is the only information the principal need to implement the index contest. Verifiability is important here: the same result cannot be obtained with cheap talk communication only.

	\section{Conclusion}\label{sec:conclusion}
	
	\paragraph{} I study the design of centralized dynamic contests in a general environment. Workers are heterogeneous and strategic. They have to be incentivized to exert effort, and their types evolve (stochastically) when they work. I showed that despite the richness of the model, the solution is simple and takes the form of an \emph{index contest}. 
	
	My analysis is limited to the specific extension of the multi-armed bandit model I consider, and I do not suggest that my findings would hold in different environments. Some of the assumptions, such as the independence of the type processes, appear crucial and very hard to relax in a significant manner (although one could consider a particular form of conditional independence for multi-parameter processes, known as condition F4, see \cite{el1997synchronization} or \cite{walsh1981optional} for example). However, the intuition behind the result is valid in other environments. For example, when the information about the project's success is private and cannot be credibly communicated but the uncertainty is small, results from the multi-armed bandit literature suggest that \emph{index contest} would still perform well. This can be seen directly by inspecting the principal's payoff in the index contest \eqref{eq:payoffequalitylowerenvelopeandprocesses}. When the uncertainty is small, the lower envelopes of the index processes associated with the case in which the principal observes the workers' types directly or observes a signal are close. Still, characterizing the specific form of the optimal mechanism when the workers have private information about the outcome of the delegation process would be interesting.
	
	More generally, the idea that the endogenous allocation of opportunities or the endogenous acquisition of information affects the final decision when allocating an asset or promoting a worker is very natural and deserves more attention in future research.

	\pagebreak
	
	\bibliography{biblio}
	
	\pagebreak

	\setcounter{section}{0}

	\section{Appendix A}\label{app:appendixA}
	
	\subsection{Jump-Diffusion Processes}\label{app:jumpdiffusion}
	
	\paragraph{} On a probability space $\left( \Omega, \mathcal{F} , \mathbb{P} \right)$, for all $i \in \left\{ 1,\dots, N\right\}$, let $B^i = \left\{ B^i_t\right\}$ be Brownian motion adapted to $\mathcal{F}^i$ and $P^i = \left\{ P^i_t\right\}$ be a homogeneous Poisson point process adapted to $\mathcal{F}^i$, such that $P^i$ and $B^i$ are independent, and the $\left( B^i, P^i \right)$ and $\left(B^j, P^j\right)$ are mutually independent for $i \neq j$. Let $\tilde{N}^i(ds,dz) = N^i(ds,dz) - m^i(dz)ds$ be the Poisson martingale measure generated by $P^i$, where $N^i(ds,dz)$ is the (homogeneous) Poisson counting measure generated by $P^i$ and $m^i(dz)$ is the L\'{e}vy measure on $\mathbb{R}\setminus \{0\}$ generated by $P^i$.
	
	The type of worker $i$, $X^i$, is a stochastic process with values in the open set $\mathcal{X}^i \subseteq \mathbb{R}$\footnote{For simplicity, assume that $X^i$ either does not reach the boundary of the set or that they are absorbing.} and evolves according to the (potentially degenerate) jump-diffusion stochastic differential equation:
	\begin{align}\label{eq:typesdynamics}
	X^i_{t^i} = x^i + \int_{0}^{t^i} \mu^i(X^i_{s}) ds + \int_{0}^{t^i} \sigma^i(X^i_{s}) dB^{i}_{s} + \int_{0}^{t^i} \int_{\mathbb{R}_+} k^i(X^i_{s^-}, z)\tilde{N}^i(ds, dz), 
	\end{align}
	where $x^i \in \mathcal{X}^i$, and $t^i$ is the cumulative effort worker $i$ has put into the project up to time $t$: $t^i = T^i(t) = \int_{0}^{t} \alpha^i_s a^i_s dt$. If $k^i = 0$, \eqref{eq:typesdynamics} is a continuous stochastic differential equation and the type of worker $i$ is a diffusion process. The $\sigma$-field $\mathcal{F}^i_{t^i}$ contains all the information accumulated on worker $i$ when he has put total effort $t^i$ into the project.
	
	I will make the following assumptions:
	\begin{assumption}\label{assumption:sdediffusioncoefficients}
		For all $i\in \left\{1,\dots, N\right\}$, $\mu^i: \mathcal{X}^i \to \mathbb{R}$, $\sigma^i: \mathcal{X}^i \to \mathbb{R}$ is locally Lipschitz continuous and grows at most linearly.
	\end{assumption}
	\begin{assumption}\label{assumption:sdejumpcoefficients}
		For all $i \in \left\{ 1,\dots, N\right\}$,
		\begin{itemize}
			\item $\int_{\mathbb{R} \setminus \{0\}} \frac{\left|z\right|^2}{1+\left|z\right|^2} d m^i(z) < \infty$;
			\item There exists $\rho: \mathbb{R} \setminus \{0\} \to \mathbb{R}_+$ such that $\int_{\mathbb{R} \setminus \{0\}} \rho(z)^2 d\pi(z) <\infty$ and such that $\left| k^i(x,z) - k^i(y,z) \right| \leq \rho(z) \left| x-y \right|$; and
			\item For all $x,y \in Q$ and $z \in \mathbb{R} \setminus \{0\}$, $x+ k^i(x,z) \geq y + k^i(y,z)$.
		\end{itemize}
	\end{assumption}
	The first condition in Assumption \ref{assumption:sdejumpcoefficients} is a restriction on the set of possible jump processes. It is standard in the theory of jump-diffusion and is satisfied by all stable processes. It is satisfied if $m$ is a finite measure which admits a second-order moment, for example.
	
	Assumptions \ref{assumption:sdediffusioncoefficients} and \ref{assumption:sdejumpcoefficients} are sufficient for the existence of a strong solution of \eqref{eq:typesdynamics} and the validity of a comparison theorem.
	
	\begin{lemma}\label{lemma:strongsolution}
		Under Assumptions \ref{assumption:sdediffusioncoefficients} and \ref{assumption:sdejumpcoefficients}, the stochastic differential equation \eqref{eq:typesdynamics} has a unique strong solution.
	\end{lemma}
	
	\begin{proof}[Proof of Lemma \ref{lemma:strongsolution}]
		This follows from Theorem 310 in \cite{rong2006theory}.
	\end{proof}
	An immediate consequence of Lemma \ref{lemma:strongsolution} and Proposition 2.1 in \cite{wang2010regularity} is that the process $X^i$ is Feller.
	
	Next, I show that it also satisfy the other Assumptions I made in Section \ref{subsec:dynamics}, under some additional conditions. Assumption \ref{assumption:comparisontheorem} holds by Theorem 295 in \cite{rong2006theory}. Assumption \ref{assumption:onesidedjumps} is satisfied if, for all $i \in \left\{ 1,\dots, N \right\}$, for all $x \in \mathcal{X}^i$, (i) either $k^i(x,z) \geq 0$; or (ii) $k^i(x,z) \leq 0$. Finally, Assumption \ref{assumption:potentialforimprovment} holds if $\sigma^i(x) > 0$ for all $x \in \mathcal{X}^i$ or if $\mu^i(x)>0$ for all $x$ such that $\sigma^i(x) = 0$. This can be seen from an application of Girsanov's theorem, Bluementhal's 0-1 law, and the Dvoretzky, Erdos, and Kakutani theorem (Theorem 9.13 in \cite{karatzas1998brownian}).
	
	\subsection{Proof of Lemma \ref{lemma:nonnegativecontinuations}}\label{app:nonnegativecontinuations}
	
	I first show that in any implementable promotion contest, each worker's continuation value after any history is nonnegative. I prove the contrapositive. Let $(T,\tau,d)$ be a promotion contest and suppose that $U^i_t(T,\tau,d) < 0$ for some $i \in \{1, \dots, N\}$ and $t\geq 0$ after some history. I claim that the promotion contest $(T,\tau,d)$ is not implementable, i.e., there is no (weak) Perfect Bayesian equilibrium that generates it. To see this, note that sequential rationality is violated for worker $i$: were he to stop exerting effort forever (which is an admissible strategy by condition (ii)), his continuation payoff would be nonnegative. So $(T,\tau,d)$ is not implementable. 
	
	Next I show that any promotion contest that gives a nonnegative continuation value to each worker after any history is implementable. Let $(T,\tau,d)$ be such a promotion contest. Consider the strategies $\left\{ a^i_t = 1 \right\}_{t\geq 0} $, for all $i \in \left\{ 1,\dots, N\right\}$ and the principal's choosing the contest defined by $\left( \left\{ \alpha^i_t = \frac{dT^{s,i}(t)}{dt}\right\}_{t\geq 0}, \tau, d\right)$ as long as $a^i_t =1$ for all workers and taking her outside option immediately otherwise. The above strategy profile is admissible. By condition (i), it is feasible. Moreover, it is immediate to see that no worker has a profitable deviation: if worker $i$ deviates, he gets a continuation payoff of $0$, while, if he does not, he gets a continuation payoff of $U^i_t\geq 0$. The result then follows. \qed
	
	\subsection{Proof of Proposition \ref{prop:strategicbanditcontest}}\label{app:strategicbanditcontest}
	
	\paragraph{} I first show that the index contest can be implemented in a (weak) Perfect Bayesian equilibrium. Consider the following strategies $\left\{ a^i_t = \mathbbm{1}_{\{ t\leq \tau^0 \wedge \bigwedge_{i=1}^N \tau^{s,i} \}} + d^i_{ \tau^0 \wedge \bigwedge_{i=1}^N \tau^{s,i}} \mathbbm{1}_{\{  t> \tau^0 \wedge \bigwedge_{i=1}^N \tau^{s,i} \}} \right\}_{t\geq 0} $, for all $i \in \left\{ 1,\dots, N\right\}$, and $\left( \left\{ \alpha^i_t = \frac{dT^{s,i}(t)}{dt}\right\}_{t\geq 0},  \tau^0 \wedge \bigwedge_{i=1}^N \tau^{s,i}, d^s\right)$. 
	
	By Theorem 7.1 in \cite{el1997synchronization}, the principal has no profitable deviation. There remains to show that no worker has a profitable deviation. This follows from the structure of the index contest.
	
	The delegation rule $T^{s}$ is an index delegation rule, and, hence, each $T^{s,i}$ is flat of the set
	\begin{align*}
	\left\{ t \geq 0 \, : \, \underline{\Gamma}^{s,i}_{T^i(t)} = \bigvee_{j=1}^N \underline{\Gamma}^{s,j}_{T^j(t)} \right\} \, \mathbb{P}\text{-a.s..}
	\end{align*}
	But, for all $i =1,\dots, N$, $\Gamma^i_{T^{s,i}(t)}$ decreases only on the set $\left\{ X^i_{T^{s,i}(t)} = \underline{M}^i_{T^{s,i}(t)} \right\}$ by lemma \ref{lemma:monotonicityofVp}, and, therefore, using Assumption \ref{assumption:potentialforimprovment}, for almost every $t$, if worker $i$ is delegated the project, then $\Gamma^i_{T^{s,i}(t)} > \underline{\Gamma}^i_{T^{s,i}(t)}$. In this case, only worker $i$ is delegated the project (i.e., at most one worker exerts effort at almost every instant $t \geq 0$.).
	
	To see this, note that, by Proposition 10 in \cite{kaspi1998multi}, the sets $\mathcal{D}^i = \left\{ t\geq 0 \, : \, \sigma^i(\Gamma^i_{t^-}) > t \right\}$ are $\mathcal{F}^i$-totally inaccessible. So two arms pulled simultaneously cannot start an excursion by an upward jump of at least one of their indices (necessarily from the value of their common minimum $\underline{\Gamma}$). This can be seen this by contradiction. Suppose not, i.e., two workers $i$ and $j$ are delegated simultaneously at time $t$. This implies that both $T^i(t)$ belongs to $\mathcal{D}^i$ and $T^j(t)$ belongs to $\mathcal{D}^j$, and both $\underline{\Gamma}^i_{T^{s,i}(t)}$ and $\underline{\Gamma}^j_{T^{s,j}(t)}$ are strictly decreasing at $t$. So let $\underline{\Gamma}$ be a point from which $\Gamma^i_{T^{s,i}(t)}$ starts an excursion from its minimum. Then, if $\Gamma^j_{T^{s,j}(t)}$ jumps upward from its minimum $\underline{\Gamma}$, $\left\{ \underline{\Gamma}^j_{t^-} = \underline{\Gamma}, \, \underline{\Gamma}^j_{t -\epsilon} > \underline{\Gamma} \text{ for all } \epsilon >0 \right\} \cap \mathcal{D}^j \neq \emptyset$. But $\left\{ \underline{\Gamma}^j_{t^-} = \underline{\Gamma}, \, \underline{\Gamma}^j_{t -\epsilon} > \underline{\Gamma} \text{ for all } \epsilon >0 \right\}$ is a predictable set, and therefore $\left\{ \underline{\Gamma}^j_{t^-} = \underline{\Gamma}, \, \underline{\Gamma}^j_{t -\epsilon} > \underline{\Gamma} \text{ for all } \epsilon >0 \right\} \cap \mathcal{D}^j = \emptyset$: a contradiction. Therefore if the indices of two arms start an excursion from a level set $\underline{\Gamma}$ at the same time, it must be from a point of continuity for both ${\Gamma}^i_{T^{s,i}(t)}$ and ${\Gamma}^j_{T^{s,j}(t)}$. But, then, the priority rule defined in \eqref{eq:strategicindexpolicy} specifies which of the two arms is pulled first and until the end of its excursion.
	
	It follows that the continuation value of worker $i$ during an excursion when his type is $X$ and his minimum is $\underline{X}$ coincides with the continuation value in the single agent problem when his type is $X$ and his minimum is $\underline{X}$, which, by theorem \ref{theorem:singleagentoptimal}, is nonnegative.
	
	Finally, for all $j \neq i$, $\Gamma^j_{T^{s,j}(t)} = \underline{\Gamma}^j_{T^{s,j}(t)}$. But, on $\Gamma^j_{T^{s,j}(t)} = \underline{\Gamma}^j_{T^{s,j}(t)}$, $\left\{ X^i_{T^{s,j}(t)} = \underline{X}^j_{T^{s,j}(t)} \right\}$. So the continuation value of worker $j$ is zero by Theorem \ref{theorem:singleagentoptimal}, and no unemployed worker wants to quit. 
	
	Thus no worker has a profitable deviation, as it is easily seen that the cost from delaying the moment they get the reward by shirking is always greater than the saved effort cost when the continuation value of a worker is nonnegative as, for all $s\geq 0$,
	\begin{align*}
	e^{-rs} U^i_t \leq U^i_t,
	\end{align*}
	and the continuation value $U^i_t$ only depends on the current state of the game which remains fixed when the worker shirks.
	
	The second part of Proposition \ref{prop:strategicbanditcontest} follows from the Martingale argument in \cite{el1994dynamic}. It is reproduced in Lemma \ref{lemma:indexstrategiesexpectedvalue} in Appendix \ref{app:strategicindex} for completeness. \qed
	
	\subsection{Proof of Proposition \ref{prop:Gittinsindexpassivearms}}\label{app:Gittinsindexpassivearms}
	
	Let $c^i(\cdot) \coloneqq 0$ for all $i \in \left\{ 1,\dots, N\right\}$. Then $\bar{P}^i(x) = \sup\left\{ x \in \mathcal{X}^i \right\}$ for all $i \in \left\{ 1,\dots, N\right\}$, and the result follows from Theorem \ref{theorem:indexability}. \qed
	
	\subsection{Proof of Proposition \ref{prop:relaxation}}\label{app:constrainedoptimization}
	
	\paragraph{} I show that any implementable promotion contest $(T,\tau, d)$ is feasible for the relaxed program. Let $(T, \tau, d) \in \mathcal{P}^i$ be an implementable promotion contest. Let $i \in \{1,\dots, N\}$ and $t \geq 0$. By the law of iterated expectations, one has
	\begin{align*}
	\mathbb{E} & \left[ e^{-r \left(\tau- \tau\wedge t\right)} g \mathbbm{1}_{\{d = i\}} - \int_{\tau\wedge t}^\tau e^{-r(s-\tau\wedge t)} c d T^i(s)  \mid \mathcal{\mathcal{F}}^i_{T^i(t)} \right] \\ 
	& = \mathbb{E}\left[ \mathbb{E}\left[ e^{-r \left(\tau- \tau\wedge t\right)} g \mathbbm{1}_{\{d = i\}} - \int_{\tau\wedge t}^\tau e^{-r(s-\tau\wedge t)} c d T^i(s) \mid \mathcal{G}^T_t \right]  \mid \mathcal{\mathcal{F}}^i_{T^i(t)} \right] \\
	& = \mathbb{E}\left[ U^i_t \mid \mathcal{\mathcal{F}}^i_{T^i(t)} \right] \\
	& \geq 0;
	\end{align*}
	where the second equality is by definition and the inequality follows from lemma \ref{lemma:nonnegativecontinuations}. Therefore any promotion contest $(T, \tau, d) \in \mathcal{P}^I$ is feasible in \eqref{eq:relaxedprogram}. 
	
	By Lemma \ref{lemma:nondecreasingd^i} below, for any promotion contest, we can find a payoff equivalent promotion contest with $d \in \mathcal{C}^*$. So for any $d$ is feasible in the original problem, there is a $d$ feasible in the new problem that gives the same payoff to the principal. Finally, by Lemma \ref{lemma:stoppingtimesareextremepointsofS} below, the set of stopping times $\mathcal{T}$ is a subset of $\mathcal{S}\left( \mathcal{G}^T \right)$. Therefore the set of randomized promotion contest $\mathcal{P}^r$ is a superset of $\mathcal{P}$, and, hence, and the result follows: ${\Pi} \geq \Pi^M$. \qed
	
	\begin{lemma}\label{lemma:nondecreasingd^i}
		For all promotion contest $(T,\tau,d)$, there exists an alternative promotion contest $(T,\tau,\tilde{d})$ with $\tilde{d}^i_t$ monotone $\mathbb{P}$-a.s. for all $i \in \left\{0, 1,\dots, N \right\}$ such that $(T,\tau,d)$ and $(T,\tau,\tilde{d})$ give the same payoff to the principal and to all the workers.
	\end{lemma}
	
	\begin{proof}[Proof of Lemma \ref{lemma:nondecreasingd^i}]
		Let $(T,\tau,d)$ be a promotion contest. Define $\tilde{d}$ as follows: for all $t \geq 0$,
		\begin{align*}
		\forall i \in \left\{ 1, \dots, N\right\}, \tilde{d}^i_t = d^i_{\tau} \mathbbm{1}_{\{t \geq \tau\}} \text{ and } \tilde{d}^0_t = d^0_{\tau} \mathbbm{1}_{\{t \geq \tau\}} + \mathbbm{1}_{\{t < \tau\}}.
		\end{align*}
		Clearly, $\tilde{d}^i$ is $\mathbb{P}$-a.s. monotone. Furthermore, $(T,\tau,d)$ and $(T,\tau,\tilde{d})$ gives the same payoff to all the players since $d_{\tau} = \tilde{d}_{\tau}$. Finally $\tilde{d}$ is $\mathcal{G}^T$-adapted, and therefore $(T,\tau, \tilde{d})$ is a promotion contest.
	\end{proof}
	
	\begin{lemma}\label{lemma:stoppingtimesareextremepointsofS}
		The set of $\mathcal{G}$-stopping times $\mathcal{T}\left(\mathcal{G}\right)$ can be identified with the set of extreme points of $\mathcal{S}\left( \mathcal{G}\right)$.
	\end{lemma}
	
	\begin{proof}[Proof of Lemma \ref{lemma:stoppingtimesareextremepointsofS}]
		Any $\mathcal{G}$-stopping time $\tau$ can be identified with the random stopping time $S^{\tau}$ defined as $S^{\tau}(t) \coloneqq \mathbbm{1}_{\left\{ t \geq \tau \right\}}$, which is easily seen to be an extreme point of $\mathcal{S}\left( \mathcal{G}\right)$. On the other hand, if $S(\cdot) \neq \mathbbm{1}_{\left\{ \cdot \geq \tau \right\}}$ for some $\mathcal{G}$-stopping time, there exists $\bar{s} \in (0,1)$ such that the processes $S^1$ and $S^2$ defined by
		\begin{align*}
		{S}^1_t \coloneqq \frac{1}{\bar{s}} S(t) \wedge \bar{s} \text{ and } S^2_t \coloneqq \frac{1}{1-\bar{s}} S(t) \vee \left(1 - \bar{s}\right)
		\end{align*}
		are different elements of $\mathcal{S}\left( \mathcal{G} \right)$. But then, $S$ is not an extreme point of $\mathcal{S}\left( \mathcal{G} \right)$.
	\end{proof}
	
	\subsection{Omitted Proofs for Section \ref{subsec:singlearmproblem}}\label{app:singlearmproblem}
	
	\subsubsection{Proof of Theorem \ref{theorem:singleagentoptimal}}\label{app:singlearmproblemproof}
	
	\paragraph{Step 1:} Define $\mathcal{Q}^i \coloneqq \left\{ x^i_q, \, 0\leq q \leq Q+1  \right\} \subseteq \mathcal{X}^i$, a grid of points in the state space such that $x^i_0 \coloneqq X^i_0$ and $x^i_{q+1} < x^i_q$ for all $i = 1,\dots, N$ and $q =0, \dots Q$.
	
	I solve the stopping problem with constraints:
	\begin{align}\label{eq:relaxedprogramsinglearmdiscretized}
	{\Pi}^{i}_{\mathcal{Q}} \coloneqq \underset{(\tau, d^i)}{\sup}\, {\mathbb{E}}\left[ \int_{0}^\tau e^{-rt} \pi^i\left(X^i_{t}\right)dt + e^{-r\tau}  \left( d^i_{\tau} \int_{\tau}^{\infty} e^{-r(t -\tau)} \pi^i\left(X^i_t\right) dt + (1-d^i_{\tau}) W \right) \right] \tag{RRP$^i$}
	\end{align}
	subject to, for all $q = 0, \dots, Q$,
	\begin{align}\label{eq:RDPxiq}
	\mathbb{E} \bigg[ e^{-r \tau}  g^i d^i_{\tau} + \int_{\tau\wedge\tau^i(x^i_q)}^{\tau} e^{-rt} {c}^{i}\left( X^i_{t}\right) d t \bigg] \geq 0, \tag{RDP($x^i_q$)}
	\end{align}
	where $\tau^i(x^i_q) = \inf \left\{ t \geq 0 \, : \, X^i_t \leq x^i_q \right\}$.
	
	\begin{lemma}\label{lemma:singlearmdiscretized}
		The value of relaxed problem \eqref{eq:relaxedprogramsinglearmdiscretized} is weakly greater than the value of problem \eqref{eq:relaxedproblemi}: ${\Pi}^{i}_{\mathcal{Q}} \geq {\Pi}^i$.
	\end{lemma}
	
	\begin{proof}[Proof of lemma \ref{lemma:singlearmdiscretized}]
		Any feasible $(\tau,d)$ in \eqref{eq:relaxedproblemi} satisfies all the \eqref{eq:RDPxiq} constraints by Lemma \ref{lemma:equivalentconstraints} in Appendix \ref{app:existencesupportinglemmas}. So the choice set in \eqref{eq:relaxedprogramsinglearmdiscretized} is weakly larger than the choice set in \ref{eq:relaxedproblemi}, and ${\Pi}^{i}_{\mathcal{Q}} \geq {\Pi}^i$.
	\end{proof}
	
	\paragraph{Step 2:} To accommodate the constraints, set up the Lagrangian associated with \eqref{eq:relaxedprogramsinglearmdiscretized}.
	\begin{lemma}\label{lemma:lagrangian}
		There exists $\left(\lambda_0, \dots, \lambda_Q\right) \in \mathbb{R}^{Q+1}_+$ with ($\lambda_q >0$ if and only if \eqref{eq:RDPxiq} is binding for $\tau^i(x^i_q)$) such that problem \eqref{eq:relaxedprogramsinglearmdiscretized} is equivalent to the unconstrained pure stopping problem:
		\begin{align}\label{eq:Lagrangianstopping}
		{\Pi}^{i}_{\mathcal{Q}} = \underset{(\tau, d^i)}{\sup }\, \mathbb{E}\Bigg[ & \int_{0}^{\tau} e^{-rt}\left( \pi^i(X^i_{t}) - \sum_{q=0}^Q  \lambda_q {c}^{i}\left(X^i_t\right) \mathbbm{1}_{\{ t \geq \tau(x^i_{q}) \}} \right) dt \notag \\
		& + e^{-r\tau}  d^i_{\tau}\left( \sum_{q=0}^{Q} \lambda_q g^i + \int_{\tau}^{\infty} e^{-r(t -\tau)} \pi^i(X^i_t)dt  \right) +e^{-r\tau} (1-d^i_{\tau}) W \bigg]
		\end{align}
	\end{lemma}
	Lemma \ref{lemma:lagrangian} follows from Theorem 1 in \cite{balzer2002duality}.
	
	\begin{proof}[Proof of lemma \ref{lemma:lagrangian}]
		For $\left(\tau \coloneqq \epsilon, d^i =1\right)$, $\epsilon>0$ small, all constraints are slack. So Theorem 1 in \cite{balzer2002duality} applies: there exist Lagrange multipliers $\left(\lambda_q \right)_{q=0}^Q \in \mathbb{R}^{Q+1}_+$ such that the optimal promotion time and decision rule, $\left( \tau, d^i\right)$, solve
		\begin{align*}
		{\Pi}^{i}_{\mathcal{Q}} = \underset{(\tau, d^i)}{\sup }\, \mathbb{E}& \left[ \int_{0}^{\tau} e^{-rt} \pi^i(X^i_{t}) dt + e^{-r\tau}  \left( d^i_{\tau} \int_{\tau}^{\infty} e^{-r(t -\tau)} \pi^i(X^i_t)dt + (1-d^i_{\tau}) W \right) \right] \\
		& \qquad + \sum_{q=0}^{Q} \lambda_q \mathbb{E}\left[ e^{-r \tau} g^i d^i_{\tau} - \int_{\tau\wedge\tau^i(x^i_q)}^{\tau} e^{-rt} {c}^{i}\left(X^i_t\right) d t  \right] 
		\end{align*}
		Rearranging yields \eqref{eq:Lagrangianstopping}.
	\end{proof}
	
	\paragraph{Step 3:} Identify a promotion contest that gives the principal a payoff weakly higher than the value of \eqref{eq:relaxedprogramsinglearmdiscretized}. 
	
	Let $\tau^i_q \coloneqq \inf\left\{ t\geq 0 \, : \, X^i_t \not \in \left( x^i_{q+1}, x^i_q \right) \right\}$. Define $\tilde{g}^i \coloneqq g^i$ if $X^i$ only jumps up and 
	\begin{align}
	\bar{g}^i \coloneqq \inf\left\{ \tilde{g} \geq 0 \, : \,  \underset{q \in \left\{0, \dots, Q-1\right\}}{\inf } \underset{x \in \left( x^i_{q+1}, x^i_{q} \right)}{\inf }  \mathbb{E}\left[e^{-r\tau^i_q} \tilde{g} - \int_{0}^{\tau^i_q} e^{-rt} c^i\left(X^i_t\right)dt \mid x\right] \geq g^i \right\}.
	\end{align}
	if $X^i$ only jumps down. Finally, let $\tilde{P}^i(\underline{x})$ be defined as $\bar{P}^i(\underline{x})$ but for $g^i$ replaced by $\tilde{g}^i$. Then
	\begin{proposition}\label{prop:structurestopping}
		Let
		\begin{align*}
		\tilde{P}^{i}_{\mathcal{Q}}\left( \underline{x} \right) \coloneqq \sum_{q=0}^{Q}\tilde{P}^i(x^i_q) \mathbbm{1}_{\{ \underline{x} \in \left( x^i_{q+1}, x^i_q \right] \}}.
		\end{align*}
		The pair $\left(\bar{\tau}_{\mathcal{Q}}, d^{\mathcal{Q}}\right) \in \mathcal{P}^i$ with
		\begin{align*}
		\tau_{\mathcal{Q}} \coloneqq \inf\left\{ t\geq 0 \, : \, X^i_t \not \in \left[ \underline{p}^i_{\mathcal{Q}}, \tilde{P}^{i}_{\mathcal{Q}}\left( \underline{X}^i_{t} \right)\right) \right\} \text{ and } d^{i, \mathcal{Q}}_{\tau_{\mathcal{Q}}} \coloneqq \mathbbm{1}_{\{ X^i_{\tau_{\mathcal{Q}}} \geq \tilde{P}^{i}_{\mathcal{Q}}\left( \underline{X}^i_{\tau_{\mathcal{Q}}}\right) \}}
		\end{align*}
		gives a weakly greater payoff to the principal than any feasible promotion contest in \eqref{eq:relaxedprogramsinglearmdiscretized}.
	\end{proposition}
	
	The intuition for the above proposition is clear. The principal always wants to wait and obtains as much information as possible before making a final and irreversible decision. Her option value of waiting is always positive. On the other hand, her cost of waiting is null since, conditionally on choosing to promote the worker, her continuation value is a martingale. The promotion contest $\left( \tau_{\mathcal{Q}}, d^{\mathcal{Q}} \right)$ guarantees that the principal waits as long as possible before making a decision.
	
	\begin{proof}[Proof of Proposition \ref{prop:structurestopping}]
		Distinguish two cases corresponding to the two cases of Assumption \ref{assumption:onesidedjumps}.
		\begin{itemize}
			\item \textbf{$X^i$ only jumps up.} Let $\left({\tau}^*, d^{*}\right)$ be the optimal promotion contest in \eqref{eq:relaxedprogramsinglearmdiscretized} given by Lemma \ref{lemma:solutiondiscretizedsinglearmupjumps} in Appendix \ref{app:singlearmproblemproofsupportinglemmas}. The payoff the principal gets from $\left( \tau_{\mathcal{Q}}, d^{\mathcal{Q}} \right)$ is weakly greater than the value of \eqref{eq:DRRPi} since $\tau_{\mathcal{Q}} \geq \tau^*$ $\mathbb{P}$-a.s. and the principal takes her outside option when $X^i$ falls below the same level $\underline{p}^i_{\mathcal{Q}}$ in both cases. This concludes the proof.
			
			\item \textbf{$X^i$ only jumps down.} Let $\bar{\tau}_{\mathcal{Q}}$ be the optimal stopping time in \eqref{eq:DRRPi} given by Lemma \ref{lemma:upperthreshold} in Appendix \ref{app:singlearmproblemproofsupportinglemmas}. By Lemma \ref{lemma:DRRPigeqRRPi} in Appendix \ref{app:singlearmproblemproofsupportinglemmas}, the value associated with $\bar{\tau}_{\mathcal{Q}}$ is weakly greater than the value of \eqref{eq:relaxedprogramsinglearmdiscretized}. But the payoff the principal gets from $\left( \tau_{\mathcal{Q}}, d^{\mathcal{Q}} \right)$ is weakly greater than the value of \eqref{eq:DRRPi} since $\tau_{\mathcal{Q}} \geq \bar{\tau}_{\mathcal{Q}}$ $\mathbb{P}$-a.s. and the principal takes her outside option when $X^i$ falls below the same level $\underline{p}^i_{\mathcal{Q}}$ in both cases. This concludes the proof.
		\end{itemize}
	\end{proof}
	
	\paragraph{Step 4:} Finally Theorem \ref{theorem:singleagentoptimal} is obtained by letting the grid $\mathcal{Q}$ become finer and finer.
	
	\begin{proof}[Proof of Theorem \ref{theorem:singleagentoptimal}]
		Let $\left(\mathcal{Q}^n\right)_{n \in \mathcal{N}} \subseteq 2^{\mathcal{X}^i}$ be a sequence of grids in $\mathcal{X}^i$ such that $\mathcal{Q}^{n} \subseteq \mathcal{Q}^{n+1}$ for all $n \in \mathbb{N}$ and such that $\underset{n \to \infty}{\lim } \, \mathcal{X}^{n}$ is dense in $\mathcal{X}^i$. Let $\left(\tau^n, d^n\right)_{n\in \mathbb{N}}$ be the pair given by Proposition \ref{prop:structurestopping}. Define 
		\begin{align*}
		\tau^* & \coloneqq \inf\left\{ t \geq 0 \, : \, X_{t} \not \in \left[\underline{p}^*, \bar{P}^i(\underline{X}_{\tau^*}) \right) \right\}, \\
		\text{ and } d^*_{\tau^*} & \coloneqq  \mathbbm{1}_{\{X_{\tau^*} = {P}(\underline{X}_{\tau^*}) \}};
		\end{align*}
		where $\underline{p}^*$ is an accumulation point of $\left( \underline{p}^i_{\mathcal{Q}^n} \right)_{n\in \mathbb{N}}$. Along a subsequence, $\left(\tau^n, d^n\right)_{n\in \mathbb{N}}$ converges to $\left(\tau^*, d^*\right)$ $\mathbb{P}$-a.s. (as $\tilde{g}^i_{\mathcal{Q}^n} \to {g}^i$). Since, for all $n \in \mathbb{N}$, $\Pi^i_{\mathcal{Q}^n} \geq \Pi^i$, it follows that $\left( \tau^*, d^* \right)$ yields a value greater than $\Pi^i$ to the principal.
		
		But, $\left( \tau^*, d^* \right)$ is feasible in \eqref{eq:relaxedproblemi}; i.e., $\left( \tau^*, d^* \right)$ satisfies the dynamic participation constraint \eqref{eq:DPCi}. This follows from Lemma \ref{lemma:continuationisnonnegative}. Therefore $\left( \tau^*, d^* \right)$ is optimal in \eqref{eq:relaxedproblemi}.
		
		There remains to show that $\underline{p}^* \coloneqq \underline{p}^i$, which follows from noting that otherwise $(\tilde{\tau}^*, \tilde{d}^*)$ with
		\begin{align*}
		\tilde{\tau}^* & \coloneqq \inf\left\{ t \geq 0 \, : \, X_{t} \not \in \left[\underline{p}^i, \bar{P}^i(\underline{X}_{\tau^*}) \right) \right\}, \\
		\text{ and } \tilde{d}^*_{\tau^*} & \coloneqq  \mathbbm{1}_{\{X_{\tau^*} = {P}(\underline{X}_{\tau^*}) \}}
		\end{align*}
		yields a greater payoff to the principal.
	\end{proof}
	
	\subsubsection{Supporting Lemmas for the proof of Theorem \ref{theorem:singleagentoptimal}}\label{app:singlearmproblemproofsupportinglemmas}
	
	\paragraph{Supporting Lemmas for Step 3:} First characterize the solution $\left( \tau_{\mathcal{Q}}, d^{\mathcal{Q}}\right)$ of \eqref{eq:relaxedproblemi}. Let
	\begin{align*}
	x^i_{W} \coloneqq \sup \left\{ x \in \mathcal{X}^i \, : \,\mathbb{E} \left[ \int_{0}^{\infty}e^{-rt} \pi^i \left(X^i_t \right)dt \mid x\right] + \sum_{q=0}^Q \lambda_q g^i \leq W \right\}.
	\end{align*}
	
	\begin{lemma}\label{lemma:structurestopping}
		There exists $\left( \tau_{\mathcal{Q}}, d^{\mathcal{Q}} \right)$ that solves \eqref{eq:Lagrangianstopping} with (i) $\tau_{\mathcal{Q}} \coloneqq \inf\left\{ t \geq 0 \, : \, X^i_{t} \not \in S\left(t\right) \right\}$, where $S\left( t \right)$ is a correspondence constant on $[\tau^i(x^i_{q}), \tau^i(x^i_{q+1}))$ for all $q \in \left\{ 0, \dots, Q\right\}$, such that $S\left( t \right) \cap \left\{ x \in \mathcal{X}^i \, : \, x < x^i_W \right\} = \left(-\infty, \underline{p}^i_{\mathcal{Q}}\right)$ for some threshold $\underline{p}^{i}_{\mathcal{Q}} \in \mathcal{X}^i$ and  ${S}(\tau^i(x^i_q)) =  {S}(\tau^i(x^i_{q+1}))$ if \eqref{eq:RDPxiq} is not binding at $\tau^i(x^i_{q+1})$, and (ii) $d^{i,\mathcal{Q}}_{\tau_{\mathcal{Q}}} = \mathbbm{1}_{\{X^i_{\tau_{\mathcal{Q}}} \geq {x}^i_W\}}$.
		
		Moreover $\mathbb{P}\left( d^i_{\tau} = 0 \right) >0$.
	\end{lemma}
	
	\begin{proof}[Proof of Lemma \ref{lemma:structurestopping}]
		Let $\bar{A} = \left( A_0, \dots, A_Q \right)$ with $A = \mathbbm{1}_{\{ \underline{X}^i_t \in \left(x^i_{q+1}, x^i_q\right] \}}$. The process $(X^i_t, \bar{A})$ on the extended state space $\left\{ (x, t_0, \dots, t_Q) \in \mathcal{X}^i \times \left\{ 0,1 \right\}^{Q+1} \, : \,  x\geq \underline{x} \right\}$ inherits the Feller property from $X^i$ under $\mathbb{P}$. The result then follows from Theorem \ref{theorem:MarkovstoppingPeskir2.4} in Appendix \ref{app:usefulresultsoptimalstopping}. In particular, for all $q \in \left\{ 0, \dots, Q\right\}$, the value function with $A^q =1$ is given by
		\begin{align}\label{eq:valuefunctionVp}
		V_q(x) \coloneqq \underset{\tau, d^i}{\sup }\, \mathbb{E}&\Bigg[ \int_{0}^{\tau\wedge \tau^i(x^i_{q+1})} e^{-rt} \left( \pi^i(X^i_{t}) - \sum_{k=0}^q \lambda_k {c}^{i}\left(X^i_t\right) \right) dt \notag \\
		&+  e^{-r\tau\wedge\tau^i(x^i_{q+1})} \Bigg\{  V_{q\left(\underline{X}^i_{\tau^{i}\left( x^i_{q+1} \right)}\right)} \left( X^i_{\tau^i(x^i_{q+1})}\right) \mathbbm{1}_{\{\tau \geq \tau^i(x^i_{q+1})\}} \notag \\ 
		& + \mathbbm{1}_{\{\tau < \tau^i(x^i_{q+1})\}} \Bigg[ d^i_{\tau }\Bigg( \sum_{q=0}^{Q} \lambda_q g^i + \int_{\tau}^{\infty} e^{-r(t -\tau)} \pi^i(X^i_t) dt \Bigg) + (1-d^i_{\tau}) W \Bigg] \Bigg\} \mid x \Bigg].
		\end{align}
		It is clear that $d^i_{\tau} = 1$ if and only if $\mathbb{E}\left[ \sum_{q=0}^{Q} \lambda_q g^i + \int_{\tau}^{\infty} e^{-r(t -\tau)} \pi^i(X^i_t) dt \mid \mathcal{F}^i_{\tau} \right] \geq W$, and therefore that $d^i$ only depends on $X^i_{\tau}$, as $X^i$ is Feller, and, hence, has the strong Markov property. So
		\begin{align*}
		V_q(x) = \underset{\tau}{\sup }\, \mathbb{E}&\Bigg[ \int_{0}^{\tau\wedge \tau^i(x^i_{q+1})} e^{-rt} \left( \pi^i(X^i_{t}) - \sum_{k=0}^q \lambda_k {c}^{i}\left(X^i_t\right) \right) dt \notag \\
		&+  e^{-r\tau\wedge\tau^i(x^i_{q+1})} \Bigg\{  V_{q\left(\underline{X}^i_{\tau^{i}\left( x^i_{q+1} \right)}\right)} \left( X^i_{\tau^i(x^i_{q+1})}\right) \mathbbm{1}_{\{\tau \geq \tau^i(x^i_{q+1})\}} \notag \\ 
		& + \mathbbm{1}_{\{\tau < \tau^i(x^i_{q+1})\}} \Bigg( \sum_{q=0}^{Q} \lambda_q g^i + \int_{\tau}^{\infty} e^{-r(t -\tau)} \pi^i(X^i_t) dt \Bigg) \vee W \Bigg\} \mid x \Bigg].
		\end{align*}
		From Theorem \ref{theorem:MarkovstoppingPeskir2.4} in Appendix \ref{app:usefulresultsoptimalstopping} again, the smallest optimal stopping time is given by
		\begin{align*}
		\tau^{\mathcal{Q}} \coloneqq \inf\left\{ (x, q)  \, : \, V_q(x) =  \sum_{q=0}^{Q} \lambda_q g^i +\mathbb{E}\left[ \int_{0}^{\infty} e^{-rt} \pi^i\left(X^i_t \right)dt \mid x\right] \right\}.
		\end{align*}
		There remains to show that the stopping region $S_q$ on $\left[\tau^i(x^i_q), \tau^i(x^i_{q+1})\right)$ can be taken to be such that
		\begin{align*}
		S_q \cap\left\{ x \in \mathcal{X}^i \, : \, x < x^i_W \right\} & = \left\{ x \in \mathcal{X}^i \, : \, x \geq x^i_{q+1} \text{ and } V_q(x) =  \sum_{q=0}^{Q} \lambda_q g^i +\mathbb{E}\left[ \int_{0}^{\infty} e^{-rt} \pi^i\left(X^i_t \right)dt \mid x\right]  \right\} \\
		& = \left[ \underline{p}^{i}_q, \bar{P}^{i}_{q} \right).
		\end{align*}
		
		By Lemma \ref{lemma:monotonicityofVp}, $V_q(x)$ is nondecreasing in $x$. It follows that, if $x' \leq x^i_W$ is such that $x' \in S_q$, then
		\begin{align*}
		\left\{ x \in \mathcal{X}^i \, : \, x \leq x' \right\} \subseteq S_q
		\end{align*}
		To see this, note that by Lemma \ref{lemma:monotonicityofVp}, for all $x'' \leq x'$, $W = V_q(x') \leq V_q(x'') \geq W$, and, hence, $V_q(x'') = W$. Therefore, it is optimal for the principal to stop at $x''$ and take her outside option, and $S_k \cap \left\{ x \in \mathcal{X}^i \, : \, x \leq x^i_W \right\} = (-\infty, \underline{p}^{i}_k)$ or $(-\infty, \underline{p}^{i}_q]$ with $\underline{p}^{i}_q \coloneqq \sup \left\{ x \in S_q \, : \, x\leq x^i_W \right\}$. 
		
		Furthermore, I claim that $\underline{p}^{i}_q \leq \underline{p}^{i}_{q+k}$ for all $k \in \left\{ 1,\dots, Q-q \right\}$. This follows from the definition of $V^q$ since ${c}^{i}\geq 0$ and $\lambda_j \geq 0$ for all $j \geq 0$. Letting $\underline{p} \coloneqq \inf\left\{ \underline{p}^{i}_q \, : \, \underline{p}^{i}_q > -\infty \right\}$ if there exists $\underline{p}^{i}_q > -\infty$, one sees that the principal stops and takes her outside option (if she ever does) the first time $X^i_t$ enters $\left( -\infty, \underline{p}^{i}_{\mathcal{Q}} \right)$ or $\left( -\infty, \underline{p}^{i}_{\mathcal{Q}} \right]$. 
		
		Note here that $V_q(x)$ is right-continuous and increasing, hence upper semicontinuous. Therefore, on $\left( - \infty, x^i_W \right)$, the stopping region is open, and hence $S_q \cap \left( - \infty, x^i_W\right) = \left( -\infty, \underline{p}^{i}_{\mathcal{Q}}  \right)$.
		
		There remains to show that $\mathbb{P}\left( d^i_{\tau} =0 \right)>0$. The proof is by contradiction. So suppose not. Then the value of the principal is given by $\mathbb{E}\left[ \int_{0}^{\infty} e^{-rt} \pi^i\left( X^i_t \right) dt \right]$, which contradicts Assumption \ref{assumption:nontrivial}. This concludes the proof.
	\end{proof}
	
	\begin{lemma}\label{lemma:monotonicityofVp}
		The $V_q$'s defined in \eqref{eq:valuefunctionVp} are nondecreasing in $x$.
	\end{lemma}
	
	\begin{proof}[Proof of lemma \ref{lemma:monotonicityofVp}]
		The proof is by induction. By definition $V_{Q+1}(x) \coloneqq W$ for all $x$ and hence is nondecreasing. 
		
		Now let $0\leq q \leq Q$ and assume that, for $k \geq q+1$, $V_k$ is nondecreasing in $x$. I show that 
		\begin{align*}
		V_q(x) & = ess\underset{\tau \geq 0}{\sup }\, \mathbb{E}\Bigg[ \int_{0}^{\tau\wedge\tau(x^i_{q+1})} e^{-rt}\left( \pi^i(X^i_{t}) -{c}^{i}\left( X^i_{t} \right) \sum_{k=0}^{q} \lambda_k \right) dt \notag \\
		& \quad \quad  + e^{-r\tau \wedge\tau^i(x^i_{q+1})} \Bigg\{  V_{q\left(\underline{X}^i_{\tau^{i}\left( x^i_{q+1} \right)}\right)} \left( X^i_{\tau^i(x^i_{q+1})}\right) \mathbbm{1}_{\{\tau \geq \tau^i(x^i_{q+1})\}}\notag \\
		& \quad \quad  + \mathbbm{1}_{\{\tau < \tau^i(x^i_{q+1})\}} \Bigg( \sum_{q=0}^{Q} \lambda_q g^i + \int_{\tau}^{\infty} e^{-r(t -\tau)} \pi^i(X^i_t) dt \Bigg) \vee W \Bigg\} \mid x\Bigg] 
		\end{align*}
		is weakly increasing in $x$.
		
		Let $\bar{x}\geq \underline{x}$, and let $\tau^{\underline{x}}$ be an optimal stopping time associated with $V(\underline{x})$, which exists by Theorem \ref{theorem:MarkovstoppingPeskir2.4}. From the definition of $V^q$ in \eqref{eq:valuefunctionVp},
		\begin{align*}
		V_q(\bar{x}) \geq & \mathbb{E}\Bigg[ \int_{0}^{\tau^{\underline{x}}\wedge\tau^i(x^i_{q+1})} e^{-rt}\left( \pi^i(X^i_{t}) -{c}^{i}\left( X^i_{t}\right) \sum_{k=0}^{q} \lambda_k \right) dt \notag \\
		& \quad \quad  + e^{-r\tau^{\underline{x}} \wedge\tau^{\underline{x}}(x^i_{q+1})} \Bigg\{  V_{q\left(\underline{X}^i_{\tau^{\underline{x}}\left( x^{i}_{q+1} \right)}\right)} \left( X^i_{\tau^{\underline{x}}(x^i_{q+1}) }\right) \mathbbm{1}_{\{\tau^{\underline{x}} \geq \tau^{\underline{x}}(x^i_{q+1})\}}\notag \\
		& \quad \quad + \mathbbm{1}_{\{\tau^{\underline{x}} < \tau^{\underline{x}}(x^i_{q+1})\}} \Bigg( \sum_{q=0}^{Q} \lambda_q g^i + \int_{\tau^{\underline{x}}}^{\infty} e^{-r(t -\tau^{\underline{x}})} \pi^i(X^i_t) dt \Bigg) \vee W \Bigg\}  \mid \bar{x}\Bigg],
		\end{align*}
		where $\tau^{\underline{x}}(x^i_{q+1}) \coloneqq \inf\left\{ t \geq 0 \, : \, X^{i, \underline{x}}_t \leq x^i_{q+1} \right\} \leq \tau(x^i_{q+1})$ $\mathbb{P}$-a.s. by Assumption \ref{assumption:comparisontheorem}. 
		
		But, from the definition of the $V_{k}$'s, since ${c}^{i}\left( X^i_{t}\right) \geq 0$ and $\lambda_{q+k} \geq 0$, $k \in \left\{ 1, \dots, Q-q \right\}$, note that $V_q(x) \geq V_{q+k}(x)$ for all $k \in \left\{ 1, \dots, Q-q \right\}$ and all $x\geq x^i_{q+1}$. Using that $X^{i, \bar{x}}_{\tau^{\underline{x}}(x^i_{q+1})} \geq X^{i, \underline{x}}_{\tau^{\underline{x}}(x^i_{q+1})}$,
		\begin{align*}
		V_q(\bar{x}) \geq \mathbb{E}\bigg[ & \int_{0}^{\tau^{\underline{x}}\wedge \tau^i(x^i_{q+1})} e^{-rs} \pi^i(X^i_s) ds - \sum_{k=0}^{q} \lambda_k \int_{0}^{\tau^{\underline{x}}\wedge \tau^i(x^i_{k+1})} e^{-rs} {c}^{i}\left( X^i_{s}\right)  ds \\
		& + e^{-r\tau^{\underline{x}}\wedge \tau(X^{p+1})} \bigg( V_{q\left(\underline{X}^{i, \underline{x}}_{\tau^{\underline{x}}\left( x^i_{q+1} \right)}\right)}(X^i_{\tau^{\underline{x}}(x^i_{q+1})} ) \mathbbm{1}_{\{\tau^{\underline{x}}(x^i_{q+1}) \leq  \tau^{\underline{x}} \}} \\
		& + \mathbbm{1}_{\{\tau^{\underline{x}} < \tau^{\underline{x}}(x^i_{q+1}) \}} \Bigg( \sum_{q=0}^{Q} \lambda_q g^i + \int_{\tau^{\underline{x}}}^{\infty} e^{-r(t -\tau^{\underline{x}})} \pi^i(X^i_t) dt \Bigg) \vee W   \bigg) \mid \bar{x} \bigg].
		\end{align*} 
		Since the $V_{q+k}$'s are nondecreasing by the induction hypothesis, $\pi^i$ is nondecreasing and ${c}^{i}\left( X^i_{t}\right)$ is nonincreasing by Assumption \ref{assumption:comparisontheorem},
		\begin{align*}
		V_q(\bar{x}) & \geq \mathbb{E}\bigg[ \int_{0}^{\tau^{\underline{x}}\wedge \tau^i(x^i_{q+1})} e^{-rs}  \pi^i(X^i_s) ds - \sum_{k=0}^{q} \lambda_k \int_{0}^{\tau^{\underline{x}}\wedge \tau^i(x^i_{q+1})} e^{-rs} {c}^{i}\left( X^i_{s}\right)  ds \\
		& \qquad + e^{-r\tau^{\underline{x}}\wedge \tau^i(x^i_{q+1})} \bigg( V_{q\left( \underline{X}^i_{\tau^i(x^i_{q+1})} \right)}\left(X_{\tau^i(x^i_{q+1})} \right) \mathbbm{1}_{\{\tau^{i}(x^i_{q+1}) \leq  \tau^{\underline{x}} \}} \\
		& \qquad + \mathbbm{1}_{\{\tau^{\underline{x}} < \tau^{i}(x^i_{q+1}) \}} \Bigg( \sum_{q=0}^{Q} \lambda_q g^i + \int_{\tau^{\underline{x}}}^{\infty} e^{-r(t -\tau^{\underline{x}})} \pi^i(X^i_t) dt \Bigg) \vee W \bigg) \mid \underline{x} \bigg] \\
		& = V_q(\underline{x}),
		\end{align*}
		as the integrand is $\mathbb{P}$-a.s. smaller. Since $\bar{x} \geq \underline{x}$ are arbitrary, $V_q$ is nondecreasing in $x$.
		
		Thus, by induction, $V_q$ is nondecreasing for all $0 \leq q\leq Q$, and the proof is complete.
	\end{proof}
	
	Next distinguish two cases, corresponding to the two cases of Assumption \ref{assumption:onesidedjumps}. 
	
	\paragraph{$X^i$ only jumps up:}
	
	\begin{lemma}\label{lemma:upperthresholdjumpsup}
		Assume that $X^i$ only jumps up. There exists $\left( \tau_{\mathcal{Q}}, d^{\mathcal{Q}} \right)$ that solves \eqref{eq:relaxedprogramsinglearmdiscretized} with (i) $\tau_{\mathcal{Q}} \coloneqq \inf\left\{ t \geq 0 \, : \, X^i_{t} \not \in \left[ \underline{p}^{i}, \bar{P}^{i}_{\mathcal{Q}}\left(t\right) \right) \right\}$, for some threshold $\underline{p}^{i} \in \mathcal{X}^i$ and process $\bar{P}^{i}_{\mathcal{Q}}(\cdot)$ constant on $[\tau^i(x^i_{q}), \tau^i(x^i_{q+1}))$ for all $q \in \left\{ 0, \dots, Q\right\}$  with $\bar{P}^i_{\mathcal{Q}}(\tau^i(x^i_q)) = \bar{P}^i_{\mathcal{Q}}(\tau^i(x^i_{q+1}))$ if \eqref{eq:DRDPxiq} is not binding at $x^i_{q+1}$., and (ii) $d^{i,\mathcal{Q}}_{\tau_{\mathcal{Q}}} = \mathbbm{1}_{\{X^i_{\tau_{\mathcal{Q}}} \geq \bar{P}^{i}_{\mathcal{Q}}(\tau_{\mathcal{Q}})) \}}$.
	\end{lemma}
	
	\begin{proof}[Proof of Lemma \ref{lemma:upperthresholdjumpsup}]
		After applying Lemma \ref{lemma:structurestopping}, there remains to show that $S(t) \cap \left\{ x \in \mathcal{X}^i \, : \, x \geq x^i_W \right\} = \left[ \bar{P}^{i}_{\mathcal{Q}}\left(\underline{x}\right), \infty \right)$ for some $\bar{P}^{i}_{\mathcal{Q}}(\cdot)$ constant on $[\tau^i(x^i_{q}), \tau^i(x^i_{q+1}))$ for all $q \in \left\{ 0, \dots, Q\right\}$. For all $q$, define 
		\begin{align*}
		\bar{P}^{i}_q \coloneqq \inf\left\{ x\in S_q \, : \, x \geq x^i_W \right\},
		\end{align*} 
		
		But $x \geq \bar{P}^{i}_q$ implies that $x \in S_q$. To see this, note that, starting from $x \geq \bar{P}^{i}_q$, $d^i_{\tau} = 1$ $\mathbb{P}$-a.s.. So, at $x$, the principal's continuation value is
		\begin{align*}
		V_q(x) & = \mathbb{E}\Bigg[ \int_{0}^{\tau_{\mathcal{Q}}} e^{-rs}  \pi\left(X^i_s\right) ds - \sum_{k=0}^{q} \lambda_k \int_{0}^{\tau_{\mathcal{Q}}} e^{-rs} c^i\left( X^i_s\right) ds \\
		& \qquad + e^{-r\tau_{\mathcal{Q}}} \Bigg[ \int_{\tau_{\mathcal{Q}}}^{\infty} e^{-r(t - \tau_{\mathcal{Q})}} \pi^i\left( X^i_t \right)dt + \sum_{q=0}^Q \lambda_q g^i \Bigg] \, \bigg| \, x \Bigg] \\
		& \leq \mathbb{E}\left[ \int_{0}^{\infty} e^{-rt} \pi^i\left( X^i_t \right)dt \mid x \right]+ \sum_{q=0}^Q \lambda_q g^i,
		\end{align*}
		Therefore it is optimal for the principal to stop at $x$ and promote the worker.
		
		Thus the principal stops and promotes the worker the first time $X^i_t$ enters $\left(\bar{P}^{i}_{q\left( \underline{X}^i_t\right)}, \infty \right)$ or $\left[\bar{P}^{i}_{q\left( \underline{X}^i_t\right)}, \infty \right)$. Note here that $\mathbb{P}\left( \left\{\tau_{(\bar{P}^{i}_q, \infty)} = 0 \right\} \mid X^i_t= \bar{P}^{i}_q \right) =1$, the stopping times $\tau_{(\bar{P}^{i}_q, \infty)}$ and $\tau_{[\bar{P}^{i}_q, \infty)}$ are indistinguishable. Moreover, $\bar{\pi}^{i}$ is right-continuous by the Portmanteau theorem, as $\pi^i$ is increasing upper-semicontinuous and $X^i$ is Feller, and therefore the two stopping times give the same payoffs to the principal and to the worker. So, one can assume that $S^j_q \cap [x^i_W \infty) = [\bar{P}^{i}_q, \infty)$.
		
		Letting $\bar{P}^{i}_{\mathcal{Q}}\left( \underline{X}^i_{t} \right) \coloneqq \bar{P}^{i}_{q\left( \underline{X}^i_t \right)}$ yields the desired result.
	\end{proof}
	
	\begin{lemma}\label{lemma:allgridpointsarebindingupjumps}
		Every constraint \eqref{eq:RDPxiq} is binding in the problem \eqref{eq:relaxedprogramsinglearmdiscretized}.
	\end{lemma}
	
	\begin{proof}
		Let $(\tau, d)$ be the solution of \eqref{eq:relaxedprogramsinglearmdiscretized} given by Lemma \ref{lemma:upperthresholdjumpsup}. 
		
		Observe that at least one constraint is binding, for otherwise the solution would coincide with that of the unconstrained problem, i.e., the worker is never promoted, which violates all the constraints. 
		
		Next, I show that \eqref{eq:RDPxiq} for $q=0$ is binding. Let $q^*$ be the first binding constraint. If $q^* = 0$, I am done. So suppose not. Then
		\begin{align*}
		\mathbb{E} \left[ e^{-r \tau} g^i d^i_{\tau}  - \int_{0}^\tau e^{-rt} c^i\left( X^i_t \right) d t \right] >0.
		\end{align*}
		By Lemma \ref{lemma:upperthresholdjumpsup}, on the random interval $[0, \tau^i(x^i_{q^*}))$, the solution consists in a stationary promotion threshold $\bar{P}_{0}$ and a stationary threshold $\underline{p}^i_{\mathcal{Q}}$ such that the principal takes her outside option at $\underline{p}^i_{\mathcal{Q}}$. However, since \eqref{eq:RDPxiq} is not binding at $x^i_0$, there exists $P > \bar{P}_{0}$ such that 
		\begin{align*}
		\mathbb{E} \left[ e^{-r \tau^i_{[P, \infty)}\wedge \tau^i(x^i_1)} g^i \mathbbm{1}_{\{\tau^i_{[P, \infty)} < \tau^i(x^i_1) \}}  - \int_{0}^{\tau^i_{[P, \infty)}\wedge \tau^i(x^i_1)} e^{-rt} c^i\left( X^i_t\right) d t \right] =0.
		\end{align*}
		Let $\tau(P)$ be defined as $\tau(P) \coloneqq \inf\left\{ t\geq 0 \, : \, X^i_t \not \in \left[\underline{p}^i_{\mathcal{Q}},  P\right)  \right\}$ on $\left[0, \tau^i(x^i_1)\right)$ and $\tau(P) = \tau$ on $\left[\tau^i(x^i_1), \infty\right)$. Note that $\tau(P)$ is feasible in the relaxed problem \eqref{eq:relaxedprogramsinglearmdiscretized} and yields a higher payoff for the principal (strictly if $W >0$); a contradiction. So \eqref{eq:RDPxiq} for $q=0$ is binding.
		
		Similarly, one can show that \eqref{eq:RDPxiq} for $q = 1,\dots, Q$ are binding. To see this, suppose not, i.e., \eqref{eq:RDPxiq} is not binding for some $q \in \left\{ 1,\dots, Q\right\}$. Let $\underline{q}\geq 1$ be the smallest $q$ such that \eqref{eq:RDPxiq} is not binding. Let $\tilde{q}>\underline{q}$ be the next binding constraint, with $\tilde{q} = Q+1$ if all constraints $q\geq \underline{q}$ are lax. Then on the random interval $\left[ \tau^i(x^i_{\underline{q}-1}), \tau^i(x^{\tilde{q}})\right)$, by Proposition \ref{lemma:upperthresholdjumpsup}, the optimal stopping time is stationary and the worker is promoted if and only if his type exceeds $\bar{P}^i_{\mathcal{Q}}\left( x^i_{\underline{q}-1} \right)$. At $\tau^i(x^i_{\underline{q}-1})$, the continuation value of the worker is zero:
		\begin{align*}
		\mathbb{E} &\left[ e^{-r\tau \wedge \tau^i(x^i_{\tilde{q}})} {g}^i \mathbbm{1}_{\{ \tau < \tau^i(x^i_{\tilde{q}}) \wedge \tau^i(\underline{p}^i_{\mathcal{Q}} ) \}} \int_{0}^{\tau \wedge \tau^i(x^i_{\tilde{q}})} e^{-rt}c^i\left( X^{i}_t \right)dt \mid {X}^i_{\tau^i(x^i_{\underline{q}-1})}\right] =0.
		\end{align*}
		But, by Assumption \ref{assumption:comparisontheorem},
		\begin{align*}
		\mathbb{E} &\left[ e^{-r\tau\wedge \tau^i(x^i_{\tilde{q}}) } {g}^i \mathbbm{1}_{\{ \tau < \tau^i(x^i_{\tilde{q}})\wedge \tau^i(\underline{p}^i_{\mathcal{Q}} ) \}} - \int_{0}^{\tau \wedge \tau^i(x^i_{\tilde{q}})} e^{-rt}c^i\left( X^{i}_t \right)dt \mid {X}^i_{\tau^i(x^i_{\underline{q}-1})}\right] \\ 
		& \geq \mathbb{E} \left[ e^{-r\tau\wedge \tau^i(x^i_{\tilde{q}}) } {g}^i \mathbbm{1}_{\{ \tau < \tau^i(x^i_{\tilde{q}})\wedge \tau^i(\underline{p}^i_{\mathcal{Q}} ) \}} - \int_{0}^{\tau \wedge \tau^i(x^i_{\tilde{q}})} e^{-rt}c^i\left( X^{i}_t \right)dt \mid X^i_{\tau^i(x^i_{\underline{q}})}\right] \\
		& > 0,
		\end{align*}
		a contradiction. 
		
		So all \eqref{eq:RDPxiq} constraints are binding.
	\end{proof}
	The next lemma describes the solution to \eqref{eq:relaxedprogramsinglearmdiscretized}.
	\begin{lemma}\label{lemma:solutiondiscretizedsinglearmupjumps}
		Assume that $X^i$ only jumps up. Define the process
		\begin{align*}
		\bar{P}^{i}_{\mathcal{Q}}(t) \coloneqq \sum_{q}^{Q} \bar{P}^i\left( \underline{X}^i_{\tau^i(x^i_q)} \right) \mathbbm{1}_{\{ [\tau^i(x^i_q), \tau^i(x^i_{q+1}) \}}.
		\end{align*}
		The pair $\left(\tau^*, d^* \right)$ with
		\begin{align*}
		\tau^* &= \inf\left\{ t \geq 0 \, : \, X^i_{t} \not \in \left( \underline{p}^{i}, \bar{P}^i_{\mathcal{Q}}(t) \right) \right\}, \\
		\text{and } d^*_{\tau^*} & = \mathbbm{1}_{\{X^i_{\tau^*} = \bar{P}^i_{\mathcal{Q}}(\tau^*) \}}
		\end{align*}
		solves \eqref{eq:relaxedprogramsinglearmdiscretized}.
	\end{lemma}
	
	\begin{proof}
		From Lemma \ref{lemma:upperthreshold}, there remains to show that $\bar{P}^i_q = \bar{P}^i\left( X^i_{\tau^i(x^i_q)}\right)$. This follows from Lemma \ref{lemma:allgridpointsarebindingupjumps}: because the \eqref{eq:RDPxiq} constraint is binding at every $q \in \mathcal{Q}$, the continuation value of worker is zero at $\tau^i(x^i_q)$, so the worker is indifferent between quitting and continuing at $\tau^i(x^i_q)$. Observing that $\bar{P}^i(\underline{x})$ is increasing in $\underline{x}$, the result follows.
	\end{proof}
	
	\paragraph{$X^i$ only jumps down:} Introduce a new stopping problem with constraints. Here one can assume, without loss of generality, that, for all $q \in \left\{ 0,\dots, Q \right\}$, $\tilde{P}^i(x^i_q) \in \mathcal{Q}$ or $\tilde{P}^i(x^i_q) \geq x^i_0$ (using Assumption \ref{assumption:potentialforimprovment}), and that $x^i_Q < \underline{p}^i_{\mathcal{Q}}$. Otherwise, consider a sequence of grids $\mathcal{Q}^m$ with $x^{i}_{Q^n} \to \inf \mathcal{X}^i$. By the last statement of Lemma \ref{lemma:structurestopping}, eventually $x^i_{Q^n} < \underline{p}^i_{_{\mathcal{Q}^n}}$. 
	
	Consider then
	\begin{align}\label{eq:DRRPi}
	\bar{\Pi}^{i}_{\mathcal{Q}} \coloneqq \underset{\tau}{\sup}\, {\mathbb{E}}& \Bigg[ \int_{0}^{\tau\wedge\tau^i(\underline{p}^i_{\mathcal{Q}} )} e^{-rt} \pi^i\left(X^i_{t}\right)dt \notag \\
	& + e^{-r\tau\wedge \tau^i(\underline{p}^i_{\mathcal{Q}} )}  \left( \mathbbm{1}_{\{ \tau \leq \tau^i(\underline{p}^i_{\mathcal{Q}} ) \}} \tilde{\bar{\pi}}^i\left( X^i_{\tau} \right) + \mathbbm{1}_{\{ \tau > \tau^i(\underline{p}^i_{\mathcal{Q}} ) \}} W\right) \Bigg] \tag{DRRP$^i$}
	\end{align}
	subject to, for all $q = 0, \dots, Q$ such that $x^i_q \geq \underline{p}^i_{\mathcal{Q}} $,
	\begin{align}\label{eq:DRDPxiq}
	\mathbb{E} \bigg[ e^{-r \tau}  g^i \mathbbm{1}{\{\tau \leq \tau^i(\underline{p}^i_{\mathcal{Q}} )\}}+ \int_{\tau\wedge\tau^i(x^i_q)}^{\tau} e^{-rt} {c}^{i}\left( X^i_{t}\right) d t \bigg] \geq 0; \tag{DRDP($x^i_q$)}
	\end{align}
	where $\underline{p}^i_{\mathcal{Q}}$ is the threshold optimal threshold obtained in Lemma \ref{lemma:structurestopping}, and
	\begin{align*}
	\tilde{\bar{\pi}}^i \left( x \right) \coloneqq \begin{cases}
	\mathbb{E}\left[ \int_{\tau}^{\infty} e^{-r(t-\tau)} \pi^i\left( X^i_t \right)dt \mid x \right]\text{ if } x \in \mathcal{Q}, \\
	- \infty \text{ otherwise.}
	\end{cases}
	\end{align*}
	
	Exactly as in step 2 and 3, by Theorem 1 in \cite{balzer2002duality} and Theorem \ref{theorem:MarkovstoppingPeskir2.4} in Appendix \ref{app:usefulresultsoptimalstopping},
	\begin{lemma}\label{lemma:nostoppingonxiqxiq+1}
		There exists $\bar{\tau}_{\mathcal{Q}}$ that solves \eqref{eq:DRRPi} with $\bar{\tau}_{\mathcal{Q}} \coloneqq \inf\left\{ t \geq 0 \, : \, X^i_{t} \not \in S\left(t\right) \right\}$, where $S\left( t\right)$ is a correspondence constant on $[\tau^i(x^i_{q}), \tau^i(x^i_{q+1}))$ for all $q \in \left\{ 0, \dots, Q\right\}$ with ${S}(\tau^i(x^i_q)) =  {S}(\tau^i(x^i_{q+1}))$ if \eqref{eq:DRDPxiq} is not binding at $\tau^i(x^i_{q+1})$. Moreover, for all $t\geq 0$ and all $q \in \left\{ 0,\dots, Q\right\}$, $S(t) \cap \left( x^i_{q+1}, x^i_{q} \right) = \emptyset$.
	\end{lemma}
	
	The value of problem \eqref{eq:DRRPi} is weakly greater than the value of \eqref{eq:relaxedprogramsinglearmdiscretized}.
	\begin{lemma}\label{lemma:DRRPigeqRRPi}
		The value of \eqref{eq:DRRPi} is (weakly) greater than the value of \eqref{eq:relaxedprogramsinglearmdiscretized}: $\bar{\Pi}^i_{\mathcal{Q}} \geq \Pi^i_{\mathcal{Q}}$.
	\end{lemma}
	
	\begin{proof}[Proof of Lemma \ref{lemma:DRRPigeqRRPi}]
		Consider the optimal stopping time $\tau_{\mathcal{Q}}$ for problem \eqref{eq:relaxedprogramsinglearmdiscretized} given by Lemma \ref{lemma:structurestopping}. I claim that there exists a feasible stopping time in \eqref{eq:DRRPi} that gives a payoff of $\Pi^i_{\mathcal{Q}}$ to the principal. 
		
		To see this, observe that if $S_q \cap (x^i_q, x^i_{q+1}) \cap \left\{ x \in \mathcal{X}^i \, :  \, x \geq x^i_W \right\}  = \emptyset$ for all $q$, I am done. So suppose not, i.e., $S_q \cap (x^i_q, x^i_{q+1}) \cap \left\{ x \in \mathcal{X}^i \, :  \, x \geq x^i_W \right\} \not = \emptyset$ for some $q \in \left\{ 0, \dots, {Q} \right\}$. So, when $\underline{X}^i_t \in (x^i_{q+1}, x^i_q]$, the principal finds it optimal to stop at $\tilde{x} \in S_q \cap (x^i_q, x^i_{q+1}) \cap \left\{ x \in \mathcal{X}^i \, :  \, x \geq x^i_W \right\}$. Her continuation value is 
		\begin{align*}
		\mathbb{E}\left[ \int_{0}^{\infty} e^{-rt} \pi^i\left( X^i_t \right) dt \mid \tilde{x}\right],
		\end{align*}
		which is equal to
		\begin{align*}
		\mathbb{E}& \Bigg[ \int_{0}^{\tau^i_{\left( x^i_{q+1}, x^i_q \right)}\wedge \tau^i(\underline{p}^i_{\mathcal{Q}} )} e^{-rt} \pi^i\left( X^i_t \right) dt \\ 
		& + e^{-r\tau^i_{\left( x^i_{q+1}, x^i_q \right)}\wedge \tau^i(\underline{p}^i_{\mathcal{Q}} )} \int_{\tau^i_{\left( x^i_{q+1}, x^i_q \right)}\wedge \tau^i(\underline{p}^i_{\mathcal{Q}} )}^{\infty} e^{-r\left(t- \tau^i_{\left( x^i_{q+1}, x^i_q \right)}\wedge \tau^i(\underline{p}^i_{\mathcal{Q}} )\right)} \pi^i\left( X^i_t \right) dt \mid \tilde{x}\Bigg].
		\end{align*}
		Thus the the stopping time $\tilde{\tau}$ defined by
		\begin{align*}
		\tilde{\tau} \coloneqq \tau_{\mathcal{Q}} + \inf\left\{ t\geq 0\, : \, X^i_{\tau +t} \not \in \left( x^i_{q+1}, x^i_q \right) \right\} \mathbbm{1}_{\{ \underline{X}^i_{\tau} \in \left( x^i_{q+1}, x^i_q \right) \}}.
		\end{align*}
		gives the same payoff to the principal as $\tau_{\mathcal{Q}}$. Next, note that it is feasible. This follows from noting that, by definition of $\tilde{g}^i$,
		\begin{align*}
		\mathbb{E}& \left[ e^{-r\tilde{\tau}} \tilde{g}^i \mathbbm{1}_{\{X^i_{\tau_{\mathcal{Q}}} \in S\left( \underline{X}^i_{\tau_{\mathcal{Q}}} \right)  \}} - \int_{\tilde{\tau}\wedge \tau^i(x^i_q)}^{\tilde{\tau}} e^{-rt} {c}^i\left( X^i_t \right)dt  \right] \\
		& \geq \mathbb{E}\left[ e^{-r\tau_{\mathcal{Q}}} g^i \mathbbm{1}_{\{X^i_{\tau_{\mathcal{Q}}} \in S\left( \underline{X}^i_{\tau_{\mathcal{Q}}} \right)  \}} - \int_{\tau_{\mathcal{Q}}\wedge \tau^i(x^i_q)}^{\tau_{\mathcal{Q}}} e^{-rt} c^i\left( X^i_t \right)dt  \right] \\
		& \geq 0;
		\end{align*}
		where the second inequality follows from \eqref{eq:RDPxiq}.
		
		Repeating the above construction for all $q$ such that $S_q \cap (x^i_q, x^i_{q+1}) \cap \left\{ x \in \mathcal{X}^i \, :  \, x \geq x^i_W \right\} \not = \emptyset$ proves the claim.
		
		Therefore the value of \eqref{eq:DRRPi} is (weakly) greater than the value of \eqref{eq:relaxedprogramsinglearmdiscretized}: $\bar{\Pi}^i_{\mathcal{Q}} \geq \Pi^i_{\mathcal{Q}}$.
	\end{proof}
	
	Next note that the optimal stopping region in \eqref{eq:DRRPi} can be taken to be $[\bar{P}^i_q, \infty)$ for some $\bar{P}^i_q \geq x^i_q$ on each random interval $[\tau^i(x^i_q), \tau^i(x^i_{q+1}))$.
	\begin{lemma}\label{lemma:upperthreshold}
		There exists $ \bar{\tau}_{\mathcal{Q}}$ that solves \eqref{eq:DRRPi} with $\tau_{\mathcal{Q}} \coloneqq \inf\left\{ t \geq 0 \, : \, X^i_{t} \not \in \left[ \underline{p}^{i}_{\mathcal{Q}}, \bar{P}^{i}_{\mathcal{Q}}\left( t \right) \right) \right\}$, where $\bar{P}^{i}_{\mathcal{Q}}(\cdot)$ is a process constant on $[\tau^i(x^i_{q}), \tau^i(x^i_{q}))$ for all $q \in \left\{ 0, \dots, Q\right\}$ with $\bar{P}^i_{\mathcal{Q}}(\tau^i(x^i_{q})) = \bar{P}^i_{\mathcal{Q}}(\tau^i(x^i_{q+1}))$ if \eqref{eq:DRDPxiq} is not binding at $\tau^i(x^i_{q+1})$.
	\end{lemma}
	
	\begin{proof}[Proof of Lemma \ref{lemma:upperthreshold}]
		After applying Lemma \ref{lemma:nostoppingonxiqxiq+1}, there remains to show that $S(\underline{t}) \cap \left\{ x \in \mathcal{X}^i \, : \, x \geq x^i_W \right\} = \left[ \bar{P}^{i}_{\mathcal{Q}}\left(t\right), \infty \right)$ for some process $\bar{P}^{i}_{\mathcal{Q}}(\cdot)$ constant on $[\tau^i(x^i_{q}), \tau^i(x^i_{q}))$, for all $q \in \left\{ 0, \dots, Q\right\}$. 
		
		For all $q$, define 
		\begin{align*}
		\bar{P}^{i}_q \coloneqq \inf\left\{ x\in S_q \, : \, x \geq x^i_{q+1} \right\},
		\end{align*}
		By Lemma \ref{lemma:nostoppingonxiqxiq+1}, the optimal stopping time $\bar{\tau}_{\mathcal{Q}}$ is such that, on $[\tau^i(x^i_q), \tau^i(x^i_{q+1}))$, $\bar{\tau}_{\mathcal{Q}}$ is the first entry time in $S_q$, with $\bar{P}^{i}_q \coloneqq \inf\left\{ x \in S_q \, : \, x \geq x^i_{q+1} \right\} \geq x^i_q$. Therefore the first entry time into $S_q \cap \left\{ x \in \mathcal{X}^i \, : \, x \geq x^i_W \right\}$ (if it happens) occurs when $X^i$ crosses $\bar{P}^i_q$ from below (as $X^i$ only jumps down). Since $S_q$ is a subset of the finite grid $\mathcal{Q}$, the $\inf$ is attained and one can take $S_q \cap [x^i_W \infty) = [\bar{P}^{i}_q, \infty)$.
		
		Letting $\bar{P}^{i}_{\mathcal{Q}}\left( t \right) \coloneqq \bar{P}^{i}_{q\left( \underline{X}^i_t \right)}$ gives the desired result.
	\end{proof}
	
	\begin{lemma}\label{lemma:allgridpointsarebinding}
		Every constraint \eqref{eq:DRDPxiq} is binding in the problem \eqref{eq:DRRPi}.
	\end{lemma}
	
	\begin{proof}
		Let $\bar{\tau}_{\mathcal{Q}}$ be the optimal stopping time for \eqref{eq:DRRPi} given by Lemma \ref{lemma:upperthreshold}. Observe first that at least one constraint is binding, for otherwise the solution would coincide with that of the unconstrained problem, i.e., the worker is never promoted, which violates all the constraints. 
		
		Next, I show that \eqref{eq:DRDPxiq} for $q=0$ is binding. Let $q^*$ be the first binding constraint. If $q^* = 0$, I am done. So suppose not. Then
		\begin{align*}
		\mathbb{E} \left[ e^{-r \tau} \tilde{g}^i d^i_{\tau}  - \int_{0}^\tau e^{-rt} c^i\left( X^i_t \right) d t \right] >0.
		\end{align*}
		By lemma \ref{lemma:nostoppingonxiqxiq+1}, on the random interval $[0, \tau^i(x^i_{q^*}))$, the solution consists in a stationary promotion threshold $\bar{P}_{0}$. However, since \eqref{eq:DRDPxiq} is not binding at $x^i_0$, there exists $P > \bar{P}_{0}$ such that 
		\begin{align*}
		\mathbb{E} \left[ e^{-r \tau^i_{[P, \infty)}\wedge \tau^i(x^i_1)\wedge \tau^i(\underline{p}^i_{\mathcal{Q}} )} \tilde{g}^i \mathbbm{1}_{\{\tau^i_{[P, \infty)} < \tau^i(x^i_1)\wedge \tau^i(\underline{p}^i_{\mathcal{Q}} ) \}}  - \int_{0}^{\tau^i_{[P, \infty)}\wedge \tau^i(x^i_1)\wedge\tau^i(\underline{p}^i_{\mathcal{Q}} )} e^{-rt} c^i\left( X^i_t\right) d t \right] =0.
		\end{align*}
		But then, choosing
		\begin{align*}
		\tilde{\tau}(\omega) = \begin{cases}
		\inf\left\{t \geq 0 \, : \, X^i_t \not \in \left[\underline{p}^i_{\mathcal{Q}}, P \right)\right\} \text{ if } inf\left\{t \geq 0 \, : \, X^i_t \not \in \left[\underline{p}^i_{\mathcal{Q}}, P \right)\right\} \leq \tau^i(x^i_1), \\
		\bar{\tau}_{\mathcal{Q}} \text{ otherwise},
		\end{cases}
		\end{align*}
		instead of $\bar{\tau}_{\mathcal{Q}}$ is feasible in the relaxed problem \eqref{eq:DRRPi} and yields a higher payoff for the principal. So \eqref{eq:DRDPxiq} for $q=0$ is binding.
		
		Similarly, I can show that \eqref{eq:DRDPxiq} for $q = 1,\dots, Q$ such that $x^i_q \geq \underline{p}^i_{\mathcal{Q}} $ are binding. To see this, suppose not, i.e., \eqref{eq:DRDPxiq} is not binding for some $q \in \left\{ 1,\dots, Q\right\}$. Let $\underline{q}\geq 1$ be the smallest $q$ such that \eqref{eq:DRDPxiq} is not binding. Let $\tilde{q}>\underline{q}$ be the next binding constraint, with $\tilde{q} = Q+1$ if all constraints $q\geq \underline{q}$ are lax. Then on the random interval $\left[ \tau^i(x^i_{\underline{q}-1}), \tau^i(x^{\tilde{q}})\right)$, by Proposition \ref{lemma:upperthresholdjumpsup}, the optimal stopping time is stationary and the worker is promoted if and only if his type exceeds $\bar{P}^i_{\mathcal{Q}}\left( \tau^i(x^i_{\underline{q}-1}) \right)$. At $\tau^i(x^i_{\underline{q}-1})$, the continuation value of the worker is zero:
		\begin{align*}
		\mathbb{E} &\Bigg[ e^{-r\bar{\tau}_{\mathcal{Q}}\wedge \tau^i(x^i_{\tilde{q}}) \wedge \tau^i(\underline{p}^i_{\mathcal{Q}} )} \tilde{g}^i \mathbbm{1}_{\{\bar{\tau}_{\mathcal{Q}} < \tau^i(x^i_{\tilde{q}}) \wedge \tau^i(\underline{p}^i_{\mathcal{Q}} ) \}} \\
		& \qquad - \int_{0}^{\bar{\tau}_{\mathcal{Q}}\wedge \tau^i(x^i_{\tilde{q}})\wedge \tau^i(\underline{p}^i_{\mathcal{Q}} )} e^{-rt}c^i\left( X^{i}_t \right)dt \mid {X}^i_{\tau^i(x^i_{\underline{q}-1})}\Bigg] =0.
		\end{align*}
		But, by Assumption \ref{assumption:comparisontheorem},
		\begin{align*}
		\mathbb{E} &\Bigg[ e^{-r \bar{\tau}_{\mathcal{Q}}\wedge \tau^i(x^i_{\tilde{q}})\wedge \tau^i(\underline{p}^i_{\mathcal{Q}} ) } \tilde{g}^i \mathbbm{1}_{\{ \bar{\tau}_{\mathcal{Q}} < \tau^i(x^i_{\tilde{q}})\wedge \tau^i(\underline{p}^i_{\mathcal{Q}} ) \}} \\ 
		&\qquad \qquad - \int_{0}^{\bar{\tau}_{\mathcal{Q}}\wedge \tau^i(x^i_{\tilde{q}})\wedge \tau^i(\underline{p}^i_{\mathcal{Q}} ) } e^{-rt}c^i\left( X^{i}_t \right)dt \mid {X}^i_{\tau^i(x^i_{\underline{q}-1})}\Bigg] \\ 
		& \geq \mathbb{E} \Bigg[ e^{-r\bar{\tau}_{\mathcal{Q}}\wedge \tau^i(x^i_{\tilde{q}})\wedge \tau^i(\underline{p}^i_{\mathcal{Q}} )} \tilde{g}^i \mathbbm{1}_{\{ \bar{\tau}_{\mathcal{Q}}< \tau^i(x^i_{\tilde{q}}) \wedge \tau^i(\underline{p}^i_{\mathcal{Q}} ) \}} \\
		&\qquad \qquad - \int_{0}^{\bar{\tau}_{\mathcal{Q}}\wedge \tau^i(x^i_{\tilde{q}})\wedge \tau^i(\underline{p}^i_{\mathcal{Q}} )} e^{-rt}c^i\left( X^{i}_t \right)dt \mid X^i_{\tau^i(x^i_{\underline{q}})}\Bigg] \\
		& > 0,
		\end{align*}
		a contradiction. 
		
		So all \eqref{eq:DRDPxiq} constraints are binding.
	\end{proof}

	\paragraph{Supporting Lemma for Step 4:}
	
	\begin{lemma}\label{lemma:continuationisnonnegative}
		For $\underline{x} \in \left( x^i_{q+1}, x^i_q \right]$ and $x \geq x^i_q$, worker $i$'s continuation value after any history before promotion with $\left( X^i_t, \underline{X}^i_t \right) = \left( x, \underline{x} \right)$ is nonnegative.
	\end{lemma}
	
	\begin{proof}[Proof of Lemma \ref{lemma:continuationisnonnegative}]
		Worker $i$'s continuation value is
		\begin{align*}
		U^i_t \coloneqq \mathbb{E}\left[ e^{-r(\tau -t)}g^i d^i_{\tau} - \int_{t}^{\tau} e^{-r(s-t)} c^i\left(X^i_s\right) ds \mid \mathcal{F}^i_t \right].
		\end{align*}
		Since $\tau^*_{\mathcal{Q}}$ and $d^*_{\mathcal{Q}}$ only depends on $\left( X^i, \underline{X}^i \right)$ and $\left( X^i, \underline{X}^i \right)$ has the strong Markov property (as $X$ is a Feller process), the continuation value of worker $i$ is a function of $\left({X}^i_t, \underline{X}^i_t\right)$: $U^i_t \coloneq U^i\left( X^i_t, \underline{X}^i_{t} \right)$. 
		
		Moreover, by construction, for all $\underline{x}$, $U^i\left( \underline{x}, \underline{x} \right) =0$ and $x \to U^i(x , \underline{x})$ is nondecreasing on $\left[\underline{x}, \bar{P}^i\left( \underline{x} \right)\right)$. This follows from Assumption \ref{assumption:comparisontheorem} and from $c^i\left( \cdot \right)$ being nonincreasing. Therefore, after any history before promotion with $\left( X^i_t, \underline{X}^i_t \right) = \left( x, \underline{x} \right)$,
		\begin{align*}
		U^i_t = U^i\left( X^i_t, \underline{X}^i_t \right) = U^i \left( x,  \underline{x}\right) \geq U^i \left( \underline{x}, \underline{x}\right) =0.
		\end{align*}
		This concludes the proof.
	\end{proof}	
	
	\subsection{Omitted Proofs for Section \ref{subsec:measurablestopping}}\label{app:measurablestopping}
	
	\paragraph{} The proof of the first part of Proposition \ref{prop:measurablestopping} is provided in Appendix \ref{app:existence}. Appendix \ref{app:existencesupportinglemmas} presents supporting lemmas needed in the proof. The second part is proved in Appendix \ref{app:measurablestopping2}.
	
	\subsubsection{First part of Proposition \ref{prop:measurablestopping}: Existence of an optimal promotion contest in \eqref{eq:relaxedprogram}}\label{app:existence}
	
	\paragraph{} The goal of this section is to prove the first part of Proposition \ref{prop:measurablestopping}, i.e., that a (randomized) promotion contest that promotes solves \eqref{eq:relaxedprogram}. 
	
	\begin{theorem}\label{theorem:existencerelaxedprogram}
		A solution to \eqref{eq:relaxedprogram} exists.
	\end{theorem}
	
	The logic of the proof is standard. It relies on the following two properties: (i) the feasible set is compact and (ii) the objective is upper semi-continuous. However the proof is technical. The set of all randomized promotion contests is a complicated object. Showing that it is compact in a suitable topology is not immediate. In particular, because the information the principal has at time $t$, $\mathcal{G}^T_t$, is endogenous, we cannot prove existence directly from a weak$^*$ compactness argument as is done in \cite{bismut1979temps} or \cite{pennanen2018optimal} for stopping problems. We cannot guarantee that the $weak^*$ limit of the maximizing sequence of stopping times and promotion decision is adapted to the ``right'' filtration. To overcome this issue, I will use the concept of weak convergence of filtration from the theory of ``extended weak convergence'' introduced now.\footnote{See \cite{coquet2001weak} for an introduction.}
	
	Let $\mathcal{R}(\mathcal{H})$ the set of $\mathcal{H}$-\textbf{regular} processes. A process $A = \left\{ A_t\right\}_{t\geq 0}$ is \textbf{regular} if it is of class (D) and its left-continuous version $A_{-}$ and its predictable projection $^pA$ are indistinguishable.\footnote{For a more detailed presentation, see \cite{dellacherie1982probabilities}, remark 50 d), or \cite{bismut1978regularite}.} The space of $\mathcal{H}$-regular processes is a Banach space, whose dual can be identified with the space of $\mathcal{H}$-optional random measure. A formal statement is found in \cite{pennanen2018optimal}, Theorem 1.\footnote{See also the section on random measure of \cite{dellacherie1982probabilities},  Theorem 2 in \cite{bismut1978regularite}, or Proposition 1.3 in \cite{bismut1979temps}.}
	\begin{definition}\label{def:weakconvergencefiltrations}
		A sequence of filtrations $\left( \mathcal{F}^n =\left\{ \mathcal{F}^n_t\right\}_{t\geq 0} \right)_{n \in \mathbb{N}}$ \textbf{converges weakly} to a filtration $\mathcal{F} = \left\{ \mathcal{F}^n_t\right\}_{t\geq 0} $ if, for every $A \in \mathcal{F}_{\infty}$, $\mathbb{E}\left[ \mathbbm{1}_{A} \mid \mathcal{F}^n_{\cdot} \right] \to \mathbb{E}\left[ \mathbbm{1}_{A} \mid \mathcal{F}_{\cdot} \right]$ in probability for the Skorokhod topology. We write $\mathcal{F}^n \to^w \mathcal{F}$.
	\end{definition}
	I then proceed in two steps: 
	\begin{itemize}
		\item First we show that for all delegation rule $T$ and randomized promotion decision $d$, there exists an optimal $\mathcal{G}^T$-(randomized) stopping time $S^*$. This part is standard and builds on the duality results derived in \cite{bismut1978regularite},\footnote{See also \cite{dellacherie1982probabilities}.} and used in \cite{bismut1979temps} and \cite{pennanen2018optimal} to obtain both the weak compactness of the feasible set and the (weak) continuity of the objective. This is done in Lemma \ref{lemma:existenceoptimalrandomstoppingforfixedTandd}. 
		
		\item Next we construct a maximizing sequence of promotion contest $(T, \tau, d)$ that has a convergent subsequence. We show, in Lemma \ref{lemma:continuityatthelimitrelaxedproblem}, that the limit of this subsequence is a solution of \eqref{eq:relaxedprogram} using results from the theory of ``extended weak convergence'' derived in \cite{coquet2001weak} and \cite{coquet2007convergence}. This is done in the proof of Theorem \ref{theorem:existencerelaxedprogram}.
	\end{itemize}

	\begin{proof}[Proof of theorem \ref{theorem:existencerelaxedprogram}]
		The value ${\Pi}$ of problem \eqref{eq:relaxedprogram} is equal to
		\begin{align*}
		{\Pi} = \underset{(T,d) \in \mathcal{D} \times \mathcal{C}^* \text{ such that } d \text{ is } \mathcal{G}^T\text{-optional}}{\sup } \, \Pi\left( T,d \right),
		\end{align*}
		where $\Pi(T,d)$ is defined by \eqref{eq:relaxedprogramfixedTandd} in Lemma \ref{lemma:existenceoptimalrandomstoppingforfixedTandd}. 
		
		Consider a maximizing sequence $\left(T^n, d^n\right)_{n\in \mathbb{N}} \subseteq \mathcal{D} \times \mathcal{C}^*$ such that
		\begin{align*}
		\underset{n\to \infty}{\lim \, } \Pi\left( T^n,d^n \right) = {\Pi}.
		\end{align*}
		By Lemmas \ref{lemma:setofoptionalincreasingpathsiscompact} and \ref{lemma:compactnessdecisionrule}, the set $\mathcal{D} \times \mathcal{C}^*$ is (sequentially) compact in the product topology. So there exists a subsequence $\left( T^{n_k}, d^{n_k} \right)_{k\in \mathbb{N}} \subseteq \left(T^n, d^n\right)_{n\in \mathbb{N}}$ that converges to some $(T^*, d^*) \in \mathcal{D} \times \mathcal{C}^*$. Furthermore, $d^*$ is $\mathcal{G}^{T^*}$-optional. 
		
		To see this, observe that, for all $K$ and all $\bar{k} \geq K$, $d^{n_{\bar{k}}}$ is $\left( \mathcal{G}^{T^*}_t \vee \bigvee_{k \geq K} G^{T^{n_k}}_t \right)$-adapted and, therefore, $d^*$ is $\left( \mathcal{G}^{T^*} \vee \bigvee_{k \geq K} \mathcal{G}^{T^{n_k}} \right)$-adapted, for all $K$. By Proposition 1 in \cite{coquet2001weak}, $\mathcal{G}^{T^{n_k}} \to \mathcal{G}^{T^*}$ as $k \to \infty$. So  $\left( \mathcal{G}^{T^*} \vee \bigvee_{k \geq K} \mathcal{G}^{T^{n_k}} \right) \to^w \mathcal{G}^{T^*}$ as $K \to \infty$.  Then, for all $t$, $\mathbb{E}\left[ d^{*}_{t} \mid \left( \mathcal{G}^{T^*}_t \vee \bigvee_{k \geq K} G^{T^{n_k}}_t \right) \right] \to \mathbb{E}\left[ d^*_t \mid \mathcal{G}^{T^*}_t \right]$ in probability, and hence $\mathbb{P}$-a.s. along a subsequence, as $K \to \infty$. But, for all $t\geq 0$ and all $K$, $d^*_t = \mathbb{E}\left[ d^{*}_t \mid \left( \mathcal{G}^{T^*}_t \vee \bigvee_{k \geq K} G^{T^{n_k}}_t \right) \right]$. So $d^*_t = \mathbb{E}\left[ d^*_t \mid \mathcal{G}^{T^*}_t \right]$ for all $t\geq 0$. The result then follows from the optional projection theorem (Theorem 2.7 in \cite{bain2008fundamentals}) as $d^*$ is c\`{a}dl\`{a}g.
		
		Therefore, by Lemma \ref{lemma:continuityatthelimitrelaxedproblem}, 
		\begin{align*}
		\underset{n\to \infty}{\lim \, } \Pi\left( T^{n_k},d^{n_k} \right) = \Pi\left( T^*, d^* \right) = {\Pi}.
		\end{align*}
		The conclusion then follows from Lemma \ref{lemma:existenceoptimalrandomstoppingforfixedTandd}.
	\end{proof}
	
	\subsubsection{Supporting Lemma for Theorem \ref{theorem:existencerelaxedprogram}}\label{app:existencesupportinglemmas}
	
	\begin{lemma}\label{lemma:setofoptionalincreasingpathsiscompact}
		The set of optional increasing paths, $\mathcal{D}$, is (sequentially) compact for the sequential convergence defined by: for all $K$ compact subset of $\mathbb{R}_+$, all continuous function $f:\mathbb{R}_+ \to \mathbb{R}$ and all $i \in \left\{1,\dots, N \right\}$,
		\begin{align*}
		T^n \to T \Leftrightarrow \int_{K} f(t) d{T^i}^n(t) \to \int_{K} f(t) d{T^i}(t) \, \,  \mathbb{P}\text{-a.s.},
		\end{align*}
		uniformly in $K$.
	\end{lemma}
	
	\begin{proof}[Proof of Lemma \ref{lemma:setofoptionalincreasingpathsiscompact}]
		Recall that we can identify the set of feasible delegation rule with the set of $\mathcal{F}_s$-adapted multi-process $\alpha = \left( \alpha^1,\dots, \alpha^N \right)$, where $\alpha_t = \left( \alpha^1_{t}, \dots, \alpha^N_{t} \right) \in \Delta^N$ is the Radon-Nikodym derivative of $T$ evaluated at $t$: $\frac{dT(t)}{dt}$. 
		
		By Theorem 2.2 and the first Corollary in \cite{becker1969existence}, the set of progressively measurable multi-parameter random measures taking values in $\Delta^{N}$ is sequentially compact under the sequential convergence defined by: for all $K$ compact subset of $\mathbb{R}_+$, all continuous function $f:\mathbb{R}_+ \times \Delta^N \to \mathbb{R}$,
		\begin{align*}
		A^n \to A \text{ if and only if, } \forall i \in \left\{1,\dots, N \right\}, \quad \int_K \int_{0}^{1} f\left(t, \alpha_t \right) d{A^i_t}^n(\omega) dt \to \int_K f\left(t, \alpha^i \right) d{A^i_t}(\omega) dt\, \, \mathbb{P}\text{-a.s.},
		\end{align*}
		uniformly in $K$.
		
		In particular, this implies that the set of delegation rule is sequentially precompact under the sequential convergence defined by: for all $K$ compact subset of $\mathbb{R}_+$, all continuous function $f$ and all $i \in \left\{1,\dots, N \right\}$,
		\begin{align*}
		T^n \to T \Leftrightarrow \int_{K} f(t) d{T^i}^n(t) \to \int_{K} f(t) d{T^i}(t) \, \,  \mathbb{P}\text{-a.s.};
		\end{align*}
		uniformly in $K$, since any control $A \in \mathcal{D}$ generates a unique delegation rule $T$.
		
		There remains to show that it is closed: So far, we have obtained the limit $T$ in the sense of the above as a progressively measurable process. We still need to verify that the limit $T$ is an increasing optional path, i.e., that it satisfies condition 1.-3. of definition \ref{def:delegationprocess}. Conditions 2. and 3. are easily seen to hold. Condition 1. follows from the $\mathbb{P}$-almost sure convergence of $T^{i,n}(t) \to T^i(t)$ for all $t$, which is seen to hold by choosing the constant function $f(t) =1$.
	\end{proof}
	
	\begin{lemma}\label{lemma:compactnessdecisionrule}
		The set of nondecreasing randomized promotion decision $\mathcal{C}^*$ is sequentially compact for the sequential convergence defined by:
		\begin{align*}
		d^n \to d \text{ if and only if } \forall t \geq 0, i \in \{0,\dots, N\}, d^{i,n}_t(\omega) \to d^i_t(\omega), \quad \mathbb{P}\text{-a.s..}
		\end{align*}
	\end{lemma}
	
	\begin{proof}[Proof of Lemma \ref{lemma:compactnessdecisionrule}]
		By Theorem 2.2 and the first Corollary in \cite{becker1969existence}, the set of promotion decision $\mathcal{C}$ is sequentially compact for the sequential convergence defined by: for all $K$ compact subset of $\mathbb{R}_+$ and all continuous function $f:\mathbb{R}_+\times \Delta^{N+1} \to \mathbb{R}$
		\begin{align*}
		d^n \to d \text{ if and only if } \int_K \sum_{i=0}^N f(t, i) d_t^{i,n}({\omega}) dt \to \int_K \sum_{i=0}^N f(t, i)  d_t^i({\omega}) dt \, \, \mathbb{P}\text{-a.s.},
		\end{align*}
		uniformly in $K$. We first show that $\mathcal{C}^*$ is closed in $\mathcal{C}$ for the above convergence, and, hence, that $\mathcal{C}^*$ is (sequentially) compact in the above sense.
		
		Consider a sequence $\left(d^n\right)_{n\in \mathbb{N}}$ in $\mathcal{C}^*$ such that $d^n \to d$ for some $d \in \mathcal{C}$. Let $i \in \left\{0, \dots, N\right\}$. Then, for all continuous function $f:\mathbb{R}_+ \to \mathbb{R}$ and all compact set $K$, 
		\begin{align*}
		\int_K f(t)  d_t^{i,n}({\omega}) dt \to \int_K f(t)  d_t^i({\omega}) dt, \quad \mathbb{P}\text{-a.s.},
		\end{align*}
		uniformly in $K$. But $\left(d_t^{i,n}({\omega})\right)_{n \in \mathbb{N}}$ is a sequence of bounded (by 1)  c\`{a}dl\`{a}g  monotone function, since $d^n \in \mathcal{C}^*$. By Helly's selection theorem, there exists a subsequence $\left(d_t^{i,n_k}({\omega})\right)_{k \in \mathbb{N}} \subseteq \left(d_t^{i,n}({\omega})\right)_{n \in \mathbb{N}}$ such that $d^{i,n_k}_t$ converges to some nondecreasing c\`{a}dl\`{a}g function $\bar{d}^i_t(\omega)$ pointwise almost everywhere on $\mathbb{R}_+$. But then, by the dominated convergence theorem, for all $K$ compact,
		\begin{align*}
		\int_K f(t)  d_t^{i,n_k}({\omega}) dt \to \int_K f(t)  \bar{d}_t^i({\omega}) dt.
		\end{align*}
		So, by uniqueness of the limit, ${d}^i_t = \bar{d}^i_t$ (in the sense of the topology of Becker, i.e., $\mathbb{P}\times \ell$-a.e.), and therefore $d^i_t$ is nondecreasing $\mathbb{P}$-a.s.. 
		
		There remains to show that it implies that, for all $t \geq 0$, $d^{i,n}_t(\omega) \to d^i_t(\omega)$, $\mathbb{P}$-a.s.. This follows from Lebesgue differentiation theorem, the right continuity of both $d^{i,n}$ and $d^i$, the fact that the convergence is uniform in $K$, and the Moore-Osgood theorem.
	\end{proof}
	
	\begin{lemma}\label{lemma:equivalentconstraints}
		Let $\left(T,S,{d}\right)$ be a promotion contest. $(T,S,{d})$ satisfies \eqref{eq:DPC} if and only if $(T, S, {d})$ satisfies, for all $i \in \left\{ 1, \dots, N\right\}$, for all $\mathcal{F}^i$-stopping times $\tilde{\tau}$,
		\begin{align*}
		\mathbb{E} & \left[ \int_{0}^{\infty} \left( e^{-r \tau} g^i {d}^i_{\tau} - \int_{0}^\tau e^{-rt} c^i\left( X^i_{T^i(t)} \right)  d T^i(t) \right) dS(\tau) \right] \\
		& \quad \quad \quad \quad \geq \mathbb{E} \left[ \int_{0}^{\infty} \left( e^{-r\tau} g^i {d}^i_{\tau} \mathbbm{1}_{\{ \tau < \tilde{\tau}\}} - \int_{0}^{\tau\wedge \tilde{\tau}} e^{-rt} c^i\left( X^i_{T^i(t)} \right)  d t\right) dS(\tau) \right].
		\end{align*}
	\end{lemma}
	
	\begin{proof}[Proof of Lemma \ref{lemma:equivalentconstraints}]
		($\Rightarrow$) This follows from lemma \ref{lemma:optimalstoppingcharacterization}. \eqref{eq:DPC} implies that, for all $\tau' \leq \tau \in \mathcal{T}\left( \mathcal{F}^i_{T^i(\cdot)} \right)$, 
		\begin{align*}
		\mathbb{E} \left[\int_{0}^{\infty} \left( e^{-r \left(\tau- {\tau'}\right)} g^i {d}^i_{\tau} - \int_{{\tau'}}^{\tau} e^{-r(s-{\tau'})} c^i\left( X^i_{T^i(s)} \right) d T^i (s) \right) d S(\tau)  \mid \mathcal{\mathcal{F}}^i_{T(\tau')} \right] \geq 0;
		\end{align*}
		which, by lemma \ref{lemma:optimalstoppingcharacterization} implies that
		\begin{align*}
		+\infty \in \arg \underset{\tilde{\tau} \in \mathcal{T}^i}{\max } \mathbb{E} \left[ \int_{0}^{\infty} \left( e^{-r \tau \wedge \tilde{\tau}} g^i \tilde{d}^i_{\tau \wedge \tilde{\tau}} \mathbbm{1}_{\{ \tau < \tilde{\tau} \}} - \int_{0}^{\tau\wedge \tau'} e^{-rt} c^i\left( X^i_{T^i(t)} \right) d T^i(t) \right) dS(\tau)\right].
		\end{align*}
		
		($\Leftarrow$) The other direction follows directly from Lemma \ref{lemma:optimalstoppingcharacterization}.
	\end{proof}
	
	\begin{lemma}\label{lemma:existenceoptimalrandomstoppingforfixedTandd}
		Let $T$ be a delegation rule and $d$ be a $\mathcal{G}^T$-optional promotion decision. Then there exists a $\mathcal{G}^T$-(randomized) stopping time $S^*$ that solves 
		\begin{align}\label{eq:relaxedprogramfixedTandd}
		\Pi(T,d) \coloneqq \underset{S}{\sup}\, & \mathbb{E}\Bigg[ \int_{0}^{\infty} \left(  \sum_{i=1}^N\int_{0}^\tau e^{-rt} \pi^i(X^i_{T^i(t)})dT^i(t) + e^{-r\tau} \bar{\pi}\left( X_{T(\tau)}, {d}_{\tau} \right) \right) dS(\tau) \Bigg] \tag{RP(T,d)}
		\end{align}
		subject to, for all $i \in \left\{ 1,\dots, N \right\}$, for all $t \geq 0$, $\mathbb{P}$-a.s.,
		\begin{align*}
		\mathbb{E} \left[ \int_{0}^{\infty}\left( e^{-r \left(\tau- \tau\wedge t\right)} g^i {d}^i_{\tau}  - \int_{\tau\wedge t}^\tau e^{-r(s-\tau\wedge t)} c^i\left( X^i_{T^i(s)} \right)  d T^i(s) \right) dS(\tau)\mid \mathcal{\mathcal{F}}^i_{T^i(t)} \right] \geq 0. \tag{DPC}
		\end{align*}
	\end{lemma}
	
	\begin{proof}[Proof of Lemma \ref{lemma:existenceoptimalrandomstoppingforfixedTandd}]
		The set of $\mathcal{G}^T$-randomized stopping times $\mathcal{S}\left( \mathcal{G}^T \right)$ is $\sigma\left(\mathcal{M}^{\infty}(\mathcal{G}^T), \mathcal{R}(\mathcal{G}^T)\right)$-compact by Lemma 2 in \cite{pennanen2018optimal}, where $\mathcal{R}(\mathcal{G}^T)$ is the set of regular processes equipped with the norm $\left\| y \right\| = \underset{\tau}{\sup} \, \mathbb{E}\left[ y_{\tau} \right]$ and $\mathcal{M}^{\infty}(\mathcal{G}^T)$ is the space of random stopping time. So the set of feasible $\mathcal{G}^T$-randomized stopping times is $\sigma\left(\mathcal{M}^{\infty}(\mathcal{G}^T), \mathcal{R}(\mathcal{G}^T)\right)$-compact as a closed subset of a $\sigma\left(\mathcal{M}^{\infty}(\mathcal{G}^T), \mathcal{R}(\mathcal{G}^T)\right)$-compact set. To see this, consider $\left(S^n \right)_{n \in \mathbb{N}}$ a sequence of feasible $\mathcal{G}^T$-randomized stopping time that converges to some $S$.\footnote{It is enough to prove sequential closeness since the dual of a normed space is a Banach space by Theorem 6.8 in \cite{guide2006infinite} and the Eberlein-\v{S}mulian Theorem (Theorem 6.34 in \cite{guide2006infinite}) then implies that the set of randomized stopping time is also sequentially compact.} 
		
		Let $\left(\mathbbm{1}^a_{\{ \cdot <0 \}}\right)_{a \in \mathbb{N}}$ be a sequence of continuous functions such that, for all $a$, $\mathbbm{1}^a_{\{ \cdot <0 \}} \leq \mathbbm{1}_{\{ \cdot <0 \}}$ and $\mathbbm{1}^a_{\{ \cdot <0 \}} \to \mathbbm{1}_{\{ \cdot <0 \}}$ pointwise.\footnote{It is easily seen that such a sequence exists.} Lemma \ref{lemma:equivalentconstraints} implies that, for all $a \in \mathbb{N}$, for all $i \in \left\{1,\dots, N\right\}$ and all $\mathcal{F}^i_{T^i(\cdot)}$-stopping time $\tilde{\tau}$,
		\begin{align*}
		\underset{n \to \infty}{\lim } \,\mathbb{E} & \left[ \int_0^{\infty} \left( e^{-r \tau} g^i {{d}^{i}_{\tau}} - \int_{0}^\tau e^{-rt} c^i\left( X^{i}_{T^{i}(t)} \right) d T^{i}(t) \right)dS^n(\tau)\right] \\
		&\geq \underset{n \to \infty}{\lim } \, \mathbb{E} \left[ \int_0^{\infty} \left( e^{-r \tau} g^i {d}^{i}_{\tau} \mathbbm{1}^a_{\{ \tau- \tilde{\tau}<0 \}} - \int_{0}^{\tau\wedge \tilde{\tau}} e^{-rt} c^i\left( X^{i}_{T^{i}(t)} \right) d T^{i}(t) \right) dS^n(\tau) \right].
		\end{align*}
		But, for all $(T, d) \in \mathcal{D} \times \mathcal{C}^*$ with $d$ $\mathcal{G}^T$-optional and all $a \in \mathbb{N}$, both the processes
		\begin{align*}
		& e^{-r t } g^i {d}^i_{t} - \int_{0}^t e^{-rs} c^i\left( X^i_{T^{i}(s)} \right) d T^{i}(s), \\
		\text{ and } & e^{-r t } g^i {d}^i_{t} \mathbbm{1}^a_{\{ t- \tilde{\tau}<0 \}}  - \int_{0}^{t\wedge \tilde{\tau}} e^{-rs} c^i\left( X^i_{T^{i}(s)} \right) d T^{i}(s)
		\end{align*}
		have continuous paths $\mathbb{P}$-a.s. and are $\mathcal{G}^T$-optional. Therefore they belong to $\mathcal{R}\left( \mathcal{G}^T \right)$. Theorem 1 in \cite{pennanen2018optimal} and Theorem 6.39 in \cite{guide2006infinite} implies that the bilinear form 
		\begin{align*}
		\mathcal{R}\left( \mathcal{G}^T \right) \times \mathcal{M}^{\infty}_+\left( \mathcal{G}^T \right) & \to \mathbb{R} \\
		(Y, S)& \to \mathbb{E}\left[ \int_{0}^{\infty} Y_\tau dS({\tau}) \right]
		\end{align*}
		is continuous. So, we have
		\begin{align*}
		\mathbb{E} & \left[ \int_0^{\infty} \left( e^{-r \tau} g^i {d}^i_{\tau} - \int_{0}^{\tau} e^{-rt} c^i\left( X^{i}_{T^{i}(s)} \right) d T^i(s) \right)dS(\tau) \right] \\
		& \geq \mathbb{E} \left[ \int_0^{\infty} \left( e^{-r \tau} g^i {d}^i_{\tau}\mathbbm{1}^a_{\{ \tau- \tilde{\tau}<0\}}  - \int_{0}^{\tau \wedge \tilde{\tau}} e^{-rt} c^i\left( X^{i}_{T^{i}(s)} \right) d T^i(s) \right)d S(\tau)\right].
		\end{align*}
		Taking the limit of the right-hand side as $a\to \infty$, by Lebesgue dominated convergence theorem (applied twice), we obtain
		\begin{align*}
		\mathbb{E} & \left[ \int_0^{\infty} \left( e^{-r \tau} g^i {d}^i_{\tau} - \int_{0}^{\tau} e^{-rt} c^i\left( X^{i}_{T^{i}(s)} \right) d T^i(s) \right)dS(\tau) \right] \\
		& \geq \mathbb{E} \left[ \int_0^{\infty} \left( e^{-r \tau} g^i {d}^i_{\tau}\mathbbm{1}_{\{ \tau< \tilde{\tau}\}}  - \int_{0}^{\tau\wedge \tilde{\tau}} e^{-rt} c^i\left( X^{i}_{T^{i}(s)} \right) d T^i(s) \right)d S(\tau)\right].
		\end{align*}
		Lemma \ref{lemma:equivalentconstraints} again implies that $(T, S, d)$ is a feasible promotion contest; and thus the set of all feasible $\mathcal{G}^T$-randomized stopping time is closed, hence compact. 
		
		To conclude, there remains to show that the objective function is continuous in $S$. This follows from the same argument as above by Theorem 1 in \cite{pennanen2018optimal} and Theorem 6.39 in \cite{guide2006infinite}. 
		
		Thus, by Weierstrass's maximum theorem, a solution to \eqref{eq:relaxedprogram} exists (since the feasible set is nonempty).
	\end{proof}
	
	\begin{lemma}[Theorem 5 in \cite{coquet2007convergence}]\label{lemma:continuityatthelimitrelaxedproblem}
		Let $\left(T^n, d^n\right)_{ n \in \mathbb{N}} \subseteq \mathcal{P}^{T^*}$ be a sequence of pairs of delegation rules and promotion decision such that $T^n \to T$ in the sense of Lemma \ref{lemma:setofoptionalincreasingpathsiscompact} and $d^n \to d$ in the sense of Lemma \ref{lemma:compactnessdecisionrule}. Suppose that $d$ $\mathcal{G}^T$-optional. Then $\Pi\left( T^n, d^n \right) \to \Pi(T,d)$.
	\end{lemma}
	
	\begin{proof}[Proof of Lemma \ref{lemma:continuityatthelimitrelaxedproblem}]
		The proof follows from the proof of the second case of Theorem 5 in \cite{coquet2007convergence}. To see this, observe that the process $Y^n \coloneqq \left( X^1_{T^{1,n}(t)}, \dots, X^N_{T^{N,n}(t)} \right)$ is quasi-left continuous for all $n \in \mathbb{N}$, and, therefore, Aldous' criterion for tightness\footnote{See equation (1) in \cite{coquet2007convergence} for a definition.} holds by Proposition 3 in \cite{coquet2007convergence}. Moreover, for all $t\geq 0$, $T^n(t) \to T(t)$ $\mathbb{P}$-a.s.. So Proposition 1 in \cite{coquet2001weak} implies that $\mathcal{G}^{T^n} \to^w \mathcal{G}^T$. Finally, since each $X^i$ is continuous in probability, we have $Y^n \to Y \coloneqq \left( X^1_{T^{1}(t)}, \dots, X^N_{T^{N}(t)} \right)$ in probability. 
		
		Therefore we obtain the desired result by Theorem 5 in \cite{coquet2007convergence}, upon noting that the proof applies to our constrained stopping problem \eqref{eq:relaxedprogramfixedTandd} provided that:
		
		\textbf{(i)} If $S^n \to S$ for the $\sigma\left(\mathcal{M}^{\infty}\left( \mathcal{G}^T\vee \bigvee_n \mathcal{G}^{T^n}\right), \mathcal{R}\left( \mathcal{G}^T\vee \bigvee_n \mathcal{G}^{T^n}\right)\right)$-convergence, then $S$ is feasible in the \eqref{eq:relaxedprogramfixedTandd}, i.e., $\left( T, S, d \right)$ satisfies all the constraints \eqref{eq:DPC}.
		
		\textbf{(ii)} The objective function $\sum_{i=1}^N\int_{0}^\tau e^{-rt} \pi^i(X^i_{T^i(t)})dT^i_n(t) + e^{-r\tau} \bar{\pi}\left( X_{T(\tau)}, d_n \right)$ converges to $\sum_{i=1}^N\int_{0}^\tau e^{-rt} \pi^i(X^i_{T^i(t)})dT^i(t) + e^{-r\tau} \bar{\pi}\left( X_{T(\tau)}, {d} \right)$ in $\mathcal{R}\left( \mathcal{G}^T\vee \bigvee_n \mathcal{G}^{T^n} \right)$.
		
		Start with \textbf{(i)}. Consider a sequence of feasible promotion contests $\left( T^n,  S^n, d^n \right)_{n \in \mathbb{N}}$ that converges to some $\left(T, S,d\right)$, in the senses defined above. Suppose first that ${c}^i$ is continuous. The result for general $c^i$'s then follows by an approximation argument. Let $T \geq 0$. For all $i \in \{1,\dots, N\}$ and all stopping times $\tau \leq T$, 
		\begin{align*}
		e^{-r \tau} g {{d}^{i,n}_{\tau}} - \int_{0}^\tau e^{-rs} c^i\left( X^i_{T^i_n(s)} \right) d T^i_n(s) \to e^{-r \tau} g^i {{d}^i_{\tau}} - \int_{0}^\tau e^{-rs} c^i\left( X^i_{T^i(s)} \right) d T^i(s),
		\end{align*}
		uniformly over $\tau \leq T$ $\mathbb{P}$-a.s.. Let $\epsilon >0$. By Egorov theorem (Theorem 10.39 in \cite{guide2006infinite}), the convergence is uniform on a set $\mathcal{O} \subseteq \Omega$ with $\mathbb{P}\left( \omega \in \mathcal{O}\right) \geq 1-\frac{\epsilon}{C_1}$, where $C_1 > 2 \left(g+\frac{\sup c^i}{r}\right) $. So, there exists $N$ such that for all $n \geq N$ and  all $\tau \leq T$,
		\begin{align*}
		\left| e^{-r \tau} g^i {{d}^{i,n}_{\tau}} - \int_{0}^\tau e^{-rt} c^i\left( X^i_{T^i_n(t)} \right) d T^i_n(t) -\left( e^{-r \tau} g^i {{d}^i_{\tau}} - \int_{0}^\tau e^{-rt} c^i\left( X^i_{T^i(t)} \right) d T^i(t)\right) \right| < \epsilon
		\end{align*}
		on $\mathcal{O}$. But then, for all $n \geq N$ and all $\tau \leq T$
		\begin{align*}
		\mathbb{E} \Bigg[ & \left| e^{-r \tau} g^i {{d}^{i,n}_{\tau}} - \int_{0}^\tau e^{-rt} c^i\left( X^i_{T^i_n(t)} \right)  d T^i_n(t) - \left( e^{-r \tau} g {{d}^i_{\tau}} - \int_{0}^\tau e^{-rt} c^i\left( X^i_{T^i(s)} \right)  d T^i(t) \right)\right| \mathbbm{1}_{O}(\omega) \\
		& +  \left| e^{-r \tau} g^i {{d}^{i,n}_{\tau}} - \int_{0}^\tau e^{-rt} c^i\left( X^i_{T^i_n(t)} \right)  d T^i_n(t) - \left( e^{-r \tau} g^i {{d}^i_{\tau}} - \int_{0}^\tau e^{-rt} c^i\left( X^i_{T^i(t)} \right)  d T^i(t) \right)\right| \mathbbm{1}_{O^c}(\omega)  \Bigg] \\ 
		& \leq \frac{\epsilon}{2} + \frac{\epsilon}{2} =\epsilon.
		\end{align*}
		So
		\begin{align*}
		\underset{\tau \leq T}{\sup } \, \mathbb{E} \left[  \left| e^{-r \tau} g^i {{d}^{i,n}_{\tau}} - \int_{0}^\tau e^{-rt} c^i\left( X^i_{T^i_n(t)} \right)  d T^i_n(t) - \left( e^{-r \tau} g^i {{d}^i_{\tau}} - \int_{0}^\tau e^{-rt}  c^i\left( X^i_{T^i(t)} \right) d T^i(t) \right)\right| \right] \to 0,
		\end{align*}
		as $n \to \infty$. But, for all $T \geq 0$,
		\begin{align*}
		\underset{\tau}{\sup } \, &\mathbb{E} \left[ \left| e^{-r \tau} g^i {{d}^{i,n}_{\tau}} - \int_{0}^\tau e^{-rs} c^i\left( X^i_{T^i_n(s)} \right) d T^i_n(s) - \left( e^{-r \tau} g^i {{d}^i_{\tau}} - \int_{0}^\tau e^{-rs} c^i\left( X^i_{T^i(s)} \right) d T^i(s) \right) \right|  \right] \\ 
		& \leq \underset{\tau}{\sup } \, \mathbb{E} \Bigg[  \left| e^{-r \tau\wedge T} g^i \left({{d}^{i,n}_{\tau\wedge T }} - {{d}^{i}_{\tau \wedge T}} \right)  - \left( \int_{0}^{\tau\wedge T} e^{-rs} c^i\left( X^i_{T^i_n(s)} \right) dT^i_n(s) - \int_{0}^{\tau\wedge T} e^{-rs} c^i\left( X^i_{T^i(s)} \right) dT^i(s)\right) \right| \\
		& + \left| e^{-r \tau\wedge T} g^i \left({{d}^{i,n}_{\tau}} - {{d}^{i}_{\tau}}  + {{d}^{i}_{\tau\wedge T}} - {{d}^{i,n}_{\tau\wedge T}}  \right) \right|+ \left| \int_{\tau\wedge T}^\tau e^{-rs} c^i\left( X^i_{T^i_n(s)} \right) d T^i_n(s) -\int_{\tau\wedge T}^\tau e^{-rs} c^i\left( X^i_{T^i(s)} \right)  dT^i(s) \right| \Bigg] \\
		& \leq \underset{\tau \leq T}{\sup } \, \mathbb{E} \Bigg[  \bigg| e^{-r \tau\wedge T} g^i {{d}^{i,n}_{\tau \wedge T}} - e^{-r \tau\wedge T} g^i {{d}^{i}_{\tau\wedge T}} \\
		& \qquad -\left( \int_{0}^{\tau\wedge T} e^{-rt} c^i\left( X^i_{T^i_n(t)} \right) dT^i_n(t) -  \int_{0}^{\tau\wedge T} e^{-rt} c^i\left( X^i_{T^i_n(t)} \right) dT^i(t) \right) \bigg| \Bigg] + e^{-rT} C_1.
		\end{align*}
		Therefore,
		\begin{align*}
		\underset{\tau}{\sup } \, \mathbb{E} \left[ \left| e^{-r \tau} g^i {{d}^{i,n}_{\tau}} - \int_{0}^\tau e^{-rs} c^i\left( X^i_{T^i(s)} \right) d T^i_n(s) - \left( e^{-r \tau} g^i {{d}^i_{\tau}} - \int_{0}^\tau e^{-rs} c^i\left( X^i_{T^i(s)} \right) d T^i(s) \right) \right|  \right] \to 0.
		\end{align*}
		So,\footnote{All the processes have continuous paths and are adapted to $\mathcal{G}^T \vee \bigvee_{n=0}^{\infty} \mathcal{G}^{T^n}$, so they belong to $\mathcal{R}\left( \mathcal{G}^T \vee \bigvee_{n=0}^{\infty} \mathcal{G}^{T^n}\right)$.}
		\begin{align*}
		e^{-r t} g^i {{d}^{i,n}_{t}} - \int_{0}^t e^{-rs} c^i\left( X^i_{T^i(s)} \right) d T^i_n(s) \to e^{-r t} g^i {{d}^i_{t}} - \int_{0}^t e^{-rs} c^i\left( X^i_{T^i(s)} \right) d T^i(s) \text{ in } \mathcal{R}\left( \mathcal{G}^T \vee \bigvee_{n=0}^{\infty} \mathcal{G}^{T^n} \right)
		\end{align*}
		Letting $\left(\mathbbm{1}^a_{\{ \cdot <0 \}}\right)_{a \in \mathbb{N}}$ be a sequence of continuous functions such that, for all $a$, $\mathbbm{1}^a_{\{ \cdot <0 \}} \leq \mathbbm{1}_{\{ \cdot <0 \}}$ and $\mathbbm{1}^a_{\{ \cdot <0 \}} \to \mathbbm{1}_{\{ \cdot <0 \}}$ pointwise as in the proof of \ref{lemma:existenceoptimalrandomstoppingforfixedTandd}, we obtain, by Theorem 1 in \cite{pennanen2018optimal} and Theorem 6.39 in \cite{guide2006infinite}, for all $a \in \mathbb{N}$ and all $\mathcal{F}^i$-stopping time $\tilde{\tau}$,
		\begin{align*}
		\mathbb{E} & \left[ \int_0^{\infty} \left( e^{-r \tau} g^i {d}^i_{\tau} - \int_{0}^{\tau} e^{-rs} c^i\left( X^i_{T^i(s)} \right) d T^i(s) \right)dS(\tau) \right] \\
		& = \underset{n \to \infty}{\lim } \, \mathbb{E} \left[ \int_0^{\infty} \left( e^{-r \tau} g^i {{d}^{i,n}_{\tau}} - \int_{0}^\tau e^{-rs} c^i\left( X^i_{T^{i,n}(s)} \right) d T^{i,n}(s) \right)dS^n(\tau)\right] \\
		&\geq \underset{n \to \infty}{\lim } \, \mathbb{E} \left[ \int_0^{\infty} \left( e^{-r \tau} g^i {d}^{i,n}_{\tau} \mathbbm{1}^a_{\{\tau - \tilde{\tau} <0 \}}  - \int_{0}^{\tau\wedge\tilde{\tau}} e^{-rs} c^i\left( X^i_{T^{i,n}(s)} \right) d T^{i,n}(s) \right)dS^n(\tau) \right] \\
		& =\mathbb{E} \left[ \int_0^{\infty} \left( e^{-r \tau} g^i {d}^i_{\tau} \mathbbm{1}^a_{\{\tau - \tilde{\tau} <0 \}} - \int_{0}^{\tau\wedge \tilde{\tau}} e^{-rs} c^i\left( X^i_{T^i(s)} \right) d T^i(s) \right)d S(\tau) \right].
		\end{align*}
		where the inequality follows from Lemma \ref{lemma:equivalentconstraints}. Taking the limit of the right-hand side as $a\to \infty$, by Lebesgue dominated convergence theorem (applied twice), we obtain
		\begin{align*}
		\mathbb{E} & \left[ \int_0^{\infty} \left( e^{-r \tau} g^i {d}^i_{\tau} - \int_{0}^{\tau} e^{-rt} c^i\left( X^{i}_{T^{i}(s)} \right) d T^i(s) \right)dS(\tau) \right] \\
		& \geq \mathbb{E} \left[ \int_0^{\infty} \left( e^{-r \tau} g^i {d}^i_{\tau}\mathbbm{1}_{\{ \tau< \tilde{\tau}\}}  - \int_{0}^{\tau\wedge \tilde{\tau}} e^{-rt} c^i\left( X^{i}_{T^{i}(s)} \right) d T^i(s) \right)d S(\tau)\right].
		\end{align*}
		If $c^i$ is not continuous, we can find a sequence of continuous function ${c}^{i,n}$ that converge to $c^i$ pointwise such that $\int_{0}^{\tau}c^{i}(X^i_{T^{i}(s)}) dT^i(s) \to \int_{0}^{\tau}c^{i,n}(X^i_{T^{i}(s)}) dT^i(s)$ for all $\tau$ $\mathbb{P}$-a.s.. We then obtain the same inequality by the Lebesgue dominated convergence theorem. Then Lemma \ref{lemma:equivalentconstraints} implies that $(T, S, d)$ satisfies \eqref{eq:DPC}.
		
		To conclude, there remains to show \textbf{(ii)}. This is done exactly as in the first part of the proof of \textbf{(i)}.
	\end{proof}
	
	\subsubsection{Second part of Proposition \ref{prop:measurablestopping}: Characterization of the promotion time}\label{app:measurablestopping2}
	
	\begin{lemma}\label{lemma:stoppingismeasurable}	
		Suppose that the promotion contest $\left(T, \tau, d\right)$ solves \eqref{eq:relaxedprogram}. If worker $i$ is promoted at time $t$, then $\left\{\tau = t \right\} \in \mathcal{F}^i_{T^i(t)}$.
	\end{lemma}
	
	\begin{proof}
		I will prove the contrapositive. So let $(T, \tau, d)$ be an implementable promotion contest with $\{\tau = t \}$ such that $i$ is promoted. Suppose that $\{\tau = t \} \not \in \mathcal{F}^i_{T^i(t)}$. I want to show that $(T, \tau, d)$ is not optimal.
		
		Since $\left\{\tau = t \right\} \not \in \mathcal{F}^i_{T^i(t)}$, $\mathbb{P}\left( \left\{\tau = t \right\} \mid \mathcal{F}^i_{T^i(t)} \right) \in (0,1)$. But then
		\begin{align*}
		\tilde{U}^i_t \coloneqq \mathbb{E} \left[ e^{-r \left(\tau- t\right)} g^i d^i_{\tau}- \int_{t}^\tau e^{-r(s-t)} c^i\left( X^i_{T^i(s)} \right) d T^i(s)  \mid \mathcal{\mathcal{F}}^i_{T^i(t)} \right]  < g^i.
		\end{align*}
		To see this, suppose not, i.e., $\tilde{U}^i_t \geq g$. Then
		\begin{align*}
		g^i &\leq \tilde{U}^i_t = \mathbb{E} \left[ e^{-r \left(\tau- t\right)} g^i d^i_{\tau}- \int_{t}^\tau e^{-r(s-t)}  c^i\left( X^i_{T^i(s)} \right) d T^i(s)  \mid \mathcal{\mathcal{F}}^i_{T^i(t)} \right] \\
		& \leq \mathbb{E} \left[ e^{-r \left(\tau- t\right)} g^i d^i_{\tau}  \mid \mathcal{\mathcal{F}}^i_{T^i(t)} \right] \\
		\Leftrightarrow  & \mathbb{E} \left[ e^{-r \left(\tau- t\right)} d^i_{\tau}  \mid \mathcal{\mathcal{F}}^i_{T^i(t)} \right] = 1 \\ 
		\Rightarrow & \mathbb{P}\left( \{\tau = t\} \cap \{d^i_\tau=1\} \mid \mathcal{F}^i_{T^i(t)}  \right) =1.
		\end{align*}
		This is a contradiction. So $\tilde{U}^i_t < g^i$. 
		
		Consider then the promotion contest $\left( \tilde{T}, \tilde{\tau}, \tilde{d}\right)$ with $\tilde{T} \coloneqq T$ on $[0, \tau)$ and $\tilde{T}^j(t) = T^j(\tau)$ if $j \neq i$ and $\tilde{T}^j(t) = t - \tau$ if $j = i$ on $[\tau, \tilde{\tau}]$, $\tilde{d} \in \arg \underset{ d \in \Delta^{N+1} }{\max}\, \mathbb{E}\left[ \bar{\pi}^i\left(X^i_{T^i(\tilde{\tau})}\right) d^i_{\tilde{\tau}} + d^0_{\tilde{\tau}} W \mid \mathcal{G}^{\tilde{T}}_{\tilde{\tau}}\right]$ subject to $d^j_{\tau} = 0$ for all $j \not \in \left\{0,i\right\}$, and $\tilde{\tau}$ is chosen to be the optimal continuation promotion contest in the single $i$-arm problem after time $t$ starting at time $t$ with
		\begin{align*}
		\mathbb{E}\left[ e^{-r \left(\tilde{\tau}- t\right)} g^i d^i_{\tilde{\tau}} - \int_{t}^{\tilde{\tau}} e^{-r(s-t)} c^i\left( X^i_{T^i(s)} \right) d T^i(s)  \mid \mathcal{\mathcal{F}}^i_{T^i(\tau)} \right] = \tilde{U}^i_\tau,
		\end{align*}
		given by Theorem \ref{theorem:singleagentoptimal}.\footnote{I.e., set the value of the running minimum such that the above equation holds.}
		
		Finally observe that $\left( \tilde{T}, \tilde{\tau}, \tilde{d}\right)$ is feasible. The payoffs of all workers $j \neq i$ are the same after any history of the game. The payoff of player $i$ is unchanged before $\tau$ (by the law of iterated expectations and the definition of $\tau$) and nonnegative on the random interval $(\tau, \tilde{\tau}]$ as $\tilde{\tau}$ is chosen $i$-arm feasible.
		
		Theorem \ref{theorem:singleagentoptimal} then guarantees that the alternative promotion contest $\left( \tilde{T}, \tilde{\tau}, \tilde{d}\right)$ yields a strictly higher payoff for the principal than the original promotion contest $\left( {T}, {\tau}, {d}\right)$. This concludes the proof.
	\end{proof}
	
	A consequence of the above lemma is that we can think of the principal as choosing $N$ promotion times, one for each agent, and one retiring time $\tau^0$, instead of just one $\tau$. Formally,
	
	\begin{corollary}\label{corollary:stoppingismeasurable}
		Let $(T, \tau, d)$ be a promotion contest solving \eqref{eq:relaxedprogram}. Then $\tau = \left( \bigwedge_{i=1}^N \tau^i \right) \wedge \tau^0$, where $\tau^i$ is a $\mathcal{F}^i$-stopping time, $\tau^0$ is a $\mathcal{G}^T$-stopping time, and $i$ is promoted only if $\tau^i \leq \tau = \left( \bigwedge_{i=1}^N \tau^i \right) \wedge \tau^0$.
	\end{corollary}
	
	\begin{proof}
		Define, for all $i \in \{1,\dots, N\}$,
		\begin{align*}
		\tau^i(\omega) \coloneqq \begin{cases}
		\tau(\omega) \text{ if } \omega \in \{ \omega \in \Omega \, : \, d^i_{\tau} = 1 \} \\
		+ \infty \text{ otherwise}
		\end{cases};
		\end{align*}
		and
		\begin{align*}
		\tau^0(\omega) \coloneqq \begin{cases}
		\tau(\omega) \text{ if } \omega \in \{ \omega \in \Omega \, : \, d^0_{\tau} = 1 \} \\
		+ \infty \text{ otherwise}
		\end{cases}.
		\end{align*}
		By lemma \ref{lemma:stoppingismeasurable}, each $\tau^i$ is a $\mathcal{F}^i$-stopping time. The result follows.
	\end{proof}

	\subsection{Omitted Proofs for Section \ref{subsec:upperboundonRP}}\label{app:upperbound}
	
	\subsubsection{Proof of Proposition \ref{prop:upperboundonRP}}\label{app:upperboundonRP}
	
	\paragraph{} From Proposition \ref{prop:measurablestopping}, we can assume without loss of optimality that $\tau = \left( \bigwedge_{i=1}^N \bar{\tau}^i \right) \wedge \tau^0$, where ${\tau}^i$ is a $\mathcal{F}^i_{T^i(\cdot)}$-stopping time, ${\tau}^0$ is a $\mathcal{G}^T_t$-stopping time, and that $d^i_t = \mathbbm{1}_{\{{\tau}^i \leq {\tau} \leq t\}}$, $i \in \left\{0, \dots, N\right\}$. Then $(T, \tau, d)$ generates the following $\mathcal{G}^T_t$-adapted reward processes
	\begin{align*}
		h^i_t = \pi^i(X^i_t) \mathbbm{1}_{\{ t < T^i({\tau}) \}} +  r \bar{\pi}^i\left( X^i_{T^i(\tau)} \right) \mathbbm{1}_{\{ t \geq T^i({\tau}) \}}, \quad i =1,\dots, N,
	\end{align*}
	Moreover let $h^{0}_t = rW$, $t\geq 0$, i.e., I consider an alternative problem in which the outside option is an $N+1^{th}$ arm that can be pulled at any instant and gives a flow payoff of $rW$ to the principal. Let $(T, {\tau}, d)$ be a feasible promotion contest. This relaxes the principal problems. Observe that
	\begin{align*}
	\mathbb{E}\left[ \sum_{i=1}^N \int_{0}^{\tau} \pi^i\left(X^i_{T^i(t)}\right) dT^i(t) + e^{-r\tau} \bar{\pi}\left(X_{T(\tau)}, d\right) \right] = \mathbb{E} \left[ \sum_{i=0}^{N} \int_{0}^{\infty} h^i_{T^i(t)} d \tilde{T}^i(t)  \right];
	\end{align*}
	with $\tilde{T}(t) = T(t)$ if $t \leq \tau$ and, for all $i \in\left\{ 0,\dots, N\right\}$ and $t \geq \tau$, $\tilde{T}^i(t) = d^i (t-\tau) + T^i(\tau)$.
	
	Define $\bar{\tau}^i \coloneqq T^i(\tau^i)$; and let $\tau^i_0$ be the solution of
	\begin{align*}
	\underset{\tau_0}{\sup } \, \mathbb{E}\left[ e^{-r\bar{\tau}^i \wedge \tau_0} g^i d^i \mathbbm{1}_{\{ \bar{\tau}^i \leq \tau^0\}} - \int_{0}^{\bar{\tau}^i \wedge \tau_0}e^{-rt} c^i\left( X^i_{t} \right) dt\right],
	\end{align*}
	where the supremum is taken over all $\mathcal{F}^i$-stopping times. Then $\bar{\tau}^i \wedge \tau^i_0 \geq T^i(\tau)$ $\mathbb{P}$-a.s.. To see this, note that, by definition of $\tau^i_0$ and Lemma \ref{lemma:optimalstoppingcharacterization}, for all $\tau^i_0 < \tilde{\tau} \leq \bar{\tau}^i$,
	\begin{align*}
	\mathbb{E}& \left[ e^{-r {\bar{\tau}^i} \wedge \tilde{\tau}} g^i \mathbbm{1}_{ \{ \bar{\tau}^i \leq \tilde{\tau} \}} - \int_{\bar{\tau}^i_0}^{\bar{\tau}^i \wedge \tilde{\tau}} e^{-rt}c^i\left( X^i_{t} \right) dt \mid \mathcal{F}^i_{\tau^i_0} \right] < 0 \\
	\Rightarrow \mathbb{E}& \left[ e^{-r {\tau^i} \wedge T^{i,-1}(\tilde{\tau})} g^i \mathbbm{1}_{ \{ \tau^i \leq T^{i,-1}(\tilde{\tau}) \}} - \int_{T^{i,-1}(\tau^i_0)}^{\tau^i \wedge T^{i,-1}(\tilde{\tau})} e^{-rt}c^i\left( X^i_{T^i(t)} \right) dT^i(t) \mid \mathcal{F}^i_{\tau^i_0} \right] < 0,
	\end{align*}
	where $T^{i,-1}(t) \coloneqq \inf\left\{ s \, : \, T^i(s) > t \right\}$. Therefore $\bar{\tau}^i \wedge \tau^i_0 \geq T^i(\tau)$ $\mathbb{P}$-a.s.. 
	
	So, for all $i =1,\dots, N$, $\mathbb{P}$-a.s.,
	\begin{align*}
	h^i_t \leq \tilde{h}^i_t \coloneqq \pi^i\left(X^i_t\right) \mathbbm{1}_{\{ t < \bar{\tau}^i \wedge \tau^i_0) \}} +  r \bar{\pi}^i\left( X^i_{\bar{\tau}^i\wedge \tau^i_0}, \tilde{d}^i \right) \mathbbm{1}_{\{ t \geq \bar{\tau}^i\wedge \tau^i_0 \}};
	\end{align*}
	where $\tilde{d}^i_t = \mathbbm{1}_{\{ \bar{\tau}^i \leq \tau^i_0 \wedge t \}}$. Then
	\begin{align*}
	\mathbb{E}\left[ \sum_{i=1}^N \int_{0}^{\tau} \pi^i\left(X^i_{T^i(t)}\right) dT^i(t) + e^{-r\tau} \bar{\pi}\left(X_{T(\tau)}, d\right) \right] & = \mathbb{E} \left[ \sum_{i=0}^{N} \int_{0}^{\infty} h^i_{\tilde{T}^i(t)} d \tilde{T}^i(t)  \right] \\
	&\leq \mathbb{E} \left[ \sum_{i=0}^{N} \int_{0}^{\infty} \tilde{h}^i_{\tilde{T}^i(t)} d \tilde{T}^i(t)  \right].
	\end{align*}
	Moreover, $(\tau^i_0 \wedge \bar{\tau}^i, \tilde{d}^i)$ is feasible in the single $i$-arm problem. To see this, note that
	\begin{align*}
	\mathbb{E} \left[ e^{-r {\tau}} g^i \mathbbm{1}_{ \{ \tau^i \leq {\tau} \}} - \int_{0}^{\tau} e^{-rt} c^i\left( X^i_{T^i(t)}\right)  dT^i(t) \right] \geq 0
	\end{align*}
	implies that
	\begin{align*}
	\mathbb{E} & \left[ e^{-r {\tau^i_0} \wedge \bar{\tau}^i } g^i \mathbbm{1}_{ \{ \bar{\tau}^i \leq {\tau}^i_0 \}} - \int_{0}^{{\tau^i_0} \wedge \bar{\tau}^i} e^{-rt} c^i\left( X^i_t \right) d t \right] \\
	& \qquad = \underset{S}{\sup} \,  \mathbb{E} \left[ \int_{0}^{\infty} e^{-r\tilde{\tau}} g^i \mathbbm{1}_{ \{ \tau^i \leq \tilde{\tau} \}} dS(\tilde{\tau}) - \int_{0}^{\infty} \int_{0}^{\tilde{\tau}} e^{-rt} c^i\left( X^i_t \right) d t dS(\tilde{\tau}) \right] \geq 0
	\end{align*}
	since 
	\begin{align*}
	\mathbb{E} & \left[ \int_{0}^{\infty} e^{-r\tilde{\tau}} g^i \mathbbm{1}_{ \{ \bar{\tau}^i \leq \tilde{\tau} \}} dS(\tilde\tau) - \int_{0}^{\infty} \int_{0}^{\tilde{\tau}} e^{-rt} c^i\left( X^i_t \right) d t dS(\tilde{\tau}) \right] \\
	& \qquad = \mathbb{E} \left[  e^{-r{\bar{\tau}^i}} g^i \mathbbm{1}_{ \{ \bar{\tau}^i \leq \tilde{\tau} \}} - \int_{0}^{\infty} e^{-rs} c^i\left( X^i_t \right) \left(1- S(t)\right) d t \right] \\
	& \qquad =  \mathbb{E} \left[  e^{-r{\tau}} g^i \mathbbm{1}_{ \{ \tau^i \leq {\tau} \}} - \int_{0}^{\tau} e^{-rt} c^i\left( X^i_{T^i(t)} \right) d T^i(t) \right] 
	\end{align*}
	for $S(t) = 1- \tilde{q}^i(t) + \mathbbm{1}_{ \{t = \tau^i\}} \tilde{q}^i(\tau)$ and $\tilde{q}^i(t) = e^{-r\left( T^{i,-1}(t) - t \right)} \mathbbm{1}_{\{t\leq \tau\}}$ where $T^{i,-1}(t) \coloneqq \inf\left\{ s \, : \, T^i(s) > t \right\}$. Then, by definition of $\tau^i_0$ and Lemma \ref{lemma:optimalstoppingcharacterization}, for all $\tilde{\tau} \leq \tau^i_0 \wedge \bar{\tau}^i$,
	\begin{align*}
	\mathbb{E} \left[ e^{-r \bar{\tau}^i \wedge \tau^i_0} g^i \mathbbm{1}_{ \{ \bar{\tau}^i \leq {\tau^i_0} \}} - \int_{\tilde{\tau}}^{\bar{\tau}^i \wedge \tau^i_0} e^{-rt} c^i\left( X^i_t \right) d t \mid \mathcal{F}^i_{\tilde{\tau}} \right] \geq 0.
	\end{align*}
	Thus $(\bar{\tau}^i\wedge \tau^i_0, \tilde{d}^i)$ is feasible in the single $i$-arm problem.
	
	Therefore, by Proposition \ref{prop:whittlecomputation} in Appendix \ref{app:upperboundonRPsupportinglemmas} and Theorem 3.7 in \cite{el1997synchronization}, since $\underline{\Gamma}^i_{\cdot}\left( \tilde{h}^i\right)$ is $\mathcal{F}^i$-adapted,
	\begin{align*}
	\mathbb{E}\left[ \sum_{i=1}^N \int_{0}^{\tau} \pi^i\left(X^i_{T^i(t)}\right) dT^i(t) + e^{-r\tau} \bar{\pi}\left(X_{T(\tau)}, d\right) \right] & \leq \mathbb{E}\left[ \sum_{i=0}^N \int_{0}^{\infty} e^{-rt} r \underline{\Gamma}^i_{\tilde{T}^{*,i}(t)} \left( \tilde{h}^i\right) d\tilde{T}^{*,i}(t)\right] \\
	& = \mathbb{E}\left[ \int_{0}^{\infty} e^{-rt} r \bigvee_{i=0}^N \underline{\Gamma}^i_{\tilde{T}^{*,i}(t)} \left( \tilde{h}^i\right) dt\right] 
	\end{align*} 
	where $\tilde{T}^{*}$ is any delegation strategy satisfying the synchronization identity and $\underline{\Gamma}^0_t =W$ for all $t \geq 0$.
	
	Then Lemma \ref{lemma:strategiclowerenvelope} and Corollary \ref{corollary:domination} implies that, for all $\bar{W}\geq W$,
	\begin{align}\label{eq:domination}
	\mathbb{E}\left[ \int_{0}^{\infty} e^{-rt} r \underline{\Gamma}^i_{t} \left( \tilde{h}^i\right) \vee \bar{W} dt\right] & = \mathbb{E} \left[ \int_{0}^{\rho(\bar{W}; \tilde{h}^i)} e^{-rt} \tilde{h}^i_{t} d t +e^{-r \rho(W; h^i)} \bar{W} \right] \leq \mathbb{E} \left[ \int_{0}^{\infty} r e^{-rt} \underline{\Gamma}^{s,i}_t \vee \bar{W} dt \right],
	\end{align}
	where $\rho(\bar{W}; h^i) \coloneqq \inf\left\{ t\geq 0 \, : \, \underline{\Gamma}^i_t\left(  \tilde{h}^i \right) \leq W \right\}$.
	
	We conclude following the proof of indexability for superprocesses in \cite{durandard2022index}. Let
	\begin{align*}
	\underline{\Gamma}^{-i}_t(\tilde{h}^{-i}) = \vee_{j\neq i} \underline{\Gamma}^j_{T^{*,j}(t+ T^{*,i}(t))}(\tilde{h}^j) \text{ and } \bar{\underline{\Gamma}}^{-i, K}_t(\tilde{h}^{-i}) = \sum_{k=0}^{\infty} \underline{\Gamma}^{-i}_{\sigma^k}(\tilde{h}^{-i}) \mathbbm{1}_{\{ t \in [\sigma^k, \sigma^{k+1}) \}},
	\end{align*}
	with $\sigma^k = \inf \left\{ t \geq 0 \, : \, \Gamma^{-i}_t(\tilde{h}^{-i}) \leq \Gamma^{-i}_0 - \frac{k}{K} \Gamma^{-i}_0 \right\}$ for some $K$ large.
	Then 
	\begin{align*}
	\mathbb{E}& \left[ \sum_{i=1}^N \int_{0}^{\infty} e^{-rt} r\underline{\Gamma}^i_{\tilde{T}^{*,i}(t)} \left( \tilde{h}^i\right) d\tilde{T}^{*,i}(t)\right] \\
	& \leq \mathbb{E}\left[ \int_{0}^{\infty} e^{-rt} r\underline{\Gamma}^i_{\tilde{T}^{*,i}(t)} \left( \tilde{h}^i\right) d\tilde{T}^{*,i}(t) + \int_{0}^{\infty} e^{-rt} r \underline{\bar{\Gamma}}^{-i, K}_{\tilde{T}^{*,-i}(t)}(\tilde{h}^{-i})d\tilde{T}^{*,-i}(t)\right] \\
	& \leq \mathbb{E}\left[ \int_{0}^{\infty} e^{-rt} r\underline{\Gamma}^i_{\bar{T}^{K,i}(t)} \left( \tilde{h}^i\right) d\bar{T}^{K,i}(t) + \int_{0}^{\infty} e^{-rt} r\underline{\bar{\Gamma}}^{-i, K}_{\bar{T}^{K,-i}(t)}(\tilde{h}^{-i}) d\bar{T}^{K,-i}(t)\right];
	\end{align*}
	where $\bar{T}^K$ is an optimal index strategy for the two arms bandits with rewards $\underline{\Gamma}^i_t(h^i)$ and $\underline{\bar{\Gamma}}^{-i, K}_t(h^{-i})$ giving priority to arm $-i$, using Theorem 3.7 in \cite{el1997synchronization} again. Letting $\tau^k = \inf\left\{ t \geq 0 \, : \, \Gamma^i_t(\tilde{h}^i) \leq \Gamma^{-i}_0 - \frac{k-1}{K} \Gamma^{-i}_0 \right\}$ and $\tau^{0} =0$, we have
	\begin{align*}
	\mathbb{E}& \left[ \int_{0}^{\infty} e^{-rt} r \underline{\Gamma}^i_{\bar{T}^{K,i}(t)} \left( \tilde{h}^i\right) d\bar{T}^{K,i}(t) + \int_{0}^{\infty} e^{-rt} r \underline{\bar{\Gamma}}^{-i, K}_{\bar{T}^{*,-i}(t)}(\tilde{h}^{-i}) d\bar{T}^{K,-i}(t)\right] \\
	& = \mathbb{E} \left[ \sum_{k=0}^{K} e^{-r\sigma^k}\int_{\tau^k}^{\tau^{k+1}} e^{-rt} r \underline{\Gamma}^i_t(\tilde{h}^i) dt + e^{-r\tau^{k+1}} \int_{\sigma^k}^{\sigma^{k+1}} e^{-rt} r \underline{\bar{\Gamma}}^{-i, K}_{t}(\tilde{h}^{-i}) dt  \right] \\
	& = \mathbb{E} \left[ \sum_{k=0}^{K} e^{-r\sigma^k} \left(\int_{0}^{\tau^{k+1}} e^{-rt} r \underline{\Gamma}^i_t(\tilde{h}^i) dt - \int_{0}^{\tau^{k}} e^{-rt} r \underline{\Gamma}^i_t(\tilde{h}^i) dt  \right) + e^{-r\tau^{k+1}} r \underline{\bar{\Gamma}}^{-i, K}_{\sigma^k}(\tilde{h}^{-i}) \left(e^{-r\sigma^k} - e^{-r\sigma^{k+1}}\right) \right] \\
	& = \mathbb{E} \left[ \sum_{k=0}^{K} \left(e^{-r\sigma^k} -e^{-r\sigma^{k+1}}\right) \left(\int_{0}^{\tau^{k+1}} e^{-rt} r \underline{\Gamma}^i_t(\tilde{h}^i) dt \right) + e^{-r\tau^{k+1}} r \underline{\bar{\Gamma}}^{-i, K}_{\sigma^k}(\tilde{h}^{-i}) \left(e^{-r\sigma^k} - e^{-r\sigma^{k+1}}\right) \right] \\
	& \leq \mathbb{E} \Bigg[ \sum_{k=0}^{K} \left(e^{-r\sigma^k} -e^{-r\sigma^{k+1}}\right) \left(\int_{0}^{\lambda^{k+1}} e^{-rt} r \underline{\Gamma}^{s,i}_t dt + \int_{\lambda^{k+1}}^{\tau^{k+1}} e^{-rt} r \underline{\bar{\Gamma}}^{-i, K}_{\sigma^k}(\tilde{h}^{-i}) dt \right) \\& \quad  + e^{-r\tau^{k+1}} r \underline{\bar{\Gamma}}^{-i, K}_{\sigma^k}(\tilde{h}^{-i}) \left(e^{-r\sigma^k} - e^{-r\sigma^{k+1}}\right) \Bigg] \\
	& = \mathbb{E} \left[ \sum_{k=0}^{\infty} \left(e^{-r\sigma^k} -e^{-r\sigma^{k+1}}\right) \int_{0}^{\lambda^{k+1}} e^{-rt} r \underline{\Gamma}^{s,i}_t dt + e^{-r\lambda^{k+1}} r \underline{\bar{\Gamma}}^{-i, K}_{\sigma^k}(\tilde{h}^{-i}) \left(e^{-r\sigma^k} - e^{-r\sigma^{k+1}}\right) \right] \\
	& \leq \mathbb{E}\left[ \int_{0}^{\infty} e^{-rt} r \underline{\Gamma}^{s,i}_{{T}^{K,i}(t)} d{T}^{K,i}(t) + \int_{0}^{\infty} e^{-rt} r \underline{\bar{\Gamma}}^{-i, K}_{{T}^{K,-i}(t)}(\tilde{h}^{-i}) d{T}^{K,-i}(t)\right],
	\end{align*}
	where $\lambda^k = \inf\left\{ t \geq 0 \, : \, \Gamma^{s,i}_t \leq \Gamma^{-i}_0 - \frac{k-1}{K} \Gamma^{-i}_0 \right\}$, $\lambda^0 =0$, and $T^{K}$ is an optimal strategy for the two arms bandits with rewards $\underline{\Gamma}^{s,i}_t$ and $\underline{\bar{\Gamma}}^{-i, K}_t(h^{-i})$. The first inequality follows from \eqref{eq:domination} and the independence of $\Gamma^{s,i}$ and $\Gamma^{-i,K}$ (as $\Gamma^i$ is $\mathcal{F}^i$-adapted and $\Gamma^{-i,K}$ is $\mathcal{F}^{-i}$-adapted and the filtrations are independent). 
	
	As $K$ goes to infinity, $\underline{\Gamma}^{-i, K}_t(\tilde{h^{-i}}) \to \underline{\Gamma}^{-i}_t(\tilde{h}^{-i})$ for all $t \geq 0$, $\mathbb{P}$-a.s.. By Lemma \ref{lemma:setofoptionalincreasingpathsiscompact}, $T^{K}$ converges to some $T$ along a subsequence. By Lebesgue dominated convergence theorem (applied twice) and Theorem 3.7 in \cite{el1997synchronization}, we obtain
	\begin{align*}
	\mathbb{E}& \left[ \sum_{i=0}^N \int_{0}^{\infty} e^{-rt} r \underline{\Gamma}^{s,i}_{\tilde{T}^{*,i}(t)} \left( \tilde{h}^i\right) d\tilde{T}^{*,i}(t)\right]\\
	&  \leq \mathbb{E}\left[ \int_{0}^{\infty} e^{-rt} r \underline{\Gamma}^{s,i}_{{T}^{*,i}(t)} d{T}^{*,i}(t) + \int_{0}^{\infty} e^{-rt} r \underline{\bar{\Gamma}}^{-i}_{{T}^{*,-i}(t)}(\tilde{h}^{-i})d{T}^{*,-i}(t)\right],
	\end{align*}
	where ${T}^*$ is an optimal index strategy for the two arms bandits with rewards $\underline{\Gamma}^i_t$ and $\underline{\bar{\Gamma}}^{-i}_t(\tilde{h}^{-i})$.
	
	Reproducing the same argument for all $j \neq i$, we have
	\begin{align*}
	\mathbb{E}\left[ \sum_{i=1}^N \int_{0}^{\tau} \pi^i\left(X^i_{T^i(t)}\right) dT^i(t) + e^{-r\tau} \bar{\pi}\left(X_{T(\tau)}, d\right) \right] & \leq \mathbb{E}\left[ \sum_{0=1}^N \int_{0}^{\infty} e^{-rt} r \underline{\Gamma}^{s,i}_{{T}^{s,i}(t)}  d{T}^{s,i}(t)\right] \\
	& = \mathbb{E}\left[ \int_{0}^{\infty} e^{-rt} r \bigvee_{i=1}^N \underline{\Gamma}^{s,i}_{{T}^{s,i}(t)} dt\right].
	\end{align*}
	This concludes the proof. \qed
	
	\subsubsection{Supporting Lemmas for the proof of Proposition \ref{prop:upperboundonRP}}\label{app:upperboundonRPsupportinglemmas}
	
	\paragraph{} Let $\left( \tau^i, d^i \right) \in \mathcal{P}^{I,r,i}$. As above, define the process $h^i$ as
	\begin{align*}
	h^i_t \coloneqq \pi^i(X^i_t) \mathbbm{1}_{\{ t < \tau^i \}} +  r \bar{\pi}^i\left( X^i_{\tau^i}, d^i \right) \mathbbm{1}_{\{ t \geq \tau^i \}}, \quad t\geq 0
	\end{align*}
	Consider the family of stopping problems: for all $t \geq 0$,
	\begin{align}\label{eq:stoppingstrategicindex}
	\tilde{V}^i\left(t, W; \tau^i, d^i \right) & \coloneqq  ess \underset{\rho \in \mathcal{T}^s(t ; {\tau}^i, d^i)}{ \sup} \, \mathbb{E} \left[ \int_{t}^{\rho} e^{-r(s-t)}  {h}^i_{s} ds + e^{-r (s-t)} W \mid \mathcal{F}^i_{t} \right] \notag \\
	\end{align}
	where
	\begin{align*}
	\mathcal{T}^{s}(t; \tau^i, d^i) = \left\{ s \geq t \, : \, U^i_{s}(\tau^i, d^i) = 0 \right\},
	\end{align*}
	and 
	\begin{align*}
	U^i_s(\tau^i, d^i) = \mathbb{E}\left[ e^{-r\tau^i} g^i d^i - \int_{s}^{\tau^i} e^{-rt} c^i\left(X^i_{t}\right) d t \mid \mathcal{F}^i_{s}  \right]
	\end{align*}
	is the continuation value of worker $i$. Let $\rho(t, {W}; \tau^i, d^i)$ be the smallest optimal stopping time in the problem \eqref{eq:stoppingstrategicindex} with outside option ${W}$.\footnote{Existence of an optimal stopping time follows from the standard Snell envelope argument.} In what follows, we will abuse a little notation and denote $\rho(0, {W}; \tau^i, d^i)$ as $\rho({W}; \tau^i, d^i)$.
	
	\begin{lemma}\label{lemma:tildeVisconvex}
		The mapping
		\begin{align*}
		W \to \tilde{V}^i\left(0, W; \tau^i, d^i \right) \coloneqq \underset{\rho \in \mathcal{T}^s(0; \tau^i, d^i)}{ \sup} \, \mathbb{E} \left[ \int_{0}^{\rho} e^{-r(s-t))} {h}^i_{s} ds + e^{-r \rho} W \right]
		\end{align*}
		is convex, nondecreasing, and locally Lipschitz, with
		\begin{align*}
		\underset{ W \to \infty}{\lim} \, \tilde{V}^i\left(0, W; \tau^i, d^i \right) - W = 0,
		\end{align*}
	\end{lemma}
	
	\begin{proof}[Proof of Lemma \ref{lemma:tildeVisconvex}]
		We first show that the $W \to \tilde{V}^i\left(0,  W; \tau^i, d^i\right)$ is nondecreasing. Observe that for all $W' \geq W \geq 0$ and all $\rho \in \mathcal{T}^s(0; \tau^i, d^i)$, we have
		\begin{align*}
		\mathbb{E} \left[ \int_{0}^{\rho} e^{-r s} {h}^i_{s} d s + e^{-r \rho} W' \right] \geq  \mathbb{E} \left[ \int_{0}^{\rho} e^{-r s} {h}^i_{s} d s + e^{-r \rho} W \right].
		\end{align*}
		It follows that $W \to \tilde{V}^i\left(0,  W;  \tau^i, d^i\right)$ is nondecreasing.
		
		Next we show that $W \to \tilde{V}^i\left(0,  W; \tau^i, d^i\right)$ is convex. Let $\alpha \in (0,1)$ and $W, W' \geq 0$. We have
		\begin{align*}
		\tilde{V}^i&\left(0, \alpha W + (1- \alpha)W'; \tau^i, d^i\right) \\
		& = ess \underset{\rho \in \mathcal{T}^s(0; \tau^i, d^i)}{ \sup} \, \mathbb{E} \left[ \int_{0}^{\rho} e^{-r s} {h}^i_{s} d s + e^{-r \rho} \left( \alpha W + (1- \alpha)W' \right) \right] \\
		& \leq \alpha ess \underset{\rho \in \mathcal{T}^s(0; \tau^i, d^i)}{ \sup} \, \mathbb{E} \left[ \int_{0}^{\rho} e^{-r s} {h}^i_{s} ds + e^{-r \rho} W \right] \\
		& \qquad + (1-\alpha) ess \underset{\rho \in \mathcal{T}^s(t;  \tau^i, d^i)}{ \sup} \, \mathbb{E} \left[ \int_{0}^{\rho} e^{-r s} {h}^i_{s} d s + e^{-r \rho} W' \right] \\
		& = \alpha \tilde{V}^i\left(0,  W;  \tau^i, d^i\right) + (1-\alpha) \tilde{V}^i\left(0, W';  \tau^i, d^i\right);
		\end{align*}
		where we used that the supremum of the sum is less than the sum of the supremum. So $W \to \tilde{V}^i\left(0,  W; \tau^i, d^i\right)$ is convex on $[0, \infty)$.
		
		Taken together, these first two results implies that $W \to \tilde{V}^i\left(0,  W; \tau^i, d^i\right)$ is locally Lipschitz, as a convex function is locally Lipschitz in the interior of its domain and $W \to \tilde{V}^i\left(0,  W; \tau^i, d^i\right)$ being nondecreasing implies that it is continuous at zero. 
		
		There remains to show that $\tilde{V}^i(0, W; \tau^i, d^i) - W  \to 0$. Note that
		\begin{align*}
		\tilde{V}^i(0, W; \tau^i, d^i) - W & \leq \mathbb{E} \left[ \int_{0}^{\infty} e^{-rt} \left( \pi^i\left(X^i_t\right) -W \right)^+ dt \right] \\
		& \to 0,
		\end{align*}
		as $W \to \infty$ by the monotone convergence theorem. Since $\tilde{V}^i(0, W; \tau^i, d^i) \geq 0$ (as $U^i_{0^-}( \tau^i, d^i) =0$ by convention), we have the desired result.
	\end{proof}
	
	\begin{lemma}\label{lemma:derivativeoftildeV}
		The mapping $W \to \tilde{V}^i\left(0,  W; \tau^i, d^i\right)$ is differentiable almost everywhere $\mathbb{P}$-a.s. with
		\begin{align}\label{eq:derivativetildeV}
		\frac{\partial \tilde{V}^i\left( 0, W; \tau^i, d^i \right)}{\partial W} = \mathbb{E}\left[ e^{-r \rho\left(W;\tau^i, d^i \right)} \right], \quad \text{a.e..}
		\end{align}
	\end{lemma}
	
	\begin{proof}[Proof of Lemma \ref{lemma:derivativeoftildeV}]
		Let $\delta > 0$. Observe that
		\begin{align*}
		\tilde{V}(0, W+\delta; \tau^i, d^i) &= \mathbb{E} \left[ \int_{0}^{\rho(W+\delta; \tau^i, d^i)} e^{-rt} h^i_{s} ds + e^{-\rho(W+\delta; \tau^i, d^i)} (W+\delta) \right] \\
		& \geq \mathbb{E} \left[\int_{0}^{\rho(W; \tau^i, d^i)} e^{-rt} h^i_{s} ds + e^{-\rho(W; \tau^i, d^i)} (W+\delta) \right] \\
		& = \tilde{V}(0, W; \tau^i, d^i) + \delta \mathbb{E}\left[ e^{-r \rho(W; \tau^i, d^i)}\right].
		\end{align*}
		Similarly,
		\begin{align*}
		\tilde{V}(0, W -\delta; \tau^i, d^i) & \geq \mathbb{E} \left[ \int_{0}^{\rho(W; \tau^i, d^i)} e^{-rt} h^i_{s} ds + e^{-\rho(W; \tau^i, d^i)} (W-\delta)\right] \\
		& = \tilde{V}(0, W; \tau^i, d^i) - \delta \mathbb{E}\left[ e^{-r \rho(W; \tau^i, d^i)}\right].
		\end{align*}
		Therefore
		\begin{align*}
		\frac{\tilde{V}^i(0, W; \tau^i, d^i) - \tilde{V}^i(0, W-\delta;\tau^i, d^i)}{\delta} &\leq \mathbb{E}\left[ e^{-r \rho(W;\tau^i, d^i)}\right] \\ 
		& \qquad \leq \frac{\tilde{V}^i(0, W+\delta;\tau^i, d^i) - \tilde{V}^i(0, W;\tau^i, d^i)}{\delta}.
		\end{align*}
		By Alexandrov's Theorem (Theorem 7.28 in \cite{guide2006infinite}), $W \to \tilde{V}^i(0, W + \delta; \tau^i, d^i)$ is differentiable almost everywhere and it derivative is continuous almost everywhere. Letting $\delta \to 0$, for almost every $W$, we then have
		\begin{align*}
		\frac{\partial \tilde{V}^i(0, W; \tau^i, d^i)}{\partial W} \leq \mathbb{E}\left[ e^{-r \rho(W; \tau^i, d^i)}\right] \leq \frac{\partial \tilde{V}^i(0, W; \tau^i, d^i)}{\partial W}.
		\end{align*}
		This concludes the proof.
	\end{proof}
	
	\paragraph{} For all $i \in \left\{1,\dots, N\right\}$, define the $h^i$-index processes
	\begin{align*}
	{\Gamma}^i_{t} \left( {h}^i \right) \coloneqq \inf\left\{ W \geq 0 \, : \, \tilde{V}^i\left( u, W; \tau^i, d^i \right) =W \text{ for } u = \sup\left\{ y \leq t \, : \, U^i_{y}(\tau^i, d^i) = 0\right\} \right\},
	\end{align*}
	and their lower envelope $\underline{\Gamma}^i_{t} \left( {h}^i\right) \coloneqq \underset{0 \leq s \leq t}{\inf }\,{\Gamma^i_s(h^i)}$.
	
	\begin{lemma}\label{lemma:strategiclowerenvelope}
		For all $W$, 
		\begin{align*}
		\tilde{V}^i\left( 0, W;\tau^i, d^i \right) = \mathbb{E} \left[ \int_{0}^{\rho({W}; \tau^i, d^i)} {h}^i_{s} d s + e^{-r \rho({W}; \tau^i, d^i)} W\right] = \mathbb{E}\left[ \int_{0}^{\rho({W}; \tau^i, d^i)} e^{-rt} r \underline{\Gamma}^i_{t} \left( {h}^i\right) dt + e^{-r\rho({W}; \tau^i, d^i)} W\right].\\
		\end{align*}
	\end{lemma}
	
	\begin{proof}[Proof of Lemma \ref{lemma:strategiclowerenvelope}]
		By Lemma \ref{eq:derivativetildeV}, 
		\begin{align*}
		\frac{\partial \tilde{V}^i\left( 0, W;\tau^i, d^i \right)}{\partial W} = \mathbb{E}\left[ e^{-r \rho\left( W; \tau^i, d^i \right)} \right] \text{a.e.}.
		\end{align*}
		By Lemma \ref{lemma:tildeVisconvex}, $\tilde{V}^i\left( 0, W; \tau^i, d^i \right)$ is locally Lipschitz, hence absolutely continuous. Therefore
		\begin{align*}
		\tilde{V}^i\left( 0, \bar{W};\tau^i, d^i \right) - \bar{W} - \tilde{V}^i\left( 0, W; \tau^i, d^i \right) +W & = \int_{W}^{\bar{W}}\left( \mathbb{E}\left[ e^{-r \rho\left( \tilde{W}; \tau^i, d^i \right)} \right] -1\right) d\tilde{W}.
		\end{align*}
		Letting $\bar{W} \to \infty$ and using Lemma \ref{lemma:tildeVisconvex}, we get
		\begin{align*}
		\tilde{V}^i\left( 0, W; \tau^i, d^i \right) -W  = \int_{W}^{\infty}\left(1 - \mathbb{E}\left[ e^{-r \rho\left( \tilde{W}; \tau^i, d^i \right)} \right]\right) d\tilde{W}.
		\end{align*}
		Then
		\begin{align*}
		\tilde{V}^i\left( 0, W;\tau^i, d^i \right) -W & = \mathbb{E} \left[ \int_{W}^{\infty} \left( 1 - e^{-r \rho(\tilde{W}; \tau^i, d^i)} \right) d \tilde{W} \right] \\
		& = \mathbb{E} \left[ \int_{W}^{\infty} r \int_{0}^{\infty} e^{-r t} \mathbbm{1}_{\{ t \leq \rho\left(\tilde{W}; \tau^i, d^i \right)\}}  dt d \tilde{W} \right]\\
		& = \mathbb{E} \left[\int_{0}^{\infty} r e^{-r t} \int_{W}^{\infty}  \mathbbm{1}_{\{ t \leq \rho\left(\tilde{W}, \tau^i, d^i \right)\}}  d \tilde{W} dt \right],
		\end{align*}
		by Fubini's theorem (twice). But,
		\begin{align*}
		t \leq \rho(\tilde{W}; \tau^i, d^i) & \Leftrightarrow \forall s \leq t \text{ such that } U^i_s(\tau^i, d^i) = 0, \, \tilde{V}^i\left( s, \tilde{W}; \tau^i, d^i \right) > \tilde{W} \\
		& \Leftrightarrow  \forall s \leq t, \, \Gamma^i_s\left( h^i\right) > \tilde{W} \\
		& \Leftrightarrow \underline{\Gamma}^i_{t} \left( h^i\right) > \tilde{W}.
		\end{align*}
		Therefore
		\begin{align*}
		\tilde{V}^i\left( 0, W;\tau^i, d^i \right) - W& = \mathbb{E} \left[ \int_{0}^{\infty} r e^{-r t} \int_{W}^{\infty} \mathbbm{1}_{\{ \underline{\Gamma}^i_{t}(h^i) > \tilde{W} \}} d\tilde{W} dt \right] \\
		& = \mathbb{E} \left[ \int_{0}^{\infty} r e^{-r t} \left(\underline{\Gamma}^i_{t}(h^i) - W\right)^+ dt \right];
		\end{align*}
		and the results follows.
	\end{proof} 
	
	Finally,	
	\begin{proposition}[Whittle Computation]\label{prop:whittlecomputation}
		\begin{align*}
		\Phi(0,W) & \coloneqq \underset{(T,\bar{\tau}^1, \dots, \bar{\tau}^N), T^i(\bar{\tau}^i) \in \mathcal{T}^s\left( 0; \tau^i, d^i \right) }{\sup} \mathbb{E}\left[ \sum_{i=1}^N \int_{0}^{\bar{\tau}^i} e^{-rt} h^i_{T^i(t)}dT^i(t) + e^{-r \bigvee \bar{\tau}^i} W \right] \\ 
		& \leq \underset{T}{\sup } \, \mathbb{E}\left[ \sum_{i=1}^N \int_{0}^{\infty} e^{-rt} r \underline{\Gamma}^i_{T^i(t)} \left( h^i \right) \vee W dT^i(t)  \right].
		\end{align*}
	\end{proposition}
	
	\begin{proof}[Proof of Proposition \ref{prop:whittlecomputation}]
		This follows from the double inequality: on $\left\{ U^i_t\left( \tau^i, d^i \right) = 0 \right\}$,
		\begin{align*}
		\Phi^{-i}(\bar{t}^{-i}, W) \vee \tilde{V}^i\left( t^i, W; \tau^i, d^i \right) \leq \Phi(\bar{t}, W) \leq \Phi^{-i}(\bar{t}^{-i}, W) + \tilde{V}^i\left( t^i, W; \tau^i, d^i \right) - W.
		\end{align*}
		But then, on $\left\{ \tilde{V}^i\left( t,W; \tau^i, d^i\right) =W \right\} \cap \left\{ U^i_t\left( \tau^i, d^i \right) = 0 \right\}$, $\Phi^{-i}(\bar{t}^{-i}, W) = \Phi(\bar{t}, W)$, and it is optimal to retire arm $i$. It follows that the optimal stopping time $\tau(W)$ is weakly smaller than $\sum_{i}^N \rho^i\left( W; \tau^i, d^i \right)$. Then, by the same argument as in the proof of Lemma \ref{lemma:derivativeoftildeV},
		\begin{align*}
		\frac{\partial \Phi(0, W) }{\partial W} & = \mathbb{E} \left[ e^{-r \tau(W)} \right].
		\end{align*}
		Integrating, we get
		\begin{align*}
		\Phi(0, W) -W & = \int_{W}^{\infty} \left( 1-  \mathbb{E}\left[e^{-r \tau(\tilde{W})}\right] \right) d\tilde{W} \\
		& \leq \mathbb{E}\left[ \int_{W}^{\infty} \left( 1- e^{-r \sum_{i}^N \rho^i\left( \tilde{W}; \tau^i, d^i \right)} \right) d\tilde{W} \right]
		\end{align*}
		where the inequality follows from $\tau(W) \leq \sum_{i}^N \rho^i\left( W; \tau^i, d^i \right)$. By Theorem 3.7 in \cite{el1997synchronization}, 
		\begin{align*}
		\mathbb{E}\left[ \int_{W}^{\infty} \left( 1- e^{-r \sum_{i}^N \rho^i\left( \tilde{W}; \tau^i, d^i \right)} \right) d\tilde{W} \right] + W
		\end{align*}
		is the value of the bandit problem with decreasing rewards $\Gamma^i_t\left( h^i_t \right)$, and the result follows.
	\end{proof}
	
	\section{Appendix B}\label{app:appendixB}
	
	\subsection{The strategic index policy}\label{app:strategicindex}
	
	\paragraph{} We construct the index delegation rule $T^{\Gamma}$ associated with the index processes $\left( \Gamma^1,\dots, \Gamma^N \right)$ following \cite{el1997synchronization}. We define $T^{\Gamma}$ pointwise on $\Omega$. Let $\omega \in \Omega$, and define
	\begin{itemize}
		\item $\sigma^i(W) = \inf\left\{ t \geq 0 \, : \, \Gamma^i_t \leq W \right\}$.
		\item ${D}^i$ is the set of discontinuities of the function $W \to \sigma^i(W)$. ${D} \coloneqq \bigcup_{i=1}^N {D}^i$.
		\item $\mathcal{D}^i = \left\{ t \geq 0 \, : \, \sigma^i\left( \underline{\Gamma}^i_{t^-}\right) > t \right\} = \bigcup_{W \in {D}^i} [ \sigma^i(W), \sigma^i(W^-))$. The intervals in $\mathcal{D}^i$ are the flat stretches of the function $\underline{\Gamma}^i_t$. $\mathcal{D} \coloneqq \bigcup_{i=1}^N \mathcal{D}^i$.
		\item $B^i$ is the set of discontinuities of the function $t \to \underline{\Gamma}^i_t$. ${B} \coloneqq \bigcup_{i=1}^N {B}^i$.
		\item $\mathcal{B}^i \coloneqq \left\{ W > 0 \, : \, \underline{\Gamma}^i_{\sigma^i(W^-)} < W \right\} = \bigcup_{t \in B^i} ( \underline{\Gamma}^i_{t}, \underline{\Gamma}^i_{t^-}]$. The intervals in $\mathcal{B}^i$ are the flat stretches of the function $W\to {\sigma}^i(W)$. $\mathcal{B} \coloneqq \bigcup_{i=1}^N \mathcal{B}^i$.
		\item $\tau^0(W) = \sum_{i=1}^d \sigma^i(W)$, $0 \leq m < \infty$.
		\item $N(t) = \inf\left\{ W \geq 0 \, : \, \tau^0(W) \leq t \right\}$, $0 \leq t, W< \infty$.
	\end{itemize}
	For all $i = 1,\dots,N$, and all $t \not \in \mathcal{D}$, define
	\begin{align*}
	T^{\Gamma,i}(t) \coloneqq \sigma^i(N(t)^-).
	\end{align*}
	For $t\in \mathcal{D}$, we still need to decide which arm to pull in the case that more than one arm achieves the highest index. In that case, we specify a priority rule: if the indices of two or more workers are the same at a time point of discontinuity, the principal delegates to the worker with smallest $i$. Formally, for $t \in \mathcal{D}$, observe that $t \in [\tau^0(W), \tau^0(W^-))$ with $W = N(t) \in D$. Define then 
	\begin{align*}
	y^0 = y^0(W) \coloneqq \tau^0(W),
	\end{align*}
	and, recursively,
	\begin{align*}
	y^i \coloneqq y^i(W) \coloneqq y^{i-1}(W) - \Delta\sigma^i(W) = \sum_{j=1}^i \sigma^j(W^-) + \sum_{j=i+1}^N \sigma^j(W),
	\end{align*}
	where $\Delta \sigma^i(W) \coloneqq \sigma^i(W) - \sigma^i(W^-)$. Set $L^i(W) \coloneqq [y^{i-1}(W), y^i(W))$, so that $L(m) = \bigcup_{i=1}^N L^i(m)$. In particular, $y^N = \tau^0(W^-)$, and $L^i(W) = \emptyset$ if $\sigma^i$ is continuous at $W$. Now find the unique $k = k(t) \in \left\{ 1,\dots, N\right\}$ for which $t \in L^k(m)$, and write
	\begin{align*}
	\sum_{i=1}^N T^{\Gamma,i}(t) = (t - y^{k-1}) + y^{k-1} = \sum_{j=1}^{k-1} \sigma^j(W^-) + (t-y^{k-1} + \sigma^k(W)) \sum_{j=k+1}^N \sigma^j(W).
	\end{align*}
	
	We then take 
	\begin{align}\label{eq:strategicindexpolicy}
	T^{\Gamma,i}(t) \coloneqq \begin{cases}
	\sigma^i(W^-) \text{ if } i =1,\dots, k(t) -1 \\
	\sigma^i(W) + t - y^{k(t)-1}  \text{ if } i = k(t) \\
	\sigma^i(W) \text{ if } i = k(t) +1, \dots, N
	\end{cases}
	\end{align}
	for $t \in \mathcal{D}$.
	
	\begin{proposition}\label{prop:strategicindexiswelldefined}
		The vector $T^{\Gamma}$ is an index delegation rule associated with the index processes $\left( \Gamma^1,\dots, \Gamma^N \right)$.
	\end{proposition}
	
	\begin{proof}[Proof of proposition \ref{prop:strategicindexiswelldefined}]
		We show that $T^{\Gamma}(t)$ is flat off the set 
		\begin{align*}
		\left\{ t \geq 0 \, : \, \underline{\Gamma}^i_{T^{\Gamma,i}(t)} = \bigvee_{j=1}^N \underline{\Gamma}^j_{T^{\Gamma,j}(t)} \right\} \, \mathbb{P}\text{-a.s..}
		\end{align*}
		Observe that, by construction, for all $W \geq 0$,
		\begin{align*}
		\bigvee_{j=1}^N \underline{\Gamma}^j_{T^{\Gamma,j}(t)} \leq W & \Leftrightarrow \underline{\Gamma}^j_{T^{\Gamma,j}(t)} \leq W \text{ for all } i \Leftrightarrow \sigma^i(W) \leq T^{\Gamma,i}(t) \text{ for all } i \\
		& \Rightarrow \tau^0(W) \leq t \Leftrightarrow N(t) \leq W \text{ for all } 0\leq t.
		\end{align*}
		Moreover, by construction, we also have, for all $t \geq 0$,
		\begin{align*}
		\underline{\Gamma}^j_{\sigma^i(N(t)^-)} \leq \underline{\Gamma}^j_{T^{\Gamma,i}(t)} \leq \underline{\Gamma}^j_{\sigma^i(N(t))} \leq N(t);
		\end{align*}
		using that $T^{\Gamma,i}$ is nondecreasing, since $T^{\Gamma,i }(\tau^0(W)) = \sigma^i(W)$ for all $W \geq 0$ implies that $T^{\Gamma,i }(\tau^0(N(t))) \leq T^{\Gamma,i }(t) \leq T^{\Gamma,i }(\tau^0(N(t)^-))$. 
		
		So $\bigvee_{j=1}^N \underline{\Gamma}^j_{T^{\Gamma,j}(t)} \leq N(t)$, and, thus, $\bigvee_{j=1}^N \underline{\Gamma}^j_{T^{\Gamma,j}(t)} = N(t)$. 
		
		Then $N(t) \not \in \mathcal{B}^i \Rightarrow \underline{\Gamma}^i_{T^{\Gamma,i}(t)} = N(t)$, and, therefore,
		\begin{align*}
		0 \leq \int_{0}^{\infty} \mathbbm{1}_{\{ \underline{\Gamma}^i_{T^{\Gamma,i}(t)} < \bigvee_{j=1}^N \underline{\Gamma}^j_{T^{\Gamma,j}(t)} \}} d T^{\Gamma,i}(t) = \int_{0}^{\infty} \mathbbm{1}_{\{ \underline{\Gamma}^i_{T^{\Gamma,i}(t)} < N(t) \}} d T^{\Gamma,i}(t) \leq \int_{0}^{\infty} \mathbbm{1}_{\{ N(t) \in \mathcal{B}^i \}} d T^{\Gamma,i}(t) =0,
		\end{align*}
		where the last equality holds as $N(t) \in \mathcal{B}^i$ implies that $T^{\Gamma,i}(t^-) = \sigma^i(N(t)^-) = \sigma^i(N(t)) = T^{\Gamma,i}(t)$ is flat at $t$. Then
		\begin{align*}
		\sum_{i=1}^N \int_{0}^{\infty} \mathbbm{1}_{\{ \underline{\Gamma}^i_{T^{\Gamma,i}(t)} < \bigvee_{j=1}^N \underline{\Gamma}^j_{T^{\Gamma,j}(t)} \}} d T^{\Gamma,i}(t) =0,
		\end{align*}
		and the result follows.
	\end{proof}
	
	\paragraph{} Next we recall one classic results from the study of bandit problems, which is used in the proof of Proposition \ref{prop:strategicbanditcontest}. 
	
	\begin{lemma}\label{lemma:indexstrategiesexpectedvalue}
		For all $i = 1,\dots, N$,
		\begin{align}\label{eq:indexstrategiesexpectedvalue}
		\tilde{\mathbb{E}} \left[ \int_{0}^{\infty} e^{-rt} h^{s,i}_{T^{s,i}(t)} dT^{s,i}(t)  \right] = {\mathbb{E}} \left[ \int_{0}^{\infty} e^{-r t} r \underline{\Gamma}^{s,i}_{T^{s,i}(t)} dT^{s,i}(t) \right].
		\end{align}
	\end{lemma}
	
	A similar statement is used in the proof of Theorem 8.1 in \cite{el1994dynamic}; and a proof follows from the arguments there and from Lemma 7.5 in \cite{el1997synchronization}. We reproduce it below for completeness.
	
	\begin{proof}[Proof of lemma \ref{lemma:indexstrategiesexpectedvalue}]
		By proposition 3.2 in \cite{el1994dynamic},
		\begin{align*}
		U^i_t & = e^{-rt}\left[ V^i\left(t; \underline{\Gamma}^{s,i}_t\right) - \underline{\Gamma}^{s,i}_t\right] + \int_{0}^{t} e^{-ru}  \left( h^{s,i}_u - r \underline{\Gamma}^{s,i}_u \right) du 
		\end{align*} 
		is a $\mathcal{F}^i$-martingale with c\`{a}dl\`{a}g paths, and, hence, by lemma 4.6 in \cite{el1997synchronization} an $\tilde{\mathcal{F}}^i$-martingale. Then
		\begin{align*}
		\tilde{\mathbb{E}} \left[ \int_{0}^{\infty} e^{-r(t - T^{s,i}(t))} dU^i_{T^{s,i}(t)} \right] & = \tilde{\mathbb{E}} \bigg[ \int_{0}^{\infty} e^{-rt} \left( h^{s,i}_{T^{s,i}(t)} - r \underline{\Gamma}^i_{T^{s,i}(t)} \right) dT^{s,i}(t) \\
		& \qquad + \int_{0}^{\infty} e^{-r(t - T^{s,i}(t))} d\left( e^{-rT^{s,i}(t)} \left( V^i\left( {T^{s,i}(t)}; \underline{\Gamma}^i_{T^{s,i}(t)}  \right) - \underline{\Gamma}^i_{T^{s,i}(t)} \right)\right) \bigg]  \\
		& = 0.
		\end{align*}
		Observe that, by lemma 7.5 in \cite{el1997synchronization} and the definition of the strategic index policy,
		\begin{align*}
		\int_{0}^{\infty} e^{-r(t - T^{s,i}(t))} & d\left( e^{-rT^{s,i}(t)} \left( V^i\left( {T^{s,i}(t)}; \underline{\Gamma}^i_{T^{s,i}(t)}  \right) - \underline{\Gamma}^i_{T^{s,i}(t)} \right)\right) \\ 
		& = \sum_{m \in {D}^i}  \int_{y_{i-1}(m)}^{y_i(m)} e^{-r(t - T^{s,i}(t))} d\left( e^{-rT^{s,i}(t)} \left( V^i\left( {T^{s,i}(t)}; \underline{\Gamma}^i_{T^{s,i}(t)}  \right) - \underline{\Gamma}^i_{T^{s,i}(t)} \right)\right)\\
		& = \sum_{m \in {D}^i}  e^{-r(y_{i-1}(m) - \sigma^i(m))} \left( e^{-rT^{s,i}(t)} \left( V^i\left( {T^{s,i}(t)}; \underline{\Gamma}^i_{T^{s,i}(t)}  \right) - \underline{\Gamma}^i_{T^{s,i}(t)} \right)\right)\bigg|_{t = y_{i-1}(m)}^{t = y_i(m)} \\
		& = \sum_{m \in {D}^i}  e^{-ry_{i-1}(m)} \bigg( e^{-r\Delta \sigma^i(m)} \left( V^i\left( {T^{s,i}(y_i(m))}; \underline{\Gamma}^i_{T^{s,i}(y_i(m))}  \right) - \underline{\Gamma}^i_{T^{s,i}(y_i(m))} \right) \\
		& \qquad \qquad \qquad \qquad \qquad \qquad - \left( V^i\left( {T^{s,i}(y_{i-1}(m))}; \underline{\Gamma}^i_{T^{s,i}(y_{i-1}(m))}  \right) - \underline{\Gamma}^i_{T^{s,i}(y_{i-1}(m))} \right) \bigg).
		\end{align*}
		Lemma 7.5 in \cite{el1997synchronization} again implies that $\mathbb{P}$-a.s.
		\begin{align*}
		V^i\left( {T^{s,i}(y_{k-1}(m))}; \underline{\Gamma}^i_{T^{s,i}(y_{k-1}(m))}  \right) - \underline{\Gamma}^i_{T^{s,i}(y_{k-1}(m))} = 0 = V^i\left( {T^{s,i}(y_k(m))}; \underline{\Gamma}^i_{T^{s,i}(y_k(m))}  \right) - \underline{\Gamma}^i_{T^{s,i}(y_k(m))}.
		\end{align*}
		Therefore
		\begin{align*}
		\tilde{\mathbb{E}} \left[ \int_{0}^{\infty} e^{-r(t - T^{s,i}(t))} dU^i_{T^{s,i}(t)} \right] & = \tilde{\mathbb{E}} \left[ \int_{0}^{\infty} e^{-rt} \left( h^{*,i}_{T^{s,i}(t)} - r \underline{\Gamma}^i_{T^{s,i}(t)} \right) dT^{s,i}(t) \right] \\
		& = 0;
		\end{align*}
		and \eqref{eq:indexstrategiesexpectedvalue} holds.
	\end{proof}
	
	\subsection{Useful Results on Optimal Stopping}\label{app:usefulresultsoptimalstopping}
	
	\paragraph{} I will be interested in the following problem. Let $Y_t \in D_r$. Consider \begin{align}\label{eq:generalstoppingproblem}
	\begin{split}
	\underset{\tau\in \mathcal{T}}{\sup }\, \mathbb{E} \left[ Y_\tau \right].
	\end{split}
	\end{align} 
	
	\begin{lemma}\label{lemma:optimalstoppingcharacterization}
		$\tau$ solves (\ref{eq:generalstoppingproblem}) if and only if, for all $\tau' \leq \tau$,
		\begin{align*}
		\mathbb{E} [Y_\tau \mid \mathcal{F}_{\tau'}] \geq \mathbb{E} [Y_{\tau'} \mid \mathcal{F}_\tau'],
		\end{align*} 
		and, for all $\tau' \geq \tau$,
		\begin{align*}
		\mathbb{E} [Y_\tau \mid \mathcal{F}_{\tau}] \geq \mathbb{E} [Y_{\tau'} \mid \mathcal{F}_\tau].
		\end{align*} 
	\end{lemma}
	
	\begin{proof}[Proof of lemma \ref{lemma:optimalstoppingcharacterization}]
		($\Rightarrow$) This is immediate. To see this, observe that the contrapositive is the following. Suppose that there exists $\tau' \leq \tau$ such that
		\begin{align*}
		\mathbb{E} [Y_\tau \mid \mathcal{F}_{\tau'}] < \mathbb{E} [Y_{\tau'} \mid \mathcal{F}_\tau'],
		\end{align*} 
		or $\tau' \geq \tau$ such that
		\begin{align*}
		\mathbb{E} [Y_\tau \mid \mathcal{F}_{\tau}] < \mathbb{E} [Y_{\tau'} \mid \mathcal{F}_\tau],
		\end{align*} 
		then $\tau$ does not solve (\ref{eq:generalstoppingproblem}), which is obviously true.
		
		($\Leftarrow$) Let $\tau$ be a Markov time and suppose that, for all $\tau' \leq \tau$,
		\begin{align*}
		\mathbb{E} [Y_\tau \mid \mathcal{F}_{\tau'}] \geq \mathbb{E} [Y_{\tau'} \mid \mathcal{F}_{\tau'}],
		\end{align*} 
		and, for all $\tau' \geq \tau$,
		\begin{align*}
		\mathbb{E} [Y_\tau \mid \mathcal{F}_{\tau}] \geq \mathbb{E} [Y_{\tau'} \mid \mathcal{F}_\tau].
		\end{align*}
		Let $\tilde{\tau}$ be any Markov time. I then have
		\begin{align*}
		\mathbb{E}[Y_\tau] & = \mathbb{E}\left[Y_\tau \mathbbm{1}_{\{\tau \leq \tilde{\tau}\}} + Y_\tau \mathbbm{1}_{\{\tau > \tilde{\tau}\}} \right] \\
		& = \mathbb{E}\left[ \mathbb{E}\left[ Y_{\tau}  \mid \mathcal{F}_\tau \right] \mathbbm{1}_{\{\tau \leq \tilde{\tau}\}}+ \mathbb{E}\left[ Y_\tau \mid \mathcal{F}_{\tilde{\tau}}\right] \mathbbm{1}_{\{\tau > \tilde{\tau}\}}  \right] \\
		& \geq \mathbb{E}\left[ \mathbb{E}\left[ Y_{\tau\vee \tilde{\tau}}  \mid \mathcal{F}_\tau \right] \mathbbm{1}_{\{\tau \leq \tilde{\tau}\}}+ \mathbb{E}\left[ Y_{\tau\wedge \tilde{\tau}} \mid \mathcal{F}_{\tilde{\tau}}\right] \mathbbm{1}_{\{\tau > \tilde{\tau}\}} \right]\\
		& = \mathbb{E}\left[ \mathbb{E}\left[ Y_{\tilde{\tau}}  \mathbbm{1}_{\{\tau \leq \tilde{\tau}\}} \mid \mathcal{F}_\tau \right]+ \mathbb{E}\left[ Y_{\tilde{\tau}} \mathbbm{1}_{\{\tau > \tilde{\tau}\}} \mid \mathcal{F}_{\tilde{\tau}}\right] \right] \\
		& = \mathbb{E}\left[ Y_{\tilde{\tau}}\right]
		\end{align*}
		where the second equality follows from the law of iterated expectations and the $\mathbb{F}_{\tau} \wedge \mathcal{F}_{\tilde{\tau}}$-measurability of $\mathbbm{1}_{\{\tau > \tilde{\tau}\}}$ and $\mathbbm{1}_{\{\tau \leq \tilde{\tau}\}}$, the inequality follows by assumption, the third equality follows from the same measurability conditions, and the last equality from the law of iterated expectation again. Since $\tilde{\tau}$ was arbitrary, $\tau$ solves \eqref{eq:generalstoppingproblem}.
	\end{proof}
	
	\paragraph{}  Finally, I need the following extension of Theorem 2.4 in \cite{peskir2006optimal}, which relaxes the assumption on G and V.
	\begin{theorem}[Theorem 2.4 in \cite{peskir2006optimal} revisited]\label{theorem:MarkovstoppingPeskir2.4}
		Let $X$ be a c\`{a}dl\`{a}g Feller process with values in $\mathcal{X}$ defined on the filtered probability space $\left( \Omega, \left\{ \mathcal{F}_t \right\}_{t \geq 0}, \mathcal{F}, \mathbb{P}\right)$, where $\left\{ \mathcal{F}_t \right\}_{t \geq 0}$ satisfies the usual conditions. Let $\pi$, $G$ be measurable function from $\mathcal{X}$ to $\mathbb{R}$ such that
		\begin{align*}
		\mathbb{E} \left[ \underset{0 \leq t \leq \infty}{\sup }\int_{0}^t e^{-rt} \pi(X_t) dt + e^{-rt} G(X_t) \right] < \infty.
		\end{align*}
		Consider the family of optimal stopping problem
		\begin{align}\label{eq:generaloptimalstopping}
		V_t \coloneqq ess\underset{\tau \geq t}{\sup }\, \mathbb{E}\b igg[ \int_{t}^{\tau}e^{-r(s-t)} p(X_s) ds + e^{-r(\tau-t)} G(X_\tau)\mid \mathcal{F}_t\bigg],
		\end{align}
		where the sup is taken over all $\mathcal{F}_t$-Markov time $\tau\geq t$. Then $V_t = V(X_t)$ $\mathbb{P}$-a.s., where
		\begin{align*}
		V(X_t) \coloneqq ess \underset{\tau \geq t}{\sup } \, \mathbb{E} \left[ \int_{t}^{\tau} e^{-r(s-t)} p(X_s) ds + e^{-r(\tau-t)} G(X_{\tau}) \mid X_t \right],
		\end{align*}
		and, for all Markov time $\theta \geq t$,
		\begin{align*}
		V_t = V(X_t) = ess\underset{\tau \geq t}{\sup }\, \mathbb{E}\bigg[ \int_{t}^{\tau\wedge \theta} e^{-r(s-t)}  p(X_s) ds +e^{-r(\tau\wedge\theta -t)} \left( V(X_{\theta}) \mathbbm{1}_{\{\theta \leq \tau\}} + G(X_{\tau})  \mathbbm{1}_{\{\tau < \theta\}}  \right) \mid X_t \bigg].
		\end{align*}
		Finally the Markov time
		\begin{align*}
		\tau^*_t \coloneqq \inf\left\{ s \geq t \, : \, V(X_s) = G(X_s) \right\},
		\end{align*}
		is the smallest optimal Markov time for $V_t$.
	\end{theorem}
	
	\begin{proof}[Proof of theorem \ref{theorem:MarkovstoppingPeskir2.4}]
		Consider the family of optimal stopping problems
		\begin{align}\label{eq:familyoflagrangianstopping}
		V_t \coloneqq ess\underset{\tau \geq t}{\sup }\, \mathbb{E}\bigg[ \int_{t}^{\tau}e^{-r(s-t)} p(X_s) ds + e^{-r(\tau-t)} G(X_\tau)\mid \mathcal{F}_t\bigg],
		\end{align}
		Note that $V_t$ is a well-defined $\mathcal{F}_t$-measurable random variable by lemma 1.3 in \cite{peskir2006optimal}, so that the process $\{V_t\}_{t\geq0}$ is adapted.
		
		Define
		\begin{align*}
		Z_t \coloneqq \int_{0}^{t} e^{-rs} p(X_s) ds + e^{-rt} V_t.
		\end{align*}
		$Z_t$ is the Snell envelope of (i.e., the smallest supermartingale with càdlàg paths that dominates) the process
		\begin{align*}
		Y_t \coloneqq & \int_{0}^{t} e^{-rs}p(X^i_s) ds + e^{-rt} G(X_{t}).
		\end{align*}
		By theorem 2.2 in \cite{peskir2006optimal}, 
		\begin{align*}
		\tau^*_t & \coloneqq \inf \left\{ s \geq t \, : \, Y_s = Z_s  \right\} \\
		& = \inf \left\{ s \geq t \, : \, V_s = G(X_{s}) \right\}
		\end{align*}
		is the smallest optimal Markov time for the problem (\ref{eq:familyoflagrangianstopping}). Furthermore, 
		\begin{align*}
		\left\{ Z_{s \wedge \tau^*_t}, \, \mathcal{F}_s, \, t \leq s \leq \infty \right\}
		\end{align*}
		is a martingale. The processes $Y$ and $Z$ are c\`{a}dl\`{a}g, progressively measurable, nonnegative, and agrees at $t = \infty$. Furthermore $\mathbb{E}\left[ \underset{0\leq t \leq \infty }{\sup }Y_t \right] < \infty$. It follows that $Z$ is class $D$, hence uniformly integrable.
		
		By the optional sampling theorem for uniformly integrable c\`{a}dl\`{a}g martingale (Theorem 7.29 in \cite{kallenberg2006foundations}), for all Markov time $\theta\geq t$, 
		\begin{align}\label{eq:dynamicprogramming}
		V_t = ess\underset{\tau \geq 0}{\sup }\, \mathbb{E}\bigg[ \int_{t}^{\tau\wedge \theta}e^{-r(s-t)} p(X_s) ds + e^{-r(\tau\wedge\theta-t)} \left( V_{\theta} \mathbbm{1}_{\{\theta \leq \tau\}} +  G(X_{s}) \mathbbm{1}_{\{\tau < \theta\}}  \right) \mid \mathcal{F}_t\bigg].
		\end{align}
		
		But, by the strong Markov property for Feller processes (Theorem 19.17 in \cite{kallenberg2006foundations}), for all stopping time $\tau < \infty$,
		\begin{align*}
		V_\tau = V(X_\tau) \, \mathbb{P}\text{-a.s..}
		\end{align*}
		In particular, $V_\tau$ does not depend on the history prior to $\tau$, and equation (\ref{eq:dynamicprogramming}) becomes, for all Markov time $\theta\geq t$, $\theta \indep \mathcal{F}^1_{t^-}$,
		\begin{align*}
		V(X_t) = ess\underset{\tau \geq t}{\sup }\, \mathbb{E}\bigg[ \int_{t}^{\tau\wedge \theta} e^{-r(s-t)}  p(X_s) ds +e^{-r(\tau\wedge\theta-t)} \left( V(X_{\theta}) \mathbbm{1}_{\{\theta \leq \tau\}} + G(X_{\tau})  \mathbbm{1}_{\{\tau < \theta\}}  \right) \mid X_t \bigg].
		\end{align*}
		Finally $\tau^*_t$ becomes
		\begin{align*}
		\tau^*_t = \inf \left\{ s \geq t \, : \, V(X_s) = G(X_{\tau})  \right\}.
		\end{align*}
		This concludes the proof.
	\end{proof}
	
	\section{Online Appendix}\label{app:onlineapp}
	
	\subsection{Omitted Proofs for Section \ref{subsec:relaxingassumptionstypeprocess}}\label{app:relaxingassumptions}
	
	\begin{proof}[Proof of Corollary \ref{corollary:nontrivial}]
		Suppose that Assumption \ref{assumption:nontrivial} is violated. Then there exists $i \in \left\{ 1,\dots, N\right\}$ such that 
		\begin{align*}
		\underset{(\tau,d) \in \mathcal{P}^{I,r,i}}{\sup}\, \mathbb{E}& \left[ \int_{0}^{\tau} e^{-rt} \pi^i\left( X^i_t \right) dt + e^{-r\tau} \left( (1-d^0_{\tau}) \int_{\tau}^{\infty} e^{-r(t-\tau)}\pi^i\left( X^i_t\right)dt +d^0_{\tau} W \right) \right]\\
		& \qquad \leq \mathbb{E}\left[ \int_{0}^{\infty} e^{-rt} \pi^i\left( X^i_t \right) dt \right].
		\end{align*}
		Let $\epsilon >0$. Define $W^i$ as 
		\begin{align*}
		W^i \coloneqq \inf\left\{ W \, : \, \text{Assumption \ref{assumption:nontrivial} holds with } W  \right\},
		\end{align*}
		and let $\tilde{W} = \bigvee_{i=1}^N W^i + \epsilon > W$. For $\tilde{W}$, Assumption \ref{assumption:nontrivial} holds. So, by Theorem \ref{theorem:indexability}, the optimal implementable promotion contest is the index contest. Letting $\epsilon \to 0$, we see that the index contest is also optimal with outside option $\tilde{W}$. Finally observe that for $i$ such that $W^i =\bigvee_{i=1}^N W^i$, it must be that, for all $x$ such that $\mathbb{P}\left( \tau^i_{(-\infty, x ]} < \infty \right)>0$, with $\tau^i_{(-\infty, x ]} \coloneqq \inf\left\{ t\geq 0 \, : \, X^i_t \leq x \right\}$, 
		\begin{align*}
		\mathbb{E}\left[ \int_{0}^{\infty} e^{-rt} \pi^i\left( X^i_t \right) dt \mid X^i_0 = x \right] \geq W^i.
		\end{align*}
		But $\Gamma^{s,i}_t \geq \mathbb{E}\left[ \int_{0}^{\infty} e^{-rt} \pi^i\left( X^i_t \right) dt \mid X^i_t \right]$ for all $t\geq 0$, $\mathbb{P}$-a.s.. Therefore, the principal never takes the outside option $\tilde{W}$, and thus the index contest is optimal in the original problem too.
	\end{proof}
	
	\begin{proof}[Proof of Corollary \ref{corollary:potentialforimprovment}]
		For simplicity, suppose that Assumption \ref{assumption:potentialforimprovment} holds for all $j \neq i$. Suppose that, for all $n\in \mathbb{N}$, $X^{i,n}$ satisfies Assumption \ref{assumption:potentialforimprovment}, and that $X^i = \underset{n\to \infty}{\lim} X^{i,n}$ uniformly on compact sets $\mathbb{P}$-a.s.. Let  $\Gamma^{s,i,n}$ be the strategic index process associated with worker $i$ and $\tau^{s,i,n}$ his promotion time when his type process is given by $X^{i,n}$. Define also $\tau^{s,i} \coloneqq \inf\left\{ t\geq 0 \, :\, X^i_t \geq \bar{P}^i\left( \underline{X}^i_t \right) \right\}$. Observe then that, for almost all $t\geq 0$, 
		\begin{align*}
		\underline{\Gamma}^{s,i,n}_t \to \Gamma^{s,i}_t, \quad \mathbb{P}{-a.s.,}
		\end{align*}
		where
		\begin{align*}
		r\Gamma^{s,i}_t \coloneqq \underset{\tau >0}{\sup } \frac{\mathbb{E}\left[ \int_{t}^{\tau} e^{-r(s-t)} \pi^i\left( X^i_s \right) \mathbbm{1}_{\{ s \leq \tau^{s,i}  \}}  + \bar{\pi}^i\left( X^i_{\tau^{s,i}} \right)  \mathbbm{1}  \mid \mathcal{F}^i_t \right]}{\mathbb{E}\left[ \int_{t}^{\tau} e^{-r(s-t)} ds \mid \mathcal{F}^i_t \right]}.
		\end{align*}
		For all $n \in \mathbb{N}$, by Theorem \ref{theorem:indexability}, the principal's value is given by 
		\begin{align*}
		\mathbb{E} \left[ \int_{0}^{\tau^n} e^{-rt} \pi^i\left( X^i_{T^{i}_n(t)}\right) dT^{i}_n(t) + \sum_{j\neq i} \int_{0}^{\tau^n} e^{-rt} \pi^j\left( X^j_{T^{j}_n(t)}\right) dT^{j}_n(t) + e^{-r\tau^n} \bar{\pi}\left( X^n_{T_n(\tau^n)}, d^n \right)\right],
		\end{align*}
		where $T_{n}$ is the index rule associated with the indices $\Gamma^{s,j}$'s, $j\neq i$, and $\Gamma^{s,i,n}$, \\ $\tau^n \coloneqq \inf\left\{ t \geq 0\, : \, T^j(t) \geq \tau^{s,j} \text{ or } T^i(t) \geq \tau^{s,i,n}  \right\}$, and $d^{n}$ is the optimal promotion rule.
		
		Next, observe that if there exists $x$ such that $\bar{P}^i(x) = x$, $\tau^{s,i,n} \not \to \tau^{s,i}$. To see this, simply note that, for all $n \in \mathbb{N}$, $\mathbb{P}\left( \tau^i =0 \mid \underline{X}^i_t =x \right) =0$. But, by Lemma 2 in \cite{pennanen2018optimal}, $\tau^{s,i,n} \to \tau^i \in \mathcal{S}\left( \mathcal{F}^i \right)$, at least along a subsequence. 
		
		So, as $n \to \infty$, passing to a subsequence if necessary, $T_n \to T$ where $T$ is the index rule associated with the strategic indices $\Gamma^{s,j}$'s, $\tau^n \to \tau^*$ with $\tau^* = \tau^{i} \bigwedge_{j\neq i} \tau^{s,j}$, and $d^n \to d$ with $d^i_{t} = 1$ only if $T^{i}(t) \geq \tau^i$ and $d^j_{t} = 1$ only if $T^{j}(t) \geq \tau^{s,j}$, $j\neq i$. By Theorem 6.39 in \cite{guide2006infinite} and the Lebesgue dominated convergence theorem,
		\begin{align*}
		\mathbb{E} &\left[ \int_{0}^{\tau^n} e^{-rt} \pi^i\left( X^i_{T^{i}_n(t)}\right) dT^{i}_n(t) + \sum_{j\neq i} \int_{0}^{\tau^n} e^{-rt} \pi^j\left( X^j_{T^{j}_n(t)}\right) dT^{j}_n(t) + e^{-r\tau^n} \bar{\pi}\left( X^n_{T^n(\tau^n)}, d^n \right)\right] \\
		& \to \mathbb{E}\left[ \sum_{j=1}^N \int_{0}^{\tau^*} e^{-rt} \pi^j\left( X^j_{T^{j}(t)}\right) dT^{j}(t) + e^{-r\tau} \bar{\pi}\left( X_{T(\tau^*)}, d \right)\right].
		\end{align*}
		Note that the principal's value is continuous in $X^i$. This is easily deduced as the difference in values is bounded by
		\begin{align*}
		\mathbb{E}\left[ \int_{0}^{\infty}e^{-rt} \left| \pi^i\left(X^i_t\right) - \pi^i\left( X^{i,n}_t \right) \right| dt \right].
		\end{align*}
		Therefore, the randomized promotion contest $(T, \tau^*, d)$ is optimal. Finally, one easily deduce from both the optimality and the limit characterization that $\tau^i = \inf\{ t\geq 0 \, : \,  X^i_t > \bar{P}^i(\underline{X}^i_t) \} \wedge \tau^{p,i}$, where $\tau^{p,i}$ is the first tick of a Poisson clock that runs only when $X^i_t = \bar{P}^i(\underline{X}^i_t) \}$ with the intensity that leaves $i$ indifferent between exerting effort or not if promoted at time $\tau^{i}$.
	\end{proof}
	
	\subsection{Proof of Theorem \ref{theorem:winnertakeall}}\label{app:winnertakeall}
	
	\paragraph{} As in the proof of \ref{theorem:indexability}, consider the relaxed program:
	\begin{align}\label{eq:relaxedprogramprizedesign}
	\Pi \coloneqq \underset{(T, \left\{ \tau_k \right\}_{k=1}^N, d)\in \mathcal{P}^r}{\sup}\, & \mathbb{E}\Bigg[ \sum_{i=1}^N\int_{0}^\tau e^{-rt} \pi^i\left(X^i_{T^i(t)}\right)dT^i(t) + e^{-r\tau} \bar{\pi}\left( X_{T(\tau)}, d \right)\Bigg] \tag{RP(P-d)}
	\end{align}
	subject to, for all $i \in \left\{ 1,\dots, N \right\}$, for all $t \geq 0$, $\mathbb{P}$-a.s.,
	\begin{align}\label{eq:DPCprizedesign}
	\mathbb{E}\left[ \sum_{k=1}^{K} e^{-r(\tau_k-t)} g d^i_{\tau_k} \mathbbm{1}_{\{ t\leq \tau_k \}} - \int_{t}^{\infty} e^{-rt} (1- \sum_{k=1}^{K}d^i_{\tau_k} \mathbbm{1}_{\{t\geq \tau_k\}}) c^i dT^i(t)  \mid \mathcal{\mathcal{F}}^i_{T^i(t)} \right] \geq 0. \tag{DPC}
	\end{align}
	By the same arguments as in the proof of Proposition \ref{prop:relaxation},
	\begin{proposition}\label{prop:relaxationprize}
		The value of \eqref{eq:prizedesign} is weakly lower than the value of \eqref{eq:relaxedprogram}: $\Pi^M \leq \Pi$.
	\end{proposition}
	Next, a straightforward adaptation of the proof of Theorem \ref{theorem:existencerelaxedprogram} in Appendix \ref{app:existence} yields:
	\begin{theorem}\label{theorem:existencerelaxedprogramprizedesign}
		A solution to \eqref{eq:relaxedprogramprizedesign} exists.
	\end{theorem}
	
	\paragraph{} Finally, I show that any promotion contest that allocate the entire prize upon at once can be improved upon. This follows from Proposition \ref{prop:winnertakeall} below.
	\begin{proposition}\label{prop:winnertakeall}
		\eqref{eq:relaxedprogramprizedesign} admits a solution $(T,\left\{ \tau_k \right\}_{k=1}^K,d)$ such that $K =1$ $\mathbb{P}$-a.s..
	\end{proposition}
	Theorem \ref{theorem:winnertakeall} then follows from Theorem \ref{theorem:indexability}.
	
	\begin{proof}[Proof of Proposition \ref{prop:winnertakeall}]
		Let $\left(T, \left\{ \tau_k \right\}_{k=1}^K,d\right)$ be a solution of \eqref{eq:relaxedprogramprizedesign}, which exists by Theorem \ref{theorem:existencerelaxedprogramprize}. $\tau_1$ is the smallest promotion time. The continuation value of the principal at $\tau_1$ is
		\begin{align*}
		e^{-r\tau_i}\Pi^M_{\tau_1} \coloneqq & e^{-r \tau_1} \bar{\pi}\left( X_{T(\tau_1)}, d_{\tau_1} \right) \\
		& + \mathbb{E} \left[ \sum_{k=2}^{K} \sum_{i=1}^N \left( \int_{\tau_{k-1}}^{\tau_{k}} e^{-rt} \pi^i\left(X^i_{T^i(t)}\right) dT^i(t) +  e^{-r\tau_k} \bar{\pi}\left( X_{T(\tau_k)}, d_{\tau_k} \right)  \right) \mid \mathcal{G}^T_{\tau_1} \right].
		\end{align*}
		By Assumption \ref{assumption:submartingalepayoffs}(i), 
		\begin{align*}
		\bar{\pi}\left( X_{T(\tau_1)}, d_{\tau_1} \right) \leq d^0_{\tau_1} W + \sum_{i=1}^N d^i_{\tau_1} \bar{\pi}^i\left( X^i_{T^i(\tau_1)}\right).
		\end{align*}
		If $\sum_{i=0}^Nd^i_{\tau} =1$, we are done. So suppose not. Observe then that
		\begin{align}\label{eq:prizedesignproof}
		\mathbb{E}& \left[ \sum_{k=2}^{K} \sum_{i=1}^N \left( \int_{\tau_{k-1}}^{\tau_{k}} e^{-rt} \pi^i\left(X^i_{T^i(t)}\right) dT^i(t) +  e^{-r\tau_k} \bar{\pi}\left( X_{T(\tau_k)}, d_{\tau_k} \right)  \right) \mid \mathcal{G}^T_{\tau_1} \right]\\
		& \leq (1-\sum_{i=0}^N d^i_{\tau_1}) \underset{\left(T,\left\{\tau_k\right\},d\right) \in \mathcal{P}^{I,r}(\tau_1)}{\sup} \, \,\mathbb{E}\left[  \sum_{k=1}^{K} \sum_{i=1}^N \left( \int_{\tau_{k-1}}^{\tau_{k}} e^{-rt} \pi^i\left(X^i_{T^i(t)}\right) dT^i(t) +  e^{-r\tau_k} \bar{\pi}\left( X_{T(\tau_k)}, d_{\tau_k} \right)  \right) \mid \mathcal{G}^T_{\tau_1} \right],
		\end{align}
		where $\mathcal{P}^{I,r}(\tau_1)$ is the set of implementable continuation contest that coincides with $(T,\tau,d)$ up to time $\tau_1$.
		To see this, let ${T_*}$ be the continuation delegation process generated by $\left(T, \left\{ \tau_k \right\}_{k=1}^K,d\right)$ after time $\tau^1$ so that :
		\begin{align*}
		\sum_{k=2}^{K} & \sum_{i=1}^N \left( \int_{\tau_{k-1}}^{\tau_{k}} e^{-rt} \pi^i\left(X^i_{T^i(t)}\right) dT^i(t) +  e^{-r\tau_k} \bar{\pi}\left( X_{T(\tau_k)}, d_{\tau_k} \right)  \right) \\
		& = \sum_{i=1}^N \int_{0}^{\infty} e^{-rt} \pi^i\left(X^i_{T^i(\tau_1) + {T}_*^i(t)}\right) d{T}_*^i(t) \, \quad \mathbb{P}\text{-a.s..} 
		\end{align*}
		Then, letting $X^i_0$ be any process taking value in $\mathcal{X}^0$ and $\pi^0(x) =W$ for all $x \in \mathcal{X}^0$, 
		\begin{align*}
		\underset{\left(T,\left\{\tau_k\right\},d\right) \in \mathcal{P}^{I,r}(t)}{\sup} \, \,\mathbb{E}& \left[  \sum_{k=1}^{K} \sum_{i=1}^N \left( \int_{\tau_{k-1}}^{\tau_{k}} e^{-rt} \pi^i\left(X^i_{T^i(t)}\right) dT^i(t) +  e^{-r\tau_k} \bar{\pi}\left( X_{T(\tau_k)}, d_{\tau_k} \right)  \right) \mid \mathcal{G}^T_{\tau_1} \right] \\
		& \geq E^{-r\tau_1} \mathbb{E} \left[ \sum_{i=0}^N \int_{0}^{\infty} e^{-rt} \pi^i\left(X^i_{T^i(\tau_1) +\frac{{T}_*^i(t)}{1-\sum_{i=0}^N d^i_{\tau_1}}}\right) d\left( \frac{T^i_*(t)}{1-\sum_{i=0}^N d^i_{\tau_1}}\right) \mid \mathcal{G}^T_{\tau_1} \right], \\
		\end{align*}
		as $\frac{T^i_*(t)}{1-\sum_{i=0}^N d^i_{\tau_1}}$ is implementable by a promotion contest when the total information in the game is restricted to $\mathcal{G}^{T_*}$ and, more information benefits the principal. So 
		\begin{align*}
		(1- \sum_{i=0}^N d^i_{\tau_1}) &\underset{\left(T,\left\{\tau_k\right\},d\right) \in \mathcal{P}^{I,r}(t)}{\sup} \, \,\mathbb{E} \left[  \sum_{k=1}^{K} \sum_{i=1}^N \left( \int_{\tau_{k-1}}^{\tau_{k}} e^{-rt} \pi^i\left(X^i_{T^i(t)}\right) dT^i(t) +  e^{-r\tau_k} \bar{\pi}\left( X_{T(\tau_k)}, d_{\tau_k} \right)  \right) \mid \mathcal{G}^T_{\tau_1} \right] \\
		& \geq e^{-r\tau_1} \mathbb{E} \left[ \sum_{i=0}^N \int_{0}^{\infty} e^{-rt} \pi^i\left(X^i_{T^i(\tau_1) + \frac{{T}_*^i(t)}{1-\sum_{i=0}^N d^i_{\tau_1}}}\right) d {T^i_*(t)}\mid \mathcal{G}^T_{\tau_1} \right] 
		\end{align*} 
		But, by a time-change argument, for ${q}_*^i(t) \coloneqq e^{-r\left(  {T^i_*}^{-1}(t)-t \right)}$, where $ {T^i_*}^{-1}(\cdot)$ is the generalized inverse of $T_*^i(\cdot)$,
		\begin{align*}
		\mathbb{E} \left[ \sum_{i=0}^N \int_{0}^{\infty} e^{-rt} \pi^i\left(X^i_{T^i(\tau_1) + T_*^i(t)}\right) dT_*^i(t) \mid \mathcal{G}^T_{\tau_1} \right] = \mathbb{E} \left[ \sum_{i=0}^N \int_{0}^{\infty} e^{-rt} q_*^i(t) \pi^i\left(X^i_{T^i(\tau_1) + t}\right) dt \mid \mathcal{G}^T_{\tau_1}\right],
		\end{align*}
		and
		\begin{align*}
		\mathbb{E}& \left[ \sum_{i=0}^N \int_{0}^{\infty} e^{-rt} \pi^i\left(X^i_{T^i(\tau_1) + \frac{{T}_*^i(t)}{1-\sum_{i=0}^N d^i_{\tau_1}}}\right) d {T^i_*(t)} \mid \mathcal{G}^T_{\tau_1} \right] \\
		& = \mathbb{E} \left[ \sum_{i=0}^N \int_{0}^{\infty} e^{-rt} q^i_*(t) \pi^i\left(X^i_{T^i(\tau_1) +\frac{t}{1-\sum_{i=0}^N d^i_{\tau_1}}}\right) dt \mid \mathcal{G}^T_{\tau_1} \right].
		\end{align*}
		By definition $q^i$ is $\mathcal{G}^{{T}_*}$-adapted. Furthermore, for all $i \in \left\{ 1,\dots, N\right\}$, 
		\begin{align}\label{eq:supproofprizedesign}
		\underset{\tau \in \mathcal{T}(\mathcal{F}^i)}{\sup } \, \mathbb{E}\left[ \int_{0}^{\tau} e^{-rt} \left(  \pi^i\left(X^i_{t}\right) - \pi^i \left(X^i_{\frac{t}{1- \sum_{i=0}^N d^i_{\tau_1}}} \right) \right)dt \mid x^i_0 = X^i_{T^i(\tau)}\right] =0.
		\end{align}
		Hence, I claim that $\tau^* = 0$ is optimal in the above problem. To see this, argue by contradiction, i.e., suppose not. Then the smallest optimal stopping time, which exists by Snell's theorem, is $\tilde{\tau} >0$. By Lemma \ref{lemma:optimalstoppingcharacterization}, there exists $t\geq 0$ such that
		\begin{align*}
		\mathbb{E}\left[ \int_{t}^{\tilde{\tau}} e^{-rs} \left( \pi^i\left( X^i_s \right) - \pi^i \left(X^i_{\frac{s}{1- \sum_{i=0}^N d^i_{\tau_1}}} \right) \right) ds \mid \mathcal{F}^i_t \right] >0.
		\end{align*}
		The above inequality is equivalent to 
		\begin{align*}
		&\int_{t}^{\infty} e^{-rs} \mathbb{E}\left[ \left( \pi^i\left( X^i_s \right) - \pi^i \left(X^i_{\frac{s}{1- \sum_{i=0}^N d^i_{\tau_1}}} \right) \right) \mathbbm{1}_{\{ s\leq \tilde{\tau}\}} \mid \mathcal{F}^i_t \right]ds  >0 \\
		\Leftrightarrow & \int_{t}^{\infty} e^{-rs} \mathbb{E}\left[ \mathbb{E} \left[\left( \pi^i\left( X^i_s \right) - \pi^i \left(X^i_{\frac{s}{1- \sum_{i=0}^N d^i_{\tau_1}}} \right) \right) \mid \mathcal{F}^i_s \right] \mathbbm{1}_{\{ s\leq \tilde{\tau}\}} \mid \mathcal{F}^i_t \right]ds  > 0
		\end{align*}
		by Fubini's theorem, the law of iterated expectations, and the fact that $\tilde{\tau}$ is a $\mathcal{F}^i$-stopping time. But, for all $s \in [t,\infty)$,
		\begin{align*}
		\mathbb{E} \left[ \pi^i\left( X^i_s \right) \mid \mathcal{F}^i_s \right] \leq \mathbb{E} \left[\pi^i \left(X^i_{\frac{s}{1- \sum_{i=0}^N d^i_{\tau_1}}} \right) \mid \mathcal{F}^i_s \right]
		\end{align*}
		by Assumption \ref{assumption:submartingalepayoffs} (i): a contradiction. So, for all $i \in \left\{1,\dots, N\right\}$, \eqref{eq:supproofprizedesign} holds. Similarly, \eqref{eq:supproofprizedesign} is easily seen to hold for $i =0$. Thus by Lemma 5 in \cite{kaspi1998multi}, 
		\begin{align*}
		& \mathbb{E}\left[ \sum_{i=0}^N\int_{0}^{\infty} e^{-rt} q^i_*(t) \left(  \pi^i\left(X^i_{T^i(\tau_1) +t}\right) - \pi^i \left(X^i_{\frac{t}{1- \sum_{i=0}^N d^i_{\tau_1}}} \right) \right)dt \right] \leq 0 \\
		\Leftrightarrow\mathbb{E} &\left[ \sum_{i=0}^N \int_{0}^{\infty} e^{-rt} \pi^i\left(X^i_{T^i(\tau_1) + T_*^i(t)}\right) dT_*^i(t) \mid \mathcal{G}^T_{\tau_1} \right] \leq \mathbb{E} \left[ \sum_{i=0}^N \int_{0}^{\infty} e^{-rt} \pi^i\left(X^i_{T^i(\tau_1) + \frac{{T}_*^i(t)}{1-\sum_{i=0}^N d^i_{\tau_1}}}\right) d {T^i_*(t)} \mid \mathcal{G}^T_{\tau_1} \right].
		\end{align*}
		Thus \eqref{eq:prizedesignproof} holds. But then, the randomized promotion contest that promotes worker $i$ at time $\tau_1$ with probability $d^i_{\tau_1}$ and otherwise play the optimal continuation contest yields a higher payoffs to the principal than $(T, \left\{ \tau_k \right\}_{k=1}^K, d)$. This concludes the proof.
	\end{proof}

\end{document}